\documentclass[useAMS,usenatbib]{mn2e}
\usepackage{graphicx}
 \usepackage{subfig} 
\usepackage{lastpage}
\usepackage{txfonts}
\def\hii {H\,{\sc ii}}

\def\kms{km\,s$^{-1}$}

\def\d {$^{\circ}$}

\title[Variability of methanol masers]{Monitoring observations of 6.7\,GHz methanol masers}
\author[M. Szymczak et al.]
{M. Szymczak \thanks{E-mail: msz@astro.umk.pl},
M. Olech,
R. Sarniak,
P. Wolak
and A. Bartkiewicz
\\
Centre for Astronomy, Faculty of Physics, Astronomy and Informatics, Nicolaus Copernicus University,\\
 Grudziadzka 5, 87-100 Torun, Poland \\
}

\begin{document}


\pagerange{\pageref{firstpage}--\pageref{LastPage}} \pubyear{2017}

\maketitle

\label{firstpage}
\begin{abstract}
We report results of 6.7\,GHz methanol maser monitoring of 139 star-forming sites with the Torun 32\,m radio telescope from 
June 2009 to February 2013. The targets were observed at least once a month, with higher cadences of 2-4 measurements per week 
for circumpolar objects. Nearly 80\,per~cent of the sources display variability greater than 10\,per~cent on a time-scale between 
a week and a few years but about three quarters of the sample have only 1-3 spectral features which vary significantly.
Irregular intensity fluctuation is the dominant type of variability and only nine objects show evidence for cyclic variations with 
periods of 120 to 416\,d. Synchronised and anti-correlated variations of maser features are detected in four sources with a disc-like morphology. 
Rapid and high amplitude bursts of individual features are seen on 3-5 occasions in five sources. Long ($>$50\,d to 20 months) lasting 
bursts are observed mostly for individual or groups of features in 19 sources and only one source experienced a remarkable global flare. 
A few flaring features display a strong anti-correlation between intensity and line-width that is expected for unsaturated amplification.
There is a weak anti-correlation between the maser feature luminosity and variability measure, i.e. maser features with low luminosity
tend to be more variable than those with high luminosity. The analysis of the spectral energy distribution and continuum radio emission 
reveals that the variability of the maser features increases when the bolometric luminosity and Lyman flux of the exciting object decreases.
Our results support the concept of a major role for infrared pumping photons in triggering outburst activity of maser emission. 
\end{abstract}


\section{Introduction}
High mass young stellar objects (HMYSOs) exert significant influence on their environments creating a variety of structures and phenomena 
such as stellar winds and jets, expanding \hii\, regions, outflows and shock waves (\citealt{zinnecker07}). One of the phenomena intimately 
linked to high-mass star formation at very early stages is the methanol maser emission (e.g. \citealt{breen10}).
The most strong and widespread maser transition at 6.7\,GHz probes dense ($10^4-10^8$\,cm$^{-3}$) and relatively cool ($<$150\,K) gas 
with methanol fractional abundance greater than $10^{-7}$ (\citealt{cragg05}). The estimated lifetime of a methanol maser is 2$-$5$\times10^4$~yr
(\citealt{vanderwalt05}), which is comparable to the time-scale of the chemical evolution of methanol species in hot cores 
(e.g. \citealt{rodgers03}) and the dynamical ages of molecular outflows associated with HMYSOs (e.g. \citealt{shepherd96}). 
The radio continuum observations suggest that the methanol masers trace the earliest stages of the (proto)stellar evolution (\citealt{walsh98}; 
\citealt{beuther02}; \citealt{urquhart13a}).

The individual maser spectral features are composed of one to a few clouds of a linear size from several to tens of au. 
The median size of the methanol maser clouds observed for a sample of about 60 sources (\citealt{bartkiewicz09}, 
\citealt*{bartkiewicz14, bartkiewicz16}) is only 5.5\,au. The methanol clouds are distributed in larger structures of various shapes 
and sizes up to several hundred au. Proper motion studies of a few objects indicate that those clouds may trace molecular discs, outflows 
or even in-fall of a molecular envelope (\citealt{goddi11}; \citealt{moscadelli11b}; \citealt{sanna10a, sanna10b}; \citealt{sugiyama14}).
In this paper we report on the variability characteristics of 6.7\,GHz methanol masers most of which have been studied at high angular resolution.

Early information on the variability of methanol masers came primarily from surveys aimed at searching for new sources
(\citealt*{caswell95a}). \cite{caswell95b} observed a sample of 245 objects on 4-5 occasions over a period of 1.5\,yr, finding 
that 75\,per cent of spectral features were not significantly variable, while noticeable variability, with an amplitude of
usually less than a factor of two, was reported only in 48 sources (\citealt{caswell95a}). They have suggested that 
the variations can be related to changes in the maser path length or pump rate.
A comparison of data sets for 384 masers from the untargeted survey of the Galactic plane in the longitude range 6 to 60\d\,
taken at two epochs separated by about 2\,yr for a majority of sources (\citealt{green10}; \citealt{breen15}) implies that 
the peak flux densities differ by a factor of two or greater for 5$-$7\,per cent of the sources. In a sample of 204 masers
observed mostly in the same Galactic area but at two epochs spanned about 9\,yr the peak flux or integrated flux densities changed 
by a factor of two or greater in about 30\,per cent of the targets (\citealt{szymczak12}).  

The largest variability survey to date is that of \cite{goedhart04}, which probed the 6.7\,GHz flux density in 54 sources over 
a period of 4.2\,yr with cadences of 2-4 observations per month or daily observations of rapidly variable targets. The survey
showed a remarkable variability of different patterns on time-scales of a few days to several years for a majority 
of the targets. One of their spectacular findings was the detection of strictly periodic (132$-$520\,d) variability in seven sources. 
More periodic masers have since been found (\citealt{goedhart09, goedhart14}; \citealt{araya10}; 
\citealt{szymczak11, szymczak15}; \citealt{fujisawa14b}; \citealt{maswanganye15, maswanganye16}). This class of masers has been
intensively studied to investigate the origin of periodic variability (e.g. \citealt{vanderwalt11}; \citealt{parfenov14}). 
One of a plausible cause is periodic modulation of the accretion rate in the circumstellar or protobinary disc
(\citealt{araya10}; \citealt{parfenov14}). The interaction of colliding winds in a binary system may vary the background flux 
of the seed photons which periodically modulate the methanol maser flux density (\citealt{goedhart09}; \citealt{vanderwalt11};
\citealt{maswanganye15}).

There are several studies of peculiar variability at 6.7\,GHz in individual sources. \cite{macleod96} found a flare activity of 
G351.78$-$0.54 on time-scales of a few months superimposed on long-term variations on a time-scale of years that can be caused
by changes in the maser pumping or disturbances within the maser regions. \cite{fujisawa14c} reported episodic and short 
($<$6\,d) bursts of only one feature in the spectrum of G33.641$-$0.228. These bursts arise in a region much smaller that 70\,au 
possibly as a result of magnetic energy release. Synchronized and anti-correlated variations of the 6.7\,GHz methanol maser 
flux density of the blue- and red-shifted features were reported in Cep\,A (\citealt{sugiyama08}; \citealt{szymczak14}). 
The same source also exhibited periodic (84-87\,d) variations of individual features for about 290\,d interval and faint flare 
events of duration of 1.4$-$1.8\,yr (\citealt{szymczak14}). 

In this paper, we present full data from a monitoring of 166 sources. Our goal is to characterise the long-term behaviour 
of 6.7\,GHz masers in attempt to study possible origins for the variability and its usefulness in probing the environment of HMYSOs. 
Results of the monitoring of individual objects were partly published in \cite{szymczak11, szymczak14, szymczak15}.

\section{Observations and data analysis}
 \subsection{Sample}
   \label{sec:sample}
A sample of 139 star-forming sites with peak 6.7\,GHz flux density greater than 5\,Jy and declination above $-$11\degr\,
was selected from the Torun methanol source catalogue (TMSC, \citealt{szymczak12}). These criteria allowed observations
of luminous targets well above the elevation limit of the telescope. The target list is given in Table~\ref{sample}
which assigns a name to each site based on its Galactic coordinates. The positions of almost all (138/139) masers
in the sample are derived from various interferometric measurements. For these sites the position offset between
the interferometric coordinates and the actual pointing coordinates are given. 
The inspection of interferometric maps reveals at least 20 sites in the sample which have two or more 
compact maser groups separated each other by more than 1\arcsec. It is likely that for a typical distance of 5\,kpc 
individual maser groups separated by more than 5000\,au are associated with physically unrelated objects. 
The number of sources towards these star-forming sites is given as a superscript to their names. The data from various 
interferometers (references to Table~\ref{sample}) indicate that our sample contains at least 166 distinct maser sources.
The fifth column of Table~\ref{sample} gives the systemic velocity chosen as the peak velocity of the optically 
thin molecular lines from \cite*{szymczak07} and \cite*{shirley13} or in the worst case as the middle velocity 
of the methanol maser profile. The last column of Table~\ref{sample} provides the number of distinct sources included or excluded
from the further analysis due to causes discussed in Section~\ref{sec:offset-confusion}.

\begin{figure}   
\resizebox{\hsize}{!}{\includegraphics[angle=0]{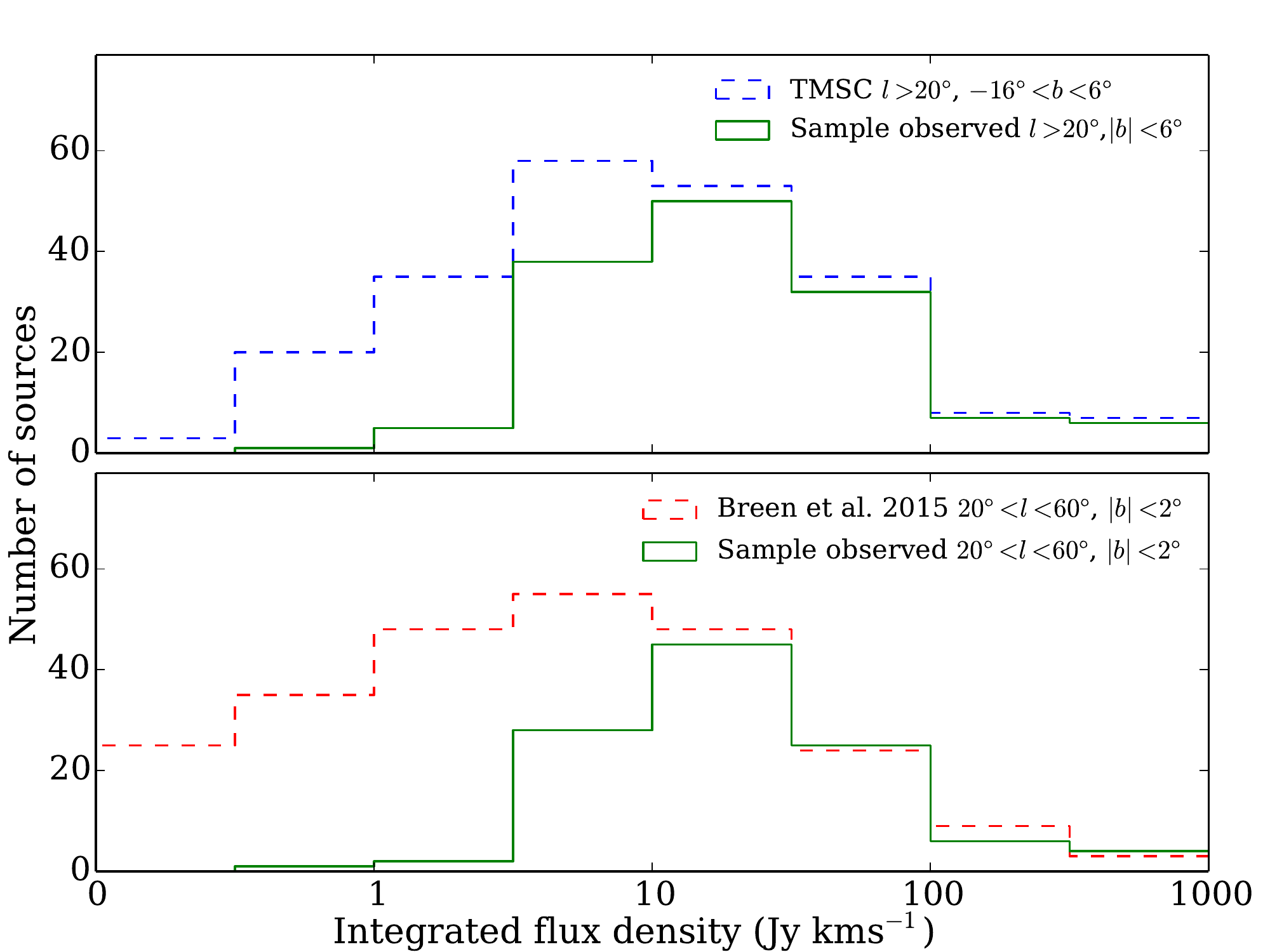}}
\caption{Histograms of the integrated  methanol flux density for sources in the present sample and two comparison samples 
from the TMSC (\citealt{szymczak12}, upper panel) and the Methanol Multibeam Survey (\citealt{breen15}, lower panel). 
\label{histo-sample}}
\end{figure}

The monitoring sample contains 63.5~per cent of TMSC sites at Galactic longitude $l>$20\degr.
92.2 and 65.5~per cent of TMSC objects with $S_{\mathrm {int}}>$10\,Jy\,\kms\, and $3<S_{\mathrm {int}}<$10\,Jy\,\kms, 
are included, respectively (Fig.~\ref{histo-sample}). Here, $S_{\mathrm {int}}$ is the integrated flux density.
\cite{breen15} made an unbiased survey the Galactic range $20^\circ<l<60^\circ$ and $|b|<$2\fdg0, finding 80 sources 
with $S_{\mathrm {int}} >$ 10\,Jy\,\kms\,  out of the 84 targets in common with our sample.
In this region 50.9~per cent of objects with $3< S_{\mathrm {int}}<$10\,Jy\,\kms\, are also in the monitoring sample
(Fig.~\ref{histo-sample}). This implies that our flux-limited sample is nearly complete in the region studied 
by \cite{breen15}. 

\begin{table}
 \caption{List of 6.7\,GHz methanol maser sites monitored. The site name superscript gives the reference for coordinates 
          derived from interferometric observations. The pointing coordinates are given for other site. Sites with multiple 
          maser sources (clusters) are marked with a $\dagger$ and the number of sources is given. The fourth column lists 
          the offset of the actual pointing from the maser position. The fifth column lists the systemic velocity. 
          The note indicates whether the object was included (Y) or excluded (N) from statistical analysis and their number
          for clusters is added.
\label{sample}}
\begin{tabular}{@{}l @{\hspace{1.5mm}}l @{\hspace{2.5mm}}r @{\hspace{2.5mm}}c @{\hspace{2.5mm}}c @{\hspace{2.5mm}}r}
\hline
Name            & RA(2000)         & Dec(2000)                   & Offset    & $V_\mathrm{s}$ & Note\\
                & (h      m      s) & ($^{\circ}$      '       ") & (')    & (km\,s$^{-1}$)  & \\
\hline
G20.237+0.065$^{a{\dagger}2}$   & 18 27 44.56      & $-$11 14 54.1               & 0.168    &  71.4$^K$ &     2Y    \\ 
G21.407$-$0.254$^b$ & 18 31 06.34      & $-$10 21 37.4               & 0.001    &  90.7$^{1,K}$  &   Y      \\ 
G21.563$-$0.033$^c$ & 18 30 36.07      & $-$10 07 10.9               & 0.469    & 113.9$^{2,K}$  &   Y      \\
G22.039+0.222$^d$   & 18 30 34.70      & $-$09 34 47.0               & 0.436    &  51.0$^{2,K}$  &   Y      \\
G22.335$-$0.155$^b$ & 18 32 29.41      & $-$09 29 29.7               & 0.004    &  30.9$^{1,K}$  &   Y      \\
G22.357+0.066$^b$   & 18 31 44.12      & $-$09 22 12.3               & 0.021    &  84.2$^{1,K}$  &   Y      \\ 
G22.435$-$0.169$^a$ & 18 32 43.82      & $-$09 24 32.8               & 0.658    &  28.3$^2$      &   Y  \\
G23.010$-$0.411$^a$ & 18 34 40.29      & $-$09 00 38.1               & 0.042    &  77.6$^{2,T}$  &   Y      \\
G23.207$-$0.378$^b$ & 18 34 55.21      & $-$08 49 14.9               & 0.015    &  79.2$^{2,T}$  &   Y      \\
G23.257$-$0.241$^c$ & 18 34 31.26      & $-$08 42 46.7               & 0.012    &  62.4          &   Y  \\
G23.389+0.185$^b$   & 18 33 14.32      & $-$08 23 57.5               & 0.001    &  75.8$^2$      &   Y  \\ 
G23.437$-$0.184$^{e{\dagger}2}$ & 18 34 39.27      & $-$08 31 39.3               & 0.074    & 106.6$^T$ &     2Y      \\
G23.484+0.097$^c$   & 18 33 44.05      & $-$08 21 20.6               & 0.345    &  83.4$^2$      &   Y  \\
G23.657$-$0.127$^b$ & 18 34 51.56      & $-$08 18 21.3               & 0.001    &  80.2$^{2,T}$  &   Y      \\
G23.707$-$0.198$^b$ & 18 35 12.37      & $-$08 17 39.4               & 0.001    &  68.9$^{1,T}$  &   Y      \\
G23.966$-$0.109$^{b{\dagger}3}$ & 18 35 22.21      & $-$08 01 22.5               & 0.001    &  72.7$^{1,K}$  &      1Y     \\
G24.148$-$0.009$^b$ & 18 35 20.94      & $-$07 48 55.7               & 0.004    &  23.1$^{1,K}$  &   Y      \\
G24.329+0.144$^f$   & 18 35 08.09      & $-$07 35 03.6               & 0.060    & 113.2$^{2,K}$  &   Y       \\
G24.494$-$0.038$^a$ & 18 36 05.73      & $-$07 31 19.2               & 0.845    & 109.2$^2$      &   Y  \\ 
G24.541+0.312$^b$   & 18 34 55.72      & $-$07 19 06.7               & 0.009    & 107.8$^{2,K}$  &   Y      \\
G24.634$-$0.324$^b$ & 18 37 22.71      & $-$07 31 42.1               & 0.163    &  42.7$^{1,K}$  &   Y      \\
G24.790+0.083$^{a{\dagger}4}$   & 18 36 12.56      & $-$07 12 10.8               & 0.036    & 109.5$^T$     &    1Y      \\
G25.411+0.105$^b$   & 18 37 16.92      & $-$06 38 30.5               & 0.042    &  96.0$^{1,K}$  &   Y      \\
G25.650+1.049$^c$   & 18 34 20.90      & $-$05 59 42.2               & 0.310    &  41.7$^K$      &   Y  \\
G25.710+0.044$^c$   & 18 38 03.15      & $-$06 24 14.9               & 0.025    &  99.2$^{2,T}$  &   Y      \\
G25.826$-$0.178$^c$ & 18 39 03.63      & $-$06 24 09.7               & 0.005    &  92.3$^{2,K}$  &   Y       \\
G26.527$-$0.267$^c$ & 18 40 40.26      & $-$05 49 12.9               & 0.090    & 105.8$^{2,K}$  &   Y      \\
G26.598$-$0.024$^b$ & 18 39 55.93      & $-$05 38 44.6               & 0.001    &  23.3$^{1,K}$  &   Y      \\
G26.601$-$0.221$^c$ & 18 40 38.57      & $-$05 44 01.6               & 0.091    & 107.5$^{2,K}$  &   Y      \\
G27.221+0.136$^b$   & 18 40 30.55      & $-$05 01 05.4               & 0.000    & 112.6$^1$      &   Y  \\
G27.286+0.151$^c$   & 18 40 34.51      & $-$04 57 14.4               & 0.017    &  31.4$^{2,K}$  &   Y      \\
G27.365$-$0.166$^c$ & 18 41 51.06      & $-$05 01 43.5               & 0.238    &  91.3$^{2,T}$  &   Y      \\
G28.011$-$0.426$^a$ & 18 43 57.97      & $-$04 34 24.1               & 1.063    &  19.3$^{2,K}$  &   Y      \\
G28.146$-$0.005$^f$ & 18 42 42.59      & $-$04 15 36.5               & 0.076    &  98.7$^{2,K}$  &   Y      \\
G28.282$-$0.359$^c$ & 18 44 13.26      & $-$04 18 04.9               & 0.036    &  48.4$^2$      &   Y  \\
G28.305$-$0.387$^c$ & 18 44 21.99      & $-$04 17 38.4               & 0.003    &  85.5$^{2,K}$  &   Y     \\
G28.817+0.365$^b$   & 18 42 37.35      & $-$03 29 40.9               & 0.002    &  87.0$^{1,K}$  &   Y      \\
G28.832$-$0.253$^c$ & 18 44 51.09      & $-$03 45 48.7               & 0.671    &  87.9$^{2,K}$  &   Y      \\
G29.863$-$0.044$^g$ & 18 45 59.57      & $-$02 45 04.4               & 0.262    & 100.9$^{2,T}$  &   Y      \\
G29.955$-$0.016$^{c{\dagger}2}$ & 18 46 03.74      & $-$02 39 22.2               & 0.000    &  99.9$^T$  &   1Y      \\
G30.198$-$0.169$^{a{\dagger}2}$ & 18 47 03.07      & $-$02 30 36.3               & 0.118    & 107.6$^K$  &   2Y      \\
G30.317+0.070$^b$   & 18 46 25.03      & $-$02 17 40.8               & 0.076    &  45.3$^{1,K}$  &   Y      \\
G30.400$-$0.296$^{b{\dagger}2}$ & 18 47 52.30      & $-$02 23 16.0               & 0.131    & 102.4$^1$  &   1Y       \\
G30.703$-$0.068$^c$ & 18 47 36.82      & $-$02 00 53.8               & 0.121    &  88.0$^2$      &   N  \\
G30.760$-$0.052$^c$ & 18 47 39.78      & $-$01 57 23.4               & 0.281    &  90.3$^K$      &   N  \\
G30.780+0.230$^c$   & 18 46 41.52      & $-$01 48 37.1               & 1.750    &  41.6$^2$      &   N  \\ 
G30.788+0.204$^c$   & 18 46 48.09      & $-$01 48 53.9               & 0.516    &  83.5$^{2.K}$  &   Y      \\%
G30.818$-$0.057$^{c{\dagger}2}$ & 18 47 46.97      & $-$01 54 26.4               & 0.293    &  94.7$^{2,K}$  &   2Y      \\
G30.898+0.161$^c$   & 18 47 09.13      & $-$01 44 11.1               & 0.471    & 105.5$^{2,K}$  &    Y     \\
G31.047+0.356$^b$   & 18 46 43.86      & $-$01 30 54.2               & 0.652    &  77.6$^{1,K}$  &    Y      \\
G31.059+0.093$^c$   & 18 47 41.35      & $-$01 37 26.2               & 0.279    &  16.8$^2$      &    Y \\
G31.158+0.046$^{c{\dagger}2}$   & 18 48 02.40      & $-$01 33 26.8               & 0.141    &  38.9$^1$  &    1Y       \\
G31.281+0.061$^c$   & 18 48 12.43      & $-$01 26 30.1               & 0.613    & 107.0$^{2,T}$  &    Y     \\
G31.412+0.307$^h$   & 18 47 34.29      & $-$01 12 45.6               & 0.356    &  98.3$^K$      &    Y \\
G31.581+0.077$^b$   & 18 48 41.94      & $-$01 10 02.5               & 0.329    &  96.0$^{1,T}$  &    Y     \\
\hline
\end{tabular}
\end{table}

\begin{table}
 \contcaption{ 
\label{tab: sample continued}}
\begin{tabular}{@{}l @{\hspace{1.5mm}}l @{\hspace{2.5mm}}r @{\hspace{2.5mm}}c @{\hspace{2.5mm}}c @{\hspace{1.5mm}}r}
\hline
Name            & RA(2000)         & Dec(2000)                   & Offset    & $V_\mathrm{s}$ & Note\\
                & (h     m      s) & ($^{\circ}$      '       ") & (')    & (km\,s$^{-1}$) & \\
\hline
G32.045+0.059$^c$   & 18 49 36.56      & $-$00 45 45.9               & 0.189    &  93.6$^{2,T}$  &    Y     \\
G32.744$-$0.076$^a$ & 18 51 21.87      & $-$00 12 05.3               & 0.024    &  36.4$^{2,K}$  &    Y     \\
G32.992+0.034$^b$   & 18 51 25.58      &    00 04 08.3               & 0.758    &  83.4$^{1,K}$  &    Y     \\
G33.093$-$0.073$^c$ & 18 51 59.58      &    00 06 35.5               & 0.891    & 105.6$^{2,K}$ &    Y      \\
G33.133$-$0.092$^c$ & 18 52 07.82      &    00 08 12.8               & 0.339    &  77.1$^{2,K}$ &    Y      \\
G33.393+0.010$^c$   & 18 52 14.62      &    00 24 52.9               & 1.059    & 102.0$^K$     &    Y  \\ 
G33.641$-$0.228$^b$ & 18 53 32.56      &    00 31 39.2               & 0.461    &  61.1$^{1,T}$ &    Y      \\
G33.980$-$0.019$^b$ & 18 53 25.02      &    00 55 26.0               & 0.022    &  61.1$^{1,K}$ &    Y      \\
G34.096+0.018$^c$   & 18 53 29.94      &    01 02 39.4               & 0.567    &  57.6$^{2,K}$ &    Y      \\
G34.245+0.134$^{a{\dagger}2}$   & 18 53 21.45      &    01 13 46.0               & 0.706    &  58.9$^K$ &     1Y      \\
G34.396+0.222$^a$   & 18 53 19.09      &    01 24 13.9               & 0.621    &  59.3$^T$     &    Y  \\
G34.751$-$0.093$^b$ & 18 55 05.22      &    01 34 36.3               & 0.136    &  51.1$^{1,K}$ &    Y      \\
G34.791$-$1.387$^c$ & 18 59 45.98      &    01 01 19.0               & 0.347    &  46.3         &    Y  \\
G35.025+0.350$^a$   & 18 54 00.66      &    02 01 19.3               & 0.159    &  52.9$^{2,T}$ &    Y      \\ 
G35.197$-$0.743$^{c{\dagger}2}$ & 18 58 13.05      &    01 40 35.7               & 0.258    &  31.5$^T$ &     1Y      \\ 
G35.200$-$1.736$^c$ & 19 01 45.54      &    01 13 32.6               & 0.080    &  43.2$^T$     &    Y  \\
G35.793$-$0.175$^b$ & 18 57 16.89      &    02 27 57.9               & 0.089    &  61.9$^{1,K}$ &    Y      \\%
G36.115+0.552$^{b{\dagger}2}$   & 18 55 16.79      &    03 05 05.4               & 0.029    &  76.0$^{1,K}$ &   1Y       \\
G36.705+0.096$^b$   & 18 57 59.12      &    03 24 06.1               & 0.080    &  59.8$^{1,K}$ &    Y      \\
G37.030$-$0.038$^{b{\dagger}2}$ & 18 59 03.64      &    03 37 45.1               & 0.019    &  80.1$^{1,K}$ &     2Y     \\
G37.430+1.518$^c$   & 18 54 14.23      &    04 41 41.1               & 1.297    &  46.4$^T$     &    N  \\
G37.479$-$0.105$^b$ & 19 00 07.14      &    03 59 53.3               & 0.003    &  59.1$^{1,K}$ &    Y      \\
G37.554+0.201$^i$   & 18 59 09.99      &    04 12 15.5               & 0.359    &  86.7$^{2,K}$ &    Y      \\
G37.598+0.425$^b$   & 18 58 26.80      &    04 20 45.5               & 0.107    &  90.0$^{1,K}$ &    Y      \\
G38.038$-$0.300$^b$ & 19 01 50.47      &    04 24 18.9               & 0.019    &  62.6$^{2,K}$ &    Y      \\
G38.119$-$0.229$^i$ & 19 01 44.15      &    04 30 37.4               & 0.883    &  83.4$^{2,K}$ &    Y      \\
G38.203$-$0.067$^b$ & 19 01 18.73      &    04 39 34.3               & 0.014    &  82.9$^{2,K}$ &    Y      \\
G38.258$-$0.074$^i$ & 19 01 26.23      &    04 42 17.3               & 0.680    &  12.1$^2$     &    Y  \\
G39.100+0.491$^b$   & 19 00 58.04      &    05 42 43.9               & 0.002    &  23.1$^1$     &    Y  \\
G40.282$-$0.219$^j$ & 19 05 41.22      &    06 26 12.7               & 0.238    &  74.1$^{2,K}$ &    Y      \\
G40.425+0.700$^j$   & 19 02 39.62      &    06 59 09.1               & 0.883    &  10.1$^K$     &    Y  \\
G40.622$-$0.138$^i$ & 19 06 01.63      &    06 46 36.2               & 0.171    &  33.1$^{2,K}$ &    Y      \\
G41.348$-$0.136$^j$ & 19 07 21.84      &    07 25 17.6               & 0.244    &  12.8$^{2,K}$ &    Y      \\
G42.034+0.190$^i$   & 19 07 28.19      &    08 10 53.5               & 0.321    &  12.4$^K$     &    Y  \\
G42.304$-$0.299$^i$ & 19 09 43.59      &    08 11 41.4               & 0.210    &  27.0$^{2,K}$ &    Y      \\
G43.038$-$0.453$^i$ & 19 11 38.98      &    08 46 30.7               & 0.182    &  57.6$^{2,K}$ &    Y      \\
G43.074$-$0.077$^i$ & 19 10 22.05      &    08 58 51.5               & 0.184    &  12.6$^{2,K}$ &    Y      \\
G43.149+0.013$^{j{\dagger}4}$   & 19 10 11.05      &    09 05 20.5               & 1.184    &   14.4$^T$ &     2Y      \\
G43.795$-$0.127$^i$ & 19 11 53.99      &    09 35 50.6               & 0.205    &  45.4$^{2,T}$ &    Y      \\
G43.890$-$0.784$^c$ & 19 14 26.39      &    09 22 36.6               & 0.843    &  50.4$^T$     &    Y  \\
G45.071+0.132$^i$   & 19 13 22.13      &    10 50 53.1               & 0.163    &  59.2$^{2,T}$ &    Y      \\
G45.467+0.053$^{j{\dagger}2}$   & 19 14 24.15      &    11 09 43.4               & 0.104    &  64.4$^{2,T}$  &   2Y      \\
G45.473+0.134$^{j{\dagger}2}$   & 19 14 07.36      &    11 12 16.0               & 0.250    &  63.1$^{2,K}$  &   2Y      \\
G45.804$-$0.356$^i$ & 19 16 31.08      &    11 16 12.0               & 0.266    &  60.1$^{2,K}$ &    Y      \\
G49.265+0.311$^i$   & 19 20 44.86      &    14 38 26.9               & 0.038    &   3.3$^{2,K}$ &    Y      \\
G49.349+0.413$^i$   & 19 20 32.45      &    14 45 45.4               & 0.190    &  68.0$^K$     &    Y  \\
G49.416+0.326$^{i{\dagger}2}$   & 19 20 59.21      &    14 46 49.7               & 0.098    &$-$21.4$^{2,K}$ &    2Y     \\
G49.482$-$0.402$^{i{\dagger}3}$ & 19 23 46.19      &    14 29 47.1               & 0.156    &  48.2$^T$   &       3N \\
G49.599$-$0.249$^i$ & 19 23 26.61      &    14 40 17.0               & 0.041    &  56.6$^{2,K}$ &    Y      \\
G50.779+0.152$^i$   & 19 24 17.41      &    15 54 01.6               & 0.137    &  42.2$^{2,K}$ &    Y      \\
G52.199+0.723$^c$   & 19 24 59.84      &    17 25 17.9               & 2.326    &   3.2     &    N  \\  
G52.663$-$1.092$^c$ & 19 32 36.08      &    16 57 38.5               & 0.212    &  62.4     &    Y  \\
G53.618+0.035$^i$   & 19 30 23.02      &    18 20 26.7               & 0.104    &  23.1$^{2,K}$ &    Y      \\
G58.775+0.644$^c$   & 19 38 49.13      &    23 08 40.2               & 1.454    &  33.1$^K$     &    N  \\
G59.783+0.065$^j$   & 19 43 11.25      &    23 44 03.3               & 0.001    &  22.5$^{2,T}$ &    Y      \\    
G59.833+0.672$^c$   & 19 40 59.33      &    24 04 46.5               & 0.128    &  37.7$^K$     &    Y  \\
G60.575$-$0.187$^k$ & 19 45 52.50      &    24 17 43.0               & 0.883    &   4.9$^{2,K}$ &    Y      \\ 
G69.540$-$0.976$^l$ & 20 10 09.07      &    31 31 35.9               & 1.120    &   7.5$^T$     &    Y  \\
G70.181+1.742$^k$   & 20 00 54.14      &    33 31 31.0               & 3.322    &$-$26.5        &    N \\
G71.522$-$0.385$^p$ & 20 12 57.91      &    33 30 26.9               & 0.008    &  10.6$^2$     &    Y  \\
G73.06+1.80     & 20 08 10.2       &    35 59 23.7               &     &   1.7          &  Y \\
G75.782+0.343$^{m{\dagger}2}$   & 20 21 44.01      &    37 26 37.5               & 1.309    &$-$1.6$^2$ &   2N        \\
G78.122+3.633$^n$   & 20 14 26.07      &    41 13 32.7               & 0.001    &$-$6.8$^T$      &   Y  \\
\hline              
\end{tabular}
\end{table}

\begin{table}
 \contcaption{ 
\label{tab: sample continued}}
\begin{tabular}{@{}l @{\hspace{1.5mm}}l @{\hspace{2.5mm}}r @{\hspace{2.5mm}}c @{\hspace{2.5mm}}c @{\hspace{1.5mm}}r}
\hline
Name            & RA(2000)         & Dec(2000)                   & Offset    & $V_\mathrm{s}$ & Note\\
                & (h     m      s) & ($^{\circ}$   '   ")            & (')       & (km\,s$^{-1}$) & \\
\hline
G79.736+0.990$^o$   & 20 30 50.67      &    41 02 27.6               & 0.052    &$-$1.7${^2,T}$  &   Y      \\
G80.861+0.383$^o$   & 20 37 00.96      &    41 34 55.7               & 2.976    &$-$1.6$^{2,T}$  &   N      \\
G81.871+0.781$^o$   & 20 38 36.43      &    42 37 34.8               & 0.005     & 10.6$^{2,T}$   &    Y      \\
G85.410+0.003$^p$   & 20 54 13.68      &    44 54 07.6               & 0.014     &$-$36.6$^2$     &    Y  \\   
G90.921+1.487$^k$   & 21 09 12.97      &    50 01 03.6               & 1.466     &$-$70.0         &    N  \\    
G94.602$-$1.796$^m$ & 21 39 58.27      &    50 14 21.0               & 1.055     &$-$42.4$^T$     &    Y  \\
G108.184+5.519$^l$  & 22 28 51.41      &    64 13 41.3               & 0.334     &$-$11.4$^T$     &    Y  \\
G109.871+2.114$^m$  & 22 56 17.90      &    62 01 49.6               & 0.130     &$-$3.5$^T$      &    Y \\
G111.256$-$0.770$^m$& 23 16 10.36      &    59 55 28.5               & 0.564     &$-$43.5$^{2,T}$ &    Y     \\
G111.542+0.777$^{m{\dagger}3}$  & 23 13 45.36      &    61 28 10.6               & 0.000     &$-$57.7$^{2,T}$ &     3Y    \\
G121.298+0.659$^l$  & 00 36 47.35      &    63 29 02.2               & 0.808     &$-$25.1$^T$     &    Y \\
G123.066$-$6.309$^l$& 00 52 24.20      &    56 33 43.2               & 0.757     &$-$32.2$^T$     &    Y \\
G136.845+1.167$^p$  & 02 49 33.59      &    60 48 27.9               & 1.360     &$-$42.3$^2$     &    N \\
G173.482+2.446$^k$  & 05 39 13.06      &    35 45 51.3               & 0.000     &$-$11.3         &    Y \\
G183.348$-$0.575$^p$& 05 51 10.95      &    25 46 17.2               & 0.040     &$-$10.0         &    Y \\
G188.794+1.031$^m$  & 06 09 06.97      &    21 50 41.4               & 0.285     &$-$1.0$^{2,T}$  &    Y     \\
G188.946+0.886$^l$  & 06 08 53.34      &    21 38 29.2               & 0.086     &  2.8$^{2,T}$   &    Y     \\
G189.030+0.784$^f$  & 06 08 40.65      &    21 31 06.9               & 1.268     &  1.7$^2$       &    N \\ 
G192.600$-$0.048$^l$& 06 12 54.02      &    17 59 23.3               & 0.001     &  9.0$^{2,T}$   &    Y     \\ 
G196.454$-$1.677$^l$& 06 14 37.05      &    13 49 36.2               & 0.014     & 14.8$^T$       &    Y \\
G232.620+0.996$^m$  & 07 32 09.78       & $-$16 58 12.8               & 0.029     & 22.6$^T$      &    Y  \\ 
\hline              
\end{tabular}

References to positions: $^a$\cite{bartkiewicz16}, $^b$\cite{bartkiewicz09},
$^c$\cite{breen15}, $^d$\cite{cyganowski09}, $^e$\cite{fujisawa14a}, $^f$\cite{caswell09},
$^g$\cite{xu09}, $^h$\cite{moscadelli13}, $^i$\cite{pandian11}, $^j$\cite{bartkiewicz14}, 
$^k$(http://bessel.vlbi-astrometry.org), $^l$\cite{rygl10}, $^m$\cite{reid14}, 
$^n$\cite{moscadelli11a}, $^o$\cite{rygl12}, $^p$\cite{hu16}\\
References to systemic velocity: $^1$\cite{szymczak07}, $^2$\cite{shirley13}. Distance adopted: $^T$ trigonometric, $^K$ kinematic. 
\end{table}

 \subsection{Observations}
  \label{sec:observations}
The 32-m radio telescope at the Torun Centre for Astronomy was used between 2009 June and 2013 February for observations of the 
6668.519\,MHz methanol maser transition. Data for each object were taken at least once a month before 2010 October, then highly 
variable targets were observed at irregular intervals of 5-10\,d, with several gaps of 3-4 weeks due to scheduling constraints. 
Several circumpolar objects were observed 3-4 times a week over the entire programme. Thus for all the targets our observations 
are sensitive to maser variability on weeks-months time-scales.

The telescope was equipped with a dual-channel HEMT receiver followed by an autocorrelation spectrometer configured to record 
4096 channels over 4\,MHz for each polarisation, yielding a spectral resolution of 0.09\,\kms\, after Hanning smoothing. 
The system temperature varied between approximately 30 and 50\,K during the observations. The telescope has a half power beam-width 
of 5\farcm8 and rms pointing errors of 25\arcsec.

The spectra were obtained in a frequency-switching mode as a series of 30\,s integrations. Most sources had an on-source
integration of approximately 30\,min. The system was regularly calibrated by observing the continuum source 3C123 (\citealt{ott94})
and daily checked using the source G32.744$-$0.076. This methanol maser source was reported as non-variable within $\sim$5\,per~cent 
(\citealt{caswell95b}). Indeed, high-cadence observations reveal that some spectral features of the source have not shown variability 
higher than 8~per cent on time-scales from weeks to a few years. Detailed analysis of the light curves of five main spectral features 
of the source was reported in \cite{szymczak14}. The features at 30.3, 38.5 and 39.1\,\kms\, were non-variable during the monitoring 
period and the standard deviation of the flux density of these features indicates that the resulting accuracy of the absolute flux 
density is better than 10\,per~cent (\citealt{szymczak14}). The system temperature for the observations was measured by a continuously 
switched noise diode. Standard procedures were used to reduce the spectra. The typical rms noise level in the final spectra was 
0.20-0.35\,Jy at 0.09\,\kms\, velocity resolution after averaging over polarisation and time. There was no detectable interference 
on the spectra of our targets with the exception of G27.286+0.151.

 \subsection{Effect of offset observation and confusion}
  \label{sec:offset-confusion}
\begin{figure}   
 \resizebox{\hsize}{!}{\includegraphics[angle=0]{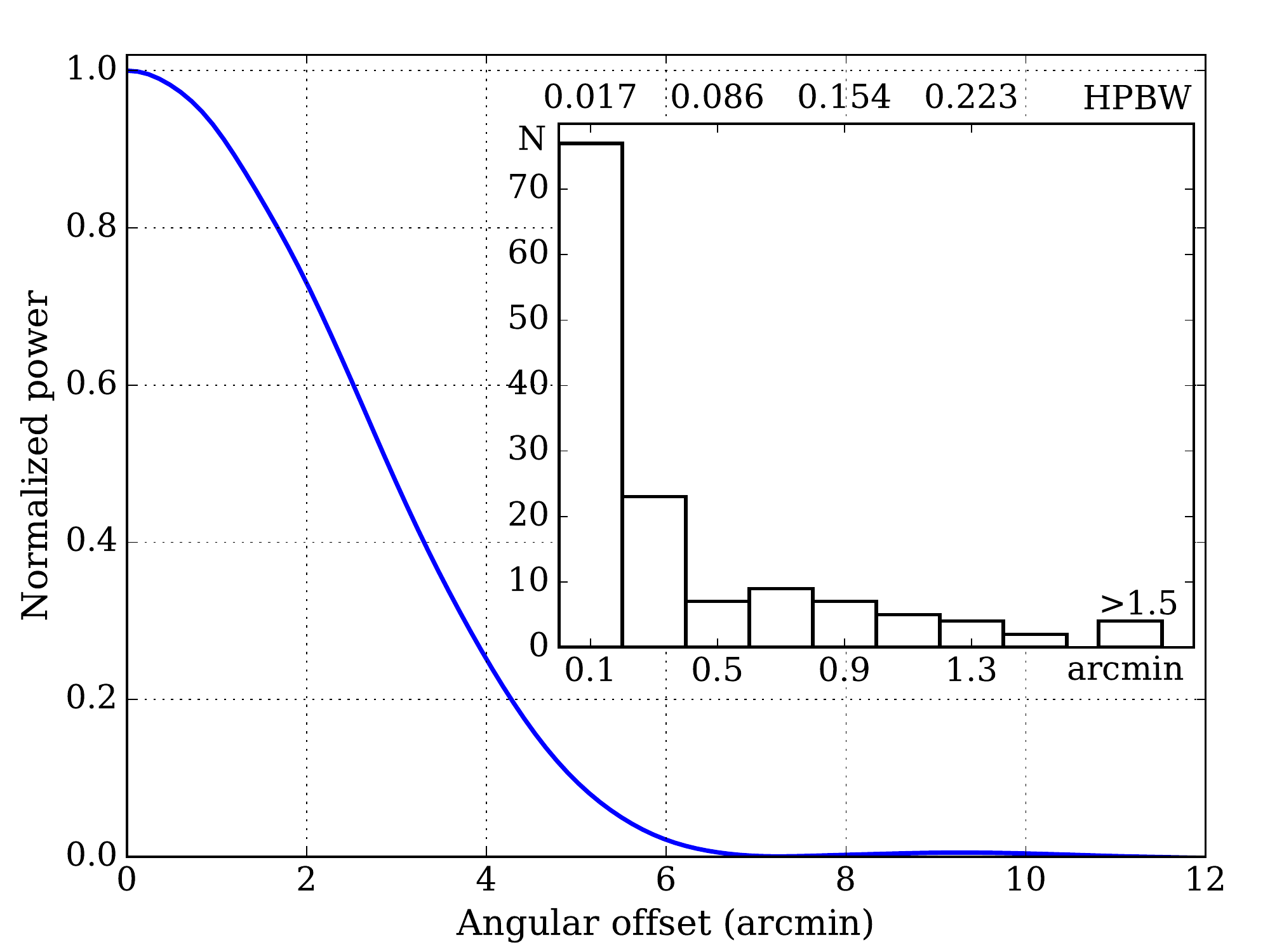}}
  \caption{Azimuthally averaged beam profile at 6.7\,GHz. The normalised power as a function of radial offset in arc-minutes is shown.
           Inset shows the histogram of the angular offset (Table~\ref{sample}) of the actual observations. 
 \label{beam-profile}}
\end{figure}

For several sites in the sample (Table~\ref{sample}) accurate positions have recently been published (e.g., \citealt{bartkiewicz14, bartkiewicz16}; 
\citealt{breen15}) which show that for a majority of the targets the observed positions deviate from the interferometric positions by less 
than 0\farcm4. These differences (the fourth column of Table~\ref{sample}) are much less or comparable to the telescope pointing errors and have 
no significant effect on the flux density measurements. The measured profile of the telescope beam (Fig.~\ref{beam-profile}) implies 
that our observations at an offset higher than 1\farcm3 systematically underestimate the flux density by 9~per cent and its uncertainty increases 
by more than 10~per cent due to the pointing errors. There are 10 sites (11 distinct sources) in the sample for which the flux density measurements are significantly 
influenced by these effects and all of them are discarded in the following statistical analysis.
Furthermore, the spectra of 18 sources (12 star-forming sites) are contaminated by confusion introduced by other sources in the telescope beam 
or side-lobe response from nearby bright sources and these sources are also excluded from the statistical analysis. Thus the variability analysis
presented from Section~\ref{sec:var-stat} is based on the spectra of 137 sources which are not affected by the above mentioned effects.

\subsection{Variability analysis}
 \label{var-measure}
The maser spectra taken with the single dish consist of a number of features with Gaussian velocity profiles
arising from localised clouds (e.g. \citealt{bartkiewicz14, bartkiewicz16}). In many sources the features are blended in 
velocity and to a lesser extent in location but in most cases the peak flux density of individual features clearly 
corresponds to the brightness of the strongest spot in the cloud as identified by Very Long Baseline Interferometry (VLBI).
In the following the light curves of individual features with an average peak flux density greater than $\sim$1\,Jy are analysed to characterise 
the variability properties of the target sources. The average flux density for three spectral channels centred at peak velocity of individual 
feature was used. The velocity alignment between epochs was done with the system Doppler track during the observations.

Firstly, we calculated the $\chi^2$ probability that the flux density of a given spectral feature remains constant.
The mean of data points is used as a model to test if the light curve is invariant. A $\chi^2$-test of the variability curve
$S_{\mathrm{i}}$ is done:
\begin{eqnarray}
\chi^{2}=\sum_{i=1}^{N}\left(\frac{S_{\rm i}-\overline S}{\sigma_{\rm i}}\right)^{2}\,\,
\label{chi2-1}
\end{eqnarray}
\noindent
The reduced $\chi^2$ statistic is given by
\begin{eqnarray}
\chi_{\rm r}^{2}=\frac{1}{N-1}\sum_{i=1}^{N}\left(\frac{S_{\rm i}-\overline S}{\sigma_{\rm i}}\right)^{2}\,\,
\label{chi2-2}
\end{eqnarray}
\noindent
where $S_{\rm i}$ denotes the individual flux density measurements of spectral feature at epoch i, 
${\overline S}$ their average in time, $\sigma_{\rm i}$ the individual measurement uncertainties and 
$N$ the number of measurements. The measurement uncertainty $\sigma$ used in Eqs. (\ref{chi2-1}) and (\ref{chi2-2}) takes into account 
the rms noise $\sigma_{\mathrm{rms}}$ and calibration error $\sigma_{\mathrm{sys}}$ and it is given 
by $\sigma^2 = \sigma_{\mathrm{rms}}^2 + \sigma_{\mathrm{sys}}^2$, where $\sigma_{\mathrm{sys}}$ is
equals to $T_{\mathrm{mas}}(\Delta T_{\mathrm{sys}}/T_{\mathrm{sys}})$. Here, $\Delta T_{\rm sys}$ is the uncertainty
in the system temperature $T_{\mathrm{sys}}$ and  $T_{\mathrm{mas}}$ is the antenna temperature at the maser peak.
For an intrinsically non-variable feature the value of $\chi_{\rm r}^2$
is expected to be one. A feature is considered to be variable if the $\chi^2$-test gives a probability 
of $\le$0.1~per cent for the assumption of constant flux density (99.9~per cent significance level for
variability). 

In order to further quantify the variability properties of spectral features of our sources 
we used the variability and fluctuation indices. The variability index given by
\begin{equation}
 VI = {(S_{\rm max} - \sigma_{\rm max}) - (S_{\rm min} + \sigma_{\rm min})\over (S_{\rm max} - \sigma_{\rm max}) + (S_{\rm min} + \sigma_{\rm min})}
\end{equation}
is a measure of the amplitude of the variability of a feature. Here $S_{\rm max}$ and $S_{\rm min}$ are 
the highest and lowest measured flux densities, respectively, and $\sigma_{\rm max}$ and $\sigma_{\rm min}$ 
are the uncertainties in these measurements. This quantity is well determined only when the variability of a spectral
feature is significantly greater than measurement errors (e.g. \citealt*{aller03}). It can be negative for faint features
or features with intrinsically little variability and is very sensitive to outliers.

The fluctuation index is defined (e.g. \citealt{aller03}) as 
\begin{equation}
 FI = {\big[}{N\over \sum_{i=1}^N\sigma_{\rm i}^2}{\big(}{\sum_{i=1}^N{S_{\rm i}^2 \sigma_{\rm i}^2 - \overline{S}\sum_{i=1}^N{S_{\rm i}\sigma_{\rm i}^2}}  \over N-1}-1{\big)}{\big]}^{0.5}/\overline{S} 
\end{equation}
\noindent
$S_{\rm i}$, ${\overline S}$, $\sigma_{\rm i}$ and $N$ are as before. This index measures the spread about 
the mean flux density and is a reliable measure of variability for spectral features with low signal-to-noise
ratio and low amplitude variability. This estimator has the advantage of being robust against outliers.

Several features in the sample switch between intervals of steady quiescence and long lasting flares (Sect.~\ref{sec:bursts}).
We note that the addition of an interval of steady intensity to a feature with a previous strong variability reduces 
the variability indices only slightly, while the variability measures of a weakly variable feature increase significantly
when it starts to vary strongly. 

\section{Results} 

Table~\ref{varindices} summarizes the variability parameters for the entire source sample.
The modified Julian Date at the start of the observations MJD$_\mathrm{s}$, the total time-span
of the observations $T_\mathrm{s}$, the total number of observations carried out $N_\mathrm{o}$ 
and the average observation cadence $C$ are given for each site. 
The peak velocity $V_\mathrm{p}$, the average flux density of peak $S_\mathrm{p}$, the variability
index $VI$,  the fluctuation index $FI$, the reduced $\chi^2$ value $\chi_\mathrm{r}^2$ are listed
for each spectral feature. The last column gives the $\chi^2$ value for a 99.9~per cent significance 
level of variability of all features of the source.

The grand average spectrum together with the spectra at high and low emission levels and the light
curves of individual maser features for each site are shown in Figure~\ref{spectra-lightcurves}.
If a given source has several features of similar intensity and variability pattern, for the sake
of clarity, only one light curve is shown. For the same reason the light curves of features in G32.744$-$0.076
contain weekly averaged measurements. In many of the sources we note correlation between the variations 
at different velocities which largely represent the final uncertainty in the absolute flux density calibration 
at each epoch.

\begin{table}
 \caption{Variability properties of 6.7\,GHz methanol maser features. MJD$_\mathrm{s}$ refers to the modified Julian Date 
 at the start of the observations,  $T_\mathrm{s}$ denotes the total time-span of the observations, $N_\mathrm{o}$
 the total number of observations, $C$ the average observation cadence, $V_\mathrm{p}$ the peak velocity, $S_\mathrm{p}$
 the average peak flux density, $VI$ the variability index, $FI$ the fluctuation index, $\chi_\mathrm{r}^2$ the reduced
 $\chi^2$ and $\chi_\mathrm{99.9\%}^2$ the corresponding value value for a 99.9~per cent significance level of variability. 
 The indices and $\chi^2$ test values of features affected by confusion are italicised.
\label{varindices}}
\begin{tabular}{@{\hspace{0.2in}} c r r r r c l}
\hline
&$V_\mathrm{p}$ (km\,s$^{-1}$) & $S_\mathrm{p}$ (Jy)& $VI$ & $FI$ & $\chi_\mathrm{r}^2$ & $\chi_\mathrm{99.9\%}^2$ \\
\hline
\multicolumn{7}{l}{G20.237+0.065 (MJD$_\mathrm{s}$=55006, $T_\mathrm{s}$= 1315, $N_\mathrm{o}$=70, $C$(month$^{-1}$)=1.597)} \\
\cline{2-7}
&60.91     &  3.9    & 0.43   &  0.21 & 4.67 & 1.62 \\
&70.26     &  6.5    &{\it 0.10}&{\it 0.09} & {\it 0.55} & \\
&71.35     & 18.4    & 0.33   &  0.18 & 4.42 & \\
&71.71     & 30.7    & 0.63   &  0.50 & 25.78 &\\
&73.16     & 16.1    & 0.17   &  0.04 & 0.94 & \\
\hline
\multicolumn{7}{l}{G21.407$-$0.254 (MJD$_\mathrm{s}$=55006, $T_\mathrm{s}$= 1315, $N_\mathrm{o}$=54, $C$(month$^{-1}$)=1.227)} \\
\hline
\end{tabular}
\end{table}

\begin{table}
\contcaption{ 
\label{tab: varindices continued}}
\begin{tabular}{@{\hspace{0.2in}} c r r r r c l}
\hline
&$V_\mathrm{p}$ (km\,s$^{-1}$) & $S_\mathrm{p}$ (Jy)& $VI$ & $FI$ & $\chi_\mathrm{r}^2$ & $\chi_\mathrm{99.9\%}^2$ \\
\hline
&29.53    &  11.8    & 0.09   &  0.05 &  0.73 & \\
&30.14    &   2.0    & 0.29   &  0.06 &  1.07 & \\
&32.69    &   2.0    & 0.47   &  0.08 &  1.42 & \\
&33.43    &   2.2    & 0.12   &  0.06 &  0.92 & \\
&35.54    &   1.6    & 0.80   &  0.49 & 14.11 & \\
&38.08    &   3.9    & 0.21   &  0.12 &  2.35 & \\
&38.53    &   6.0    & 0.08   &  0.06 &  0.72 & \\
\hline
\end{tabular}
\end{table}

\begin{table}
\contcaption{ 
\label{tab: varindices continued}}
\begin{tabular}{@{\hspace{0.2in}} c r r r r c l}
\hline
&$V_\mathrm{p}$ (km\,s$^{-1}$) & $S_\mathrm{p}$ (Jy)& $VI$ & $FI$ & $\chi_\mathrm{r}^2$ & $\chi_\mathrm{99.9\%}^2$ \\
\hline
&81.79    &   6.2   & 0.05    &  0.09 &  0.51 & \\
&82.36    &  12.6   & 0.01    &  0.09 &  0.24 & \\
&84.64    &   4.3   & 0.03    &  0.10 &  0.26 & \\
\hline
\multicolumn{7}{l}{G23.707$-$0.198 (MJD$_\mathrm{s}$=55011, $T_\mathrm{s}$= 1287, $N_\mathrm{o}$=61, $C$(month$^{-1}$)=1.419)} \\
\cline{2-7}
&72.85    &   1.5   & 0.44    &  0.01 &  1.39 & 1.66 \\
&74.50    &   2.9   & 0.52    &  0.37 & 10.10 & \\
&74.85    &   3.6   & 0.72    &  0.64 & 28.96 & \\
\hline
\end{tabular}
\end{table}

\begin{table}
\contcaption{ 
\label{tab: varindices continued}}
\begin{tabular}{@{\hspace{0.2in}} c r r r r c l}
\hline
&$V_\mathrm{p}$ (km\,s$^{-1}$) & $S_\mathrm{p}$ (Jy)& $VI$ & $FI$ & $\chi_\mathrm{r}^2$ & $\chi_\mathrm{99.9\%}^2$ \\
\hline
\multicolumn{7}{l}{G25.710+0.044 (MJD$_\mathrm{s}$=55013, $T_\mathrm{s}$= 1396, $N_\mathrm{o}$=44, $C$(month$^{-1}$)=0.937)} \\
\cline{2-7}
&89.80    &  43.6    & 0.03   &  0.08 & 0.35 & 1.80 \\
&91.94    &  22.9    & 0.03   &  0.09 & 0.28 & \\
&95.54    & 427.5    & 0.09   &  0.05 & 0.75 & \\
\hline
\multicolumn{7}{l}{G25.826$-$0.178 (MJD$_\mathrm{s}$=55014, $T_\mathrm{s}$= 1312, $N_\mathrm{o}$=53, $C$(month$^{-1}$)=1.206)} \\
\cline{2-7}
&90.75    &  31.0    & 0.05   &  0.07 & 0.54 & 1.72\\
&91.59    &  58.9    & 0.07   &  0.05 & 0.77 & \\
\hline
\end{tabular}
\end{table}

\begin{table}
\contcaption{ 
\label{tab: varindices continued}}
\begin{tabular}{@{\hspace{0.2in}} c r r r r c l}
\hline
&$V_\mathrm{p}$ (km\,s$^{-1}$) & $S_\mathrm{p}$ (Jy)& $VI$ & $FI$ & $\chi_\mathrm{r}^2$ & $\chi_\mathrm{99.9\%}^2$ \\
\hline
&104.02  &    1.9    & 0.03   &  0.10 &  0.45 & \\
&104.68  &    3.7    & 0.02   &  0.09 &  0.38 & \\
\hline
\multicolumn{7}{l}{G28.282$-$0.359 (MJD$_\mathrm{s}$=55017, $T_\mathrm{s}$= 1282, $N_\mathrm{o}$=60, $C$(month$^{-1}$)=1.401)} \\
\cline{2-7}            
&40.88   &    1.2    & 0.38   &  0.25 &  0.95 & 1.67 \\ 
&41.32   &    1.8    & 0.55   &  0.17 &  1.68 & \\
\hline
\multicolumn{7}{l}{G28.305$-$0.387 (MJD$_\mathrm{s}$=55017, $T_\mathrm{s}$= 1282, $N_\mathrm{o}$=60, $C$(month$^{-1}$)=1.401)} \\
\cline{2-7}
&79.77   &    3.2    & 0.90   &  1.28 & 85.74 & 1.67 \\
\hline
\end{tabular}
\end{table}

\begin{table}
\contcaption{ 
\label{tab: varindices continued}}
\begin{tabular}{@{\hspace{0.2in}} c r r r r c l}
\hline
&$V_\mathrm{p}$ (km\,s$^{-1}$) & $S_\mathrm{p}$ (Jy)& $VI$ & $FI$ & $\chi_\mathrm{r}^2$ & $\chi_\mathrm{99.9\%}^2$ \\
\hline
&102.94  &    2.5    &{\it 0.84} &{\it 0.47} &{\it 12.96} &  \\
&104.70  &    1.2    & 0.78   &  0.78 & 16.87 & \\
&105.00  &    1.6    & 0.61   &  0.74 & 21.35 & \\   
\hline
\multicolumn{7}{l}{G30.703$-$0.068 (MJD$_\mathrm{s}$=55008, $T_\mathrm{s}$= 1309, $N_\mathrm{o}$=44, $C$(month$^{-1}$)=1.000)} \\
\cline{2-7}
&86.02   &   10.9    &{\it 0.05} &{\it 0.09} &{\it 0.34} & 1.80 \\
&88.22   &  144.0    &{\it 0.21} &{\it 0.08} &{\it 1.62} & \\
&91.69   &   12.3    &{\it 0.41} &{\it 0.20} &{\it 4.62} & \\
\hline
\end{tabular}
\end{table}

\begin{table}
\contcaption{ 
\label{tab: varindices continued}}
\begin{tabular}{@{\hspace{0.2in}} c r r r r c l}
\hline
&$V_\mathrm{p}$ (km\,s$^{-1}$) & $S_\mathrm{p}$ (Jy)& $VI$ & $FI$ & $\chi_\mathrm{r}^2$ & $\chi_\mathrm{99.9\%}^2$ \\
\hline
\multicolumn{7}{l}{G31.412+0.307 (MJD$_\mathrm{s}$=55016, $T_\mathrm{s}$= 1394, $N_\mathrm{o}$=46, $C$(month$^{-1}$)=0.982)} \\
\cline{2-7}
&92.21   &   9.0     & 0.07   &  0.07 &  0.57 & 1.78 \\
&95.76   &  14.4     & 0.03   &  0.08 &  0.39 & \\
&99.01   &   2.7     & 0.11   &  0.10 &  0.73 & \\
&103.49  &  10.6     & 0.10   &  0.03 &  0.90 & \\
\hline
\multicolumn{7}{l}{G31.581+0.077 (MJD$_\mathrm{s}$=55021, $T_\mathrm{s}$= 1298, $N_\mathrm{o}$=66, $C$(month$^{-1}$)=1.524)} \\
\cline{2-7}
&95.54   &   4.1     & 0.33   &  0.13 &  2.78 & 1.63 \\
\hline
\end{tabular}
\end{table}

\begin{table}
\contcaption{ 
\label{tab: varindices continued}}
\begin{tabular}{@{\hspace{0.2in}} c r r r r c l}
\hline
&$V_\mathrm{p}$ (km\,s$^{-1}$) & $S_\mathrm{p}$ (Jy)& $VI$ & $FI$ & $\chi_\mathrm{r}^2$ & $\chi_\mathrm{99.9\%}^2$ \\
\hline
\multicolumn{7}{l}{G34.096+0.018 (MJD$_\mathrm{s}$=55011, $T_\mathrm{s}$= 1398, $N_\mathrm{o}$=49, $C$(month$^{-1}$)=1.045)} \\
\cline{2-7}
&55.15   &   1.5     & 0.45   &  0.43 &  6.85 & 1.75 \\
&56.00   &   4.5     & 0.18   &  0.07 &  1.43 & \\
&61.17   &   1.8     & 0.33   &  0.17 &  2.78 & \\
\hline
\multicolumn{7}{l}{G34.245+0.134 (MJD$_\mathrm{s}$=55031, $T_\mathrm{s}$= 1379, $N_\mathrm{o}$=51, $C$(month$^{-1}$)=1.104)} \\
\cline{2-7}
&54.91   &   4.4     & 0.23   &  0.05 &  1.07 & 1.73\\
&56.89   &   3.7     &{\it 0.18} &{\it 0.13} &{\it 2.17} \\
\hline
\end{tabular}
\end{table}

\begin{table}
\contcaption{ 
\label{tab: varindices continued}}
\begin{tabular}{@{\hspace{0.2in}} c r r r r c l}
\hline
&$V_\mathrm{p}$ (km\,s$^{-1}$) & $S_\mathrm{p}$ (Jy)& $VI$ & $FI$ & $\chi_\mathrm{r}^2$ & $\chi_\mathrm{99.9\%}^2$ \\
\hline
&55.15   &   1.9     & 0.77   &  0.76 & 25.25 & \\
&62.09   &   2.0     & 0.60   &  0.29 &  4.81 & \\
\hline
\multicolumn{7}{l}{G37.030$-$0.038 (MJD$_\mathrm{s}$=55014, $T_\mathrm{s}$= 1298, $N_\mathrm{o}$=55, $C$(month$^{-1}$)=1.266)} \\
\cline{2-7}
&78.52   &   4.7     & 0.32   &  0.18 &  3.99 & 1.70\\
&80.14   &   8.8     & 0.12   &  0.01 &  1.00 & \\
&81.59   &   3.0     & 0.48   &  0.26 &  5.37 & \\
&83.70   &   4.4     & 0.30   &  0.08 &  1.68 & \\
\hline
\end{tabular}
\end{table}

\begin{table}
\contcaption{ 
\label{tab: varindices continued}}
\begin{tabular}{@{\hspace{0.2in}} c r r r r c l}
\hline
&$V_\mathrm{p}$ (km\,s$^{-1}$) & $S_\mathrm{p}$ (Jy)& $VI$ & $FI$ & $\chi_\mathrm{r}^2$ & $\chi_\mathrm{99.9\%}^2$ \\
\hline
&74.34   &  19.1     & 0.10   &  0.04 &  0.81 & \\
&77.76   &   5.6     & 0.49   &  0.37 & 12.72 & \\
&80.00   &   6.0     & 0.11   &  0.06 &  0.80 & \\
&83.52   &   4.6     & 0.24   &  0.10 &  2.05 & \\
\hline
\multicolumn{7}{l}{G40.425+0.700 (MJD$_\mathrm{s}$=55023, $T_\mathrm{s}$= 1286, $N_\mathrm{o}$=61, $C$(month$^{-1}$)=1.420)} \\
\cline{2-7}
&5.50    &   2.4     & 0.17   &  0.09 &  1.06 & 1.66 \\
&6.64    &   2.9     & 0.37   &  0.14 &  2.29 & \\
\hline
\end{tabular}
\end{table}

\begin{table}
\contcaption{ 
\label{tab: varindices continued}}
\begin{tabular}{@{\hspace{0.2in}} c r r r r c l}
\hline
&$V_\mathrm{p}$ (km\,s$^{-1}$) & $S_\mathrm{p}$ (Jy)& $VI$ & $FI$ & $\chi_\mathrm{r}^2$ & $\chi_\mathrm{99.9\%}^2$ \\
\hline
\multicolumn{7}{l}{G45.071+0.132 (MJD$_\mathrm{s}$=55006, $T_\mathrm{s}$= 1331, $N_\mathrm{o}$=65, $C$(month$^{-1}$)=1.463)} \\
\cline{2-7}		
&57.76   &  39.2     & 0.05   &  0.08 &  0.44 & 1.64 \\
\hline
\multicolumn{7}{l}{G45.467+0.053 (MJD$_\mathrm{s}$=55006, $T_\mathrm{s}$= 1335, $N_\mathrm{o}$=56, $C$(month$^{-1}$)=1.254)} \\
\cline{2-7}
&49.96   &   1.0     & 0.48   &  0.31 &  3.66 & 1.69 \\
&56.04   &   3.3     & 0.34   &  0.12 &  2.28 & \\
&56.48   &   2.9     & 0.18   &  0.03 &  1.17 & \\
&57.41   &   3.3     & 0.23   &  0.11 &  1.99 & \\
\hline
\end{tabular}
\end{table}

\begin{table}
\contcaption{ 
\label{tab: varindices continued}}
\begin{tabular}{@{\hspace{0.2in}} c r r r r c l}
\hline
&$V_\mathrm{p}$ (km\,s$^{-1}$) & $S_\mathrm{p}$ (Jy)& $VI$ & $FI$ & $\chi_\mathrm{r}^2$ & $\chi_\mathrm{99.9\%}^2$ \\
\hline
\multicolumn{7}{l}{G53.618+0.035 (MJD$_\mathrm{s}$=55009, $T_\mathrm{s}$= 1312, $N_\mathrm{o}$=98, $C$(month$^{-1}$)=2.250)} \\
\cline{2-7}
&18.56    &  2.9     & 0.65   &  0.36 & 10.24 & 1.50 \\
&18.87    &  4.8     & 0.52   &  0.24 &  6.54 & \\
\hline
\multicolumn{7}{l}{G58.775+0.644 (MJD$_\mathrm{s}$=55014, $T_\mathrm{s}$= 1317, $N_\mathrm{o}$=135, $C$(month$^{-1}$)=3.097)} \\
\cline{2-7}
&33.21    &  5.3     &{\it 0.28} &{\it 0.13} &{\it 2.29} & 1.42 \\
&34.83    &  1.6     &{\it 0.27} &{\it 0.09} &{\it 1.32} & \\
&35.49    &  2.7     &{\it 0.33} &{\it 0.16} &{\it 2.12} & \\
\hline
\end{tabular}
\end{table}

\begin{table}
\contcaption{ 
\label{tab: varindices continued}}
\begin{tabular}{@{\hspace{0.2in}} c r r r r c l}
\hline
&$V_\mathrm{p}$ (km\,s$^{-1}$) & $S_\mathrm{p}$ (Jy)& $VI$ & $FI$ & $\chi_\mathrm{r}^2$ & $\chi_\mathrm{99.9\%}^2$ \\
\hline
\multicolumn{7}{l}{G81.871+0.781 (MJD$_\mathrm{s}$=55005, $T_\mathrm{s}$= 1335, $N_\mathrm{o}$=284, $C$(month$^{-1}$)=6.453)} \\
\cline{2-7}
&3.42     & 54.4     & 0.32   &  0.11 &  2.19 & 1.28\\
&4.03     &146.3     & 0.38   &  0.14 &  3.06 & \\
&4.56     &233.4     & 0.22   &  0.02 &  1.02 & \\
&5.13     & 97.4     & 0.15   &  0.03 &  0.88 & \\
&5.75     & 80.5     & 0.16   &  0.04 &  0.85 & \\
&7.20     &246.9     & 0.22   &  0.07 &  1.44 & \\
\hline
\end{tabular}
\end{table}

\begin{table}
\contcaption{ 
\label{tab: varindices continued}}
\begin{tabular}{@{\hspace{0.2in}} c r r r r c l}
\hline
&$V_\mathrm{p}$ (km\,s$^{-1}$) & $S_\mathrm{p}$ (Jy)& $VI$ & $FI$ & $\chi_\mathrm{r}^2$ & $\chi_\mathrm{99.9\%}^2$ \\
\hline
&$-$13.03 & 46.7     & 0.13   &  0.04 &  0.87 & \\
&$-$11.54 &  3.6     & 0.27   &  0.06 &  1.29 & \\
&$-$6.45  &  1.8     & 0.17   &  0.13 &  1.78 & \\
\hline
\multicolumn{7}{l}{G183.348$-$0.575 (MJD$_\mathrm{s}$=55005, $T_\mathrm{s}$= 1341, $N_\mathrm{o}$=305, $C$(month$^{-1}$)=6.898)} \\
\cline{2-7}
&$-$15.20 &  3.8     & 0.61   &  0.32 &  9.35 & 1.27 \\
&$-$14.68 &  4.9     & 0.36   &  0.17 &  3.60 & \\
&$-$4.89  &  9.7     & 0.58   &  0.22 &  5.50 & \\
\hline
\end{tabular}
\end{table}

\subsection{Individual maser sites}
  \label{sec:individuals}
In this section we provide additional details which are not adequately conveyed by Table~\ref{varindices} 
or Figure~\ref{spectra-lightcurves} for individual maser sites. In particular, we highlight sites with peculiar types of variation and sites with multiple 
sources and complex spectral structure that could be confused with side-lobe effects and pointing offsets.

\noindent
{\bf G20.237+0.065}. The emission near 60.91\,\kms\, comes from the nearby source G20.239+0.065 and is composed of two features
(\citealt{bartkiewicz16}); a feature centred at 61.00\,\kms\, decreased by a factor of 3.5, whereas that at $\sim$60.77\,\kms\, 
remained constant which resulted in a velocity drift of $-$0.26\,\kms. The feature at velocity 70.26\,\kms\, which is a superposition 
of emission from G20.239+0.065 and G20.237+0.065, was not variable. The red-shifted ($>$71.1\,\kms) emission comes from 
G20.237+0.065 and showed complex variability (Section~\ref{sec:complex-spectra}).

\noindent
{\bf G22.039+0.222}. The source does not show significant variations, except for the feature at 54.42\,\kms\, which increased by
a factor of 2.6. Faint ($\sim$1.5\,Jy) emission at 46.24\,\kms\, dropped below the sensitivity limit after MJD $\sim$55435.

\noindent
{\bf G22.357+0.066}. The blue-shifted ($<$83.5\,\kms) emission showed high amplitude variations with a period of $\sim$179\,d 
(\citealt{szymczak15}). Weak features at 85.05 and 88.43\,\kms\, showed a marginal indication of periodic variability.

\noindent
{\bf G22.435$-$0.169}. Almost all the features were not variable. The exceptions are the features at 24.70 and 38.08\,\kms\, which 
decreased by a factor of $\sim$1.8. The feature at 35.54\,\kms\, exhibited erratic variability of 4$-$5\,Jy amplitude but 
the data sampling is too sparse to estimate its time-scale (Section~\ref{sec:bursts}). 

\noindent
{\bf G23.010$-$0.411}. The prominent features showed little or no variation. The feature at 79.40\,\kms\, increased by
a factor $\sim$1.8 after MJD 55820. Comparison with the data in \cite{goedhart04} suggests that the feature at velocity 74.73\,\kms\, 
increased from $\sim$350 to $\sim$540\,Jy after 10.5\,years, but that at 72.71\,\kms\, remained at the same level of $\sim$45\,Jy.

\noindent
{\bf G23.207$-$0.378}. The feature at 79.87\,\kms\, increased by a factor of two. The feature with a peak flux density of 1.9\,Jy centred 
at 82.94\,\kms\, increased to 16\,Jy at 82.72\,\kms\, (Section~\ref{sec:drifts}). The rest of the features showed little or no variations.

\noindent
{\bf G23.257$-$0.241}. The source has several stable features. Faint ($\sim$1.1\,Jy) emission near 60.22\,\kms\, 
showed moderate variability with a velocity drift of +0.25\,\kms\, probably due to significant variability of blended features.

\noindent
{\bf G23.389+0.185}. The features were not significantly variable, except for the feature at velocity 76.84\,\kms\, which decreased 
by a factor of $\sim$2.3.

\noindent
{\bf G23.437$-$0.184}. The emission from 94.7 to 100.4\,\kms\, comes from the nearby source G23.440$-$0.182 and shows little variation.
The features at 101.72, 102.90 and 103.95\,\kms\, exhibited synchronous and anti-correlated variations (Section~\ref{sec:synchro}).

\noindent
{\bf G23.484+0.097}. All the features showed little or no variation except for the feature 86.95\,\kms\, which decreased by a factor of 2.5.

\noindent
{\bf G23.707$-$0.198}. The emission in the entire velocity range of 72.55$-$82.80\,\kms\, showed very complex variations including apparent 
velocity drifts and bursts of several features (Section~\ref{sec:complex-spectra}). 

\noindent
{\bf G24.148$-$0.009}. The strongest features at 17.42 and 17.68\,\kms\, showed regular low-amplitude ($\sim$20 per cent) variations 
in flux density with a period of $\sim$175\,d.  The feature at 19.26\,\kms exhibited the same variability pattern.

\noindent
{\bf G24.329+0.144}. This source is highly variable (Section~\ref{sec:complex-spectra}). All the features experienced strong flares lasting 200$-$400\,d (Section~\ref{sec:bursts}).

\noindent
{\bf G24.494$-$0.038}. The emission at velocities higher than 114.5\,\kms\, was significantly variable. The feature at 114.97\,\kms\, exponentially 
rose from 7 to 25\,Jy with a characteristic time of 330$\pm$40\,d and the shoulder at 115.5\,\kms\, declined from 4.3 to 1.3\,Jy.

\noindent
{\bf G24.541+0.312}. The features at velocities lower than 109.60\,\kms\, have high variability indices. The feature at 105.54\,\kms\, with 
an initial flux density of $\sim$11\,Jy dropped to $\sim$1\,Jy after 880\,d. The strongest feature at velocity 106.91\,\kms\, of 9$-$15\,Jy 
peak showed irregular variations and rapidly decreased after 880\,d then remained nearly constant at a level of 6\,Jy. 
The emission at 108.53\,\kms\, displayed  flare behaviour with a rise time of about 310\,d and a decay time of 930\,d. 
The feature at 109.93\,\kms\, was constant to within the noise. 

\noindent
{\bf G24.634$-$0.324}. All but the 35.51\,\kms\, feature were strongly variable with different patterns (Section~\ref{sec:complex-spectra}).

\noindent
{\bf G24.790+0.083}. This is very crowded site and hence the profile is largely a blend of the emission from four sources (\citealt{breen15}).
The unblended features were not variable.

\noindent
{\bf G25.411+0.105}. The two major features showed variability with a period of $\sim$245\,d (\citealt{szymczak15}).

\noindent
{\bf G25.650+1.049}. The main feature at 41.79\,\kms\, showed a decreasing trend from $\sim$120 to $\sim$95\,Jy with
a flare at $\sim$MJD 55655. The other features were not variable.

\noindent
{\bf G25.826$-$0.178}. The features at 90.75, 91.59 and 93.83\,\kms\, showed flaring variability with the same pattern. The emission at 
velocities higher than 95\,\kms\, is complex and showed uncorrelated variations (Section~\ref{sec:complex-spectra}).

\noindent
{\bf G26.527$-$0.267}. The emission at velocities lower than 109\,\kms\, was not significantly variable. The feature at 109.55\,\kms\, 
showed evidence of 'flickering' which was not  fully sampled with the average cadence of 3 weeks. This feature is weak but its variation 
is not due to calibration errors since the strongest feature at 104.19\,\kms\, and the other features did not show similar behaviour 
(Fig.~\ref{flicker}). 
The overall spectrum of the source reported in \cite{breen15} is completely different from ours; the flux density in the velocity range 
of 102$-$105\,\kms\, increased by a factor of 5 after 2$-$3 years. It is not clear whether a faint feature at 115.00\,\kms\, is a side-lobe 
response of nearby source G26.601$-$0.221. 

\noindent
{\bf G26.598$-$0.024}. This source shows weak variation, except for the emission near 23.88\,\kms\, which decreased from 8 to 2\,Jy
and is probably a blend of two features separated from each other by $\sim$0.17\,\kms\, (Section~\ref{sec:drifts}).

\noindent
{\bf G26.601$-$0.221}. The feature at 103.25\,\kms\, initially had a flux density of 8.5\, Jy and dropped to $\sim$5.2\,Jy 
during $\sim$210\,d then remained constant over 2.1\,yr and then decreased to $\sim$3\,Jy at the end of our monitoring. 
The feature at 103.78\,\kms\, showed a sinusoidal-like variation of $\sim$3\,yr period and $\sim$4\,Jy amplitude.
The emission near 114.97\,\kms\, decreased from 8 to 5\,Jy and showed evidence for a velocity drift of $\sim$0.12\,\kms\, 
(Section~\ref{sec:drifts}).

\noindent
{\bf G27.221+0.136}. The highly variable feature at 115.49\,\kms\, increased from 1 to 7.5\,Jy, while the feature at 117.11\,\kms\, 
decreased from 7 to 2\,Jy. Similar anti-correlated variations appear for strongly blended emission near 117.77 and 118.30\,\kms; 
the former feature increased from 12.5 to 9\,Jy, whereas the latter feature decreased from 10 to 15\,Jy. This synchronised 
and anti-correlated behaviour was seen over the entire observing period (Section~\ref{sec:synchro}).  

\noindent
{\bf G27.286+0.151}. No significant variations were seen, except for weak emission at 36.31\,\kms. The features near 24 and 28\,\kms\, 
reported in \cite{breen15} are omitted due to interference.

\noindent
{\bf G27.365$-$0.166}. Most of the features were not variable. The exception is the feature at 100.72\,\kms\, which linearly increased
from 21 to 34\,Jy.

\noindent
{\bf G28.011$-$0.426}. The source was relatively weak and the flux density variations were dominated by noise.

\noindent
{\bf G28.282$-$0.359}. The strongest feature at 41.32\,\kms\, showed little variation.

\noindent
{\bf G28.305$-$0.387}. All the features were variable. The emission at 79.77\,\kms\, showed  'flickering' with an amplitude 
of 6$-$25\,Jy until  MJD 55940, and at several epochs it dropped below the sensitivity limit of 0.8\,Jy (Section~\ref{sec:bursts}). A similar 'flickering' pattern but of low 
($\sim$2\,Jy) amplitude and with a quiescent state higher than 4\,Jy was seen for the feature 82.93\,\kms. The features at 81.30 and 81.83\,\kms\, 
decreased from 38 to 7\,Jy and from 34 to 22\,Jy, respectively. These features may be blended and the apparent velocity drift is about 0.17\,\kms\,
(Section~\ref{sec:drifts}). The features at 92.54 and 93.82\,\kms\, experienced a weak ($\sim$4$-$5\,Jy) flare of relatively flat profile 
with a maximum near MJD 55540.  

\noindent
{\bf G28.832$-$0.253}. The major features showed moderate variations.

\noindent
{\bf G29.863$-$0.044}. The strongest features at 100.28, 101.38 and 101.81\,\kms\, showed correlated, low ($\sim$20 per cent) amplitude variations 
with a period of $\sim$250\,d from MJD 55410 to 55950. The feature at 95.89\,\kms\, is a side-lobe response of nearby source G29.955$-$0.016 
(\citealt{breen15}).

\noindent
{\bf G29.955$-$0.016}. The strongest feature at 95.88\,\kms\, was not variable. The emission at velocities higher than 96.5\,\kms\,
is confused with a side-lobe response from G29.978$-$0.047 (\citealt{breen15}).

\noindent
{\bf G30.198$-$0.169}. The emission from this site was significantly variable. The emission ranging from 107.7 to 108.8\,\kms\, generally 
increases and shows small (3$-$4\,Jy) amplitude fluctuations. The emission at velocities higher than 109.8\,\kms\, 
comes from the nearby source G30.225$-$0.180 (\citealt{bartkiewicz16}).

\noindent
{\bf G30.317+0.070}. A new feature appeared at 30.28\,\kms\, reaching $\sim$2.5\,Jy at the end of the monitoring period. The emission near 
35.20\,\kms\, was constant ($\sim$5\,Jy) before MJD $\sim$55670 then started to diminish by a factor of 2.5 at the end of the observations. 
The emission at velocities higher than 35.8\,\kms\, showed no variation.  

\noindent
{\bf G30.400$-$0.296}. The emission in the velocity range of 104.5$-$105.3\,\kms\, showed cyclic variations with a period of 200\,d. 
These periodic variations are less pronounced for the features at 98.11 and 100.31\,\kms. The emission in the velocity range 102.1$-$103.2\,\kms\,
may be contaminated by that from G30.419$-$0.232 (\citealt{breen15}). 

\noindent
{\bf G30.703$-$0.068 and G30.760$-$0.052}. The features of these sources are spatially blended with our 5\farcm8 beam and 
share the velocity range from 85.8 to 92.8\,\kms\, (\citealt{breen15}) so that their variability indices are less reliable. 

\noindent
{\bf G30.780+0.230}. 
The source is weakly variable. Synchronised variations with similar relative amplitude appeared at all the features.
The source is discarded from further analysis because of the large pointing offset.

\noindent
{\bf G30.788+0.204}. The emission from 75.6 to 77.5\,\kms\, is probably a superposition of several features of significant
variability. Two variable features at 84.73 and 85.08\,\kms\, decreased from 10 to 2\,Jy and increased 
from 5 to 11\,Jy, respectively. A feature at 89.65\,\kms\, marginally detected near MJD 55250 increased up to 2.4\,Jy 
at the end of our monitoring.

\noindent
{\bf G30.818$-$0.057 and G30.822$-$0.053}. The emission in the velocity range of 99.3$-$110.2\,\kms\, comes from G30.818$-$0.057 
(\citealt{breen15}) and was not variable with the exception of feature 109.10\,\kms\, which showed a monotonic increase from 7 to 16\,Jy. 
In G30.822$-$0.053 the 93.17\,\kms\, feature was constant within our accuracy. Features near 88.20, 91.28 and 91.72\,\kms\, flickered 
at MJD 55220, 55553, 55785 and 56301 with amplitudes usually higher than 10\,Jy (Section~\ref{sec:bursts}). 

\noindent
{\bf G31.047+0.356}. The emission near 82.8\,\kms\, was nearly constant but the peak velocity drifted
by $-$0.15\,\kms\ (Section~\ref{sec:drifts}).

\noindent
{\bf G31.059+0.093}. All the features in the spectrum experienced a flare lasting $\sim$650\,d with a maximum near MJD 55642. 

\noindent
{\bf G31.581+0.077}. The source showed moderate variability with a flare near MJD 55340 with the relative amplitude of 0.5.

\noindent
{\bf G32.045+0.059}. The source had stable and highly variable features. The feature at 92.63\,\kms\, was moderately variable,  whereas 
those at 94.12 and 94.82\,\kms\, flared near MJD 55760. The emission from 95.2 to 100.0\,\kms\, remained constant. The features at 
100.88 and 101.14\,\kms\, gradually increased before MJD 55760 then remained stable.

\noindent
{\bf G32.744$-$0.076}. The feature at 33.41\,\kms\, exhibited moderate modulation of the flux density, but the remainder showed little 
or no variation to within the measurement accuracy of about 8$-$10 per cent (\citealt{szymczak14}).

\noindent
{\bf G32.992+0.034}. The source does not show significant variability, except for the feature 90.96\,\kms\, for which the flux density 
exponentially increases from 3 to 8\,Jy with a periodic ($\sim$370\,d) and low amplitude modulation.

\noindent
{\bf G33.093$-$0.073}. The spectrum is complex and variable. Weak ($<$4\,Jy) emission in the velocity range from 94.5 to 99.5\,\kms\, 
showed little variation. The features at 102.19 and 102.76\,\kms\, gradually increased and the latter peaked at $\sim$MJD 56122. 
The emission near 105.87\,\kms\, flared for $\sim$500\,d with a maximum at $\sim$MJD 56030.

\noindent
{\bf G33.133$-$0.092}. This source showed a low amplitude, apparently quasi-periodic variation. However, the variability indices indicate
that the source was not variable to within the noise.

\noindent
{\bf G33.393+0.010}. The shoulder near 104.21\,\kms\, decreased by a factor of 1.7. The rest of the features were not variable.

\noindent
{\bf G33.641$-$0.228}. The source is highly variable. The features at 58.87 and 59.26\,\kms\, showed anti-correlated variations
in the flux density (Section~\ref{sec:synchro}). The feature at 59.57\,\kms\, showed irregular 'flickering' behaviour; the flux density at 
a quiescent level of 12$-$30\,Jy rapidly rose up to 90$-$370\,Jy but our observations were too sparse to recover the light curves 
(Section~\ref{sec:bursts}). \cite{fujisawa14c} reported that a typical light curve of a flicker-like burst of the feature 
is characterised by a very short ($\sim$1\,d) rise phase followed by a slow ($\sim$5\,d) decline. No flickers were detected after $\sim$MJD 55750. 
The emission at velocities higher than 59.8\,\kms\, showed complex uncorrelated variations of different amplitudes.  

\noindent
{\bf G33.980$-$0.019}. The source is weak and the features at 59.04 and 64.09\,\kms\, showed little variation.

\noindent 
{\bf G34.096+0.018}. The feature at 55.15\,\kms\, decreased by a factor of three, whereas that at 61.17\,\kms\, showed little variation. 

\noindent
{\bf G34.245+0.134}. The emission at velocities higher than 56.0\,\kms\, is confused by a side-lobe response from G34.257+0.153 (\citealt{breen15}).

\noindent
{\bf G34.396+0.222}. The strongest feature at 55.62\,\kms\, increased by a factor of two. Faint (2$-$3\,Jy) emission in the velocity
range of 57.5$-$63.2\,\kms\, remained stable.

\noindent
{\bf G34.751$-$0.093}. The features at 51.86 and 52.51\,\kms\, showed variability with the same pattern; after an initial decline the flux density 
remained stable from MJD 55148 to 55900, then increased reaching a peak near MJD 556006. 

\noindent
{\bf G35.025+0.350}. The flux density of the strongest feature at 44.33\,\kms\, varied from 13 to 24\,Jy in a sinusoidal-like manner.
A new feature appeared at 43.32\,\kms\, reaching a $\sim$4.5\,Jy peak at the end of the monitoring period.

\noindent
{\bf G35.197$-$0.743}. The feature at 28.48\,\kms\, is a blend of emission from the two sources (\citealt{breen15}). The feature at 36.56\,\kms\,
was significantly variable, probably showing a low amplitude 'flickering'.

\noindent
{\bf G35.200$-$1.736}. All the features showed an exponential decline modulated with a period of 350$-$410\,d. The flux density of individual 
features dropped by a factor of 2$-$8. The source was also highly variable when monitored $\sim$10\,yr earlier for a period of about 3.5\,yr  
(\citealt{goedhart04}). Thus the source shows significant long-term ($>$ 10\,yr) variability; the overall shape of the spectrum reported by 
\cite{goedhart04} is very different from ours. The faintest feature ($\sim$15\,Jy)  observed by \cite{goedhart04}, at 46.03\,\kms\,, has now 
disappeared. 

\noindent
{\bf G35.793$-$0.175}. The source showed significant variability. The extreme blue- and red-shifted faint (4$-$5\,Jy) features at 58.74 and 
62.86\,\kms\, sometimes dropped below the detection limit.

\noindent
{\bf G36.115+0.552}. The emission at velocities lower than 73.9\,\kms\, comes from two sources separated by 1\farcs2
(\citealt{bartkiewicz09}). The strongest feature near 72.84\,\kms\, declined by a factor of two over one year then increased to 
its initial flux density of $\sim$25\,Jy. The emission near 74.55\,\kms\, was moderately variable and showed a velocity drift of 
0.26\,\kms\, (Section~\ref{sec:drifts}). The features at 82.15 and 83.86\,\kms\, varied in flux density by a factor of three, whereas the feature at 
81.36\,\kms\, increased from 2 to 5\,Jy. The features at velocity 70.38 and 71.70\,\kms\, remained constant. 

\noindent
{\bf G36.705+0.096}. The source is weak and all the features  clearly varied to different extents.

\noindent
{\bf G37.030$-$0.038}. This source has a single peak at 78.52\,\kms\, which decreased by a factor of 1.4. The emission at velocities
higher than 79.0\,\kms\, comes from G37.043$-$0.035 (\citealt{breen15}) and was not variable except for the feature at
81.59\,\kms\, which decreased by a factor of 2.2.

\noindent
{\bf G37.430+1.518}. Quasi-periodic variations of up to 25~per cent seen for all the features are probably effects of a pointing offset of
1\farcm3. The source is discarded from further analysis.

\noindent
{\bf G37.479$-$0.105}. Most of the features showed moderate variations except for the feature at 53.80\,\kms\, which increased 
by a factor of 5 and the 54.64\,\kms\, feature which showed a scatter of  1$-$7\,Jy in flux density on a time-scale of a few weeks. 

\noindent
{\bf G37.554+0.201}. \cite{araya10} reported quasi-periodic flares of the 6.7\,GHz methanol and 4.8\,GHz formaldehyde masers. 
Our data confirm variations of almost all the methanol maser features with a period of $\sim$249\,d. 

\noindent
{\bf G37.598+0.425}. The features at 85.80, 87.12 and 88.66\,\kms\, showed quasi-periodic ($\sim$420\,d) variations of decreasing amplitude. 
The emission at 91.07 and 93.13\,\kms\, synchronously flared near MJD 55373 then gradually dropped below the detection limit. 

\noindent
{\bf G38.038$-$0.300}. The feature at 58.13\,\kms\, decreased by a factor of two. There was no significant change in the other features.

\noindent
{\bf G38.119$-$0.229}. The main feature at 70.37\,\kms\, increased from 3 to 5\,Jy. A new feature appeared at 69.09\,\kms. 
The other features showed little variation or remained stable.

\noindent
{\bf G38.203$-$0.067}. The flux density ($\sim$7\,Jy) of the 79.74\,\kms\, feature was constant before MJD 55410 then dropped 
to $\sim$4\,Jy near MJD 55530 and then gradually increased to 13\,Jy at the end of observations.
The emission at 84.22\,\kms\, showed a flare lasting about 220\,d with a maximum near MJD 55400. 

\noindent
{\bf G39.100+0.491}. The feature at 14.63\,\kms\, decreased by a factor of three. The rest of the features showed moderate or no significant variations.

\noindent
{\bf G40.282$-$0.219}. The features of this source showed little or no significant variations except the feature at 77.76\,\kms\, which 
increased by a factor of three.

\noindent
{\bf G40.425+0.700}. The feature at 6.64\,\kms\, was moderately variable but the other features were stable.

\noindent
{\bf G42.304$-$0.299}. The major feature at 28.06\,\kms\, decreased by a factor of 1.5.

\noindent
{\bf G43.038$-$0.453}. The flux density of the 55.46\,\kms\, feature varied from 1 to 3\,Jy. The emission near 61.34\,\kms\, flared 
near MJD 55050 and 55590 (Section~\ref{sec:bursts}). 

\noindent
{\bf G43.149+0.013}. This well known site W49N contains at least four sources (\citealt{bartkiewicz14}; \citealt{breen15}) so some
features in the spectrum are blended. All confusion free features were not variable.

\noindent
{\bf G43.890$-$0.784}. The main features show little variation with different patterns.

\noindent
{\bf G45.071+0.132}. This source was not variable. It is interesting that our peak flux density of the main feature is similar to that 
observed with the Parkes telescope but very different from that derived from the ATCA observation (\citealt{breen15}).

\noindent
{\bf G45.467+0.053}. This source showed little or no variations to within the noise. The emission at 49.96\,\kms\, comes from the 
nearby source G45.445+0.069 (\citealt{breen15}) and showed moderate variation.

\noindent
{\bf G45.473+0.134}. The feature at 59.60\,\kms\, varied with a period of 196\,d (\citealt{szymczak15}). The emission from 
64.2 to 66.3\,\kms\, showed the same variability pattern. The feature at 62.23\,\kms\, remained constant. The emission in the velocity 
range of 55.8$-$58.3\,\kms\ comes from G45.493+0.126 (\citealt{bartkiewicz14}) and did not show variation. 

\noindent
{\bf G45.804$-$0.356}. The flux density of the main feature at 59.90\,\kms\, varied irregularly by a factor of two.

\noindent
{\bf G49.265+0.311}. The emission from $-$5.90 to $-$4.00\,\kms\, is composed of at least four blended features which showed moderate or 
small variations in flux density and drift in velocity up to $-$0.17\,\kms.

\noindent
{\bf G49.416+0.326}. All the features at velocities higher than $-$15.0\,\kms\, showed quasi-periodic (370$-$420\,d) variations with 
a relative amplitude of 0.8$-$1.5. The emission from $-$26.8 to $-$23.7\,\kms\, comes from G49.417+0.324 (\citealt{breen15}) and 
showed little variation.

\noindent
{\bf G49.482$-$0.402}. The emission of this source ranged from 44.5 to 52.0\,\kms\, and was significantly variable but it is confused
by the emission from G49.490$-$0.388 (\citealt{breen15}). The other features in the spectrum come from G49.490$-$0.338 and 
G49.489$-$0.369 (\citealt{breen15}) and are strongly blended.

\noindent
{\bf G49.599$-$0.249}. The emission in the velocity range 60.7$-$64.5\,\kms\ was highly variable (Section~\ref{sec:complex-spectra}).
The features in the velocity range 57.5$-$60.5\,\kms\, are side-lobe responses from G49.490$-$0.338 and G49.489$-$0.369 (\citealt{breen15}).

\noindent
{\bf G50.779+0.152}. The main feature at 49.01\,\kms\, increased by a factor of two.

\noindent
{\bf G52.199+0.723}. The feature at 3.19\,\kms\, increased by a factor of less than two, whereas the feature at 3.68\,\kms\, showed 
the opposite trend. This source is discarded from further analysis because of the large pointing offset.

\noindent
{\bf G52.663$-$1.092}. This source was significantly variable. The main features at 65.13 and 65.83\,\kms\, increased by a factor 
of about two. The 66.54\,\kms\, feature irregularly showed sudden increases to $\sim$4\,Jy, up from the detection limit of $\sim$0.8\,Jy. 

\noindent
{\bf G53.618+0.035}. The main feature at 18.87\,\kms\, decreased from $\sim$6 to $\sim$3\,Jy with a flare lasting $\sim$380\,d 
with a maximum near MJD 55390 (Section~\ref{sec:bursts}). The same pattern of variability was seen for the feature at 18.56\,\kms\, but with a relative 
amplitude of about three.

\noindent
{\bf G58.775+0.644}. This source was observed with a large pointing offset and is discarded from the statistical analysis.

\noindent
{\bf G59.783+0.065}. The features in the velocity range of 15.0$-$22.0\,\kms\, showed large and correlated variations. 
The strongest feature at 19.77\,\kms\, showed sinusoidal-like variations with a relative amplitude of 2.0$-$2.5 before 
MJD 55650 then dropped to a quiescent level of $\sim$18\,Jy. The peak at 27.02\,\kms\, was constant to within the noise level.

\noindent
{\bf G59.833+0.672}. The feature at 37.08\,\kms\, showed an increase from 6 to 10\,Jy modulated in a quasi-periodic manner. 
The flux density of feature at 36.08\,\kms\, has a scatter of more than 30\,per cent. 

\noindent
{\bf G69.540$-$0.976}. All the features showed weak variations.

\noindent
{\bf G70.181+1.742}. The variability indices listed in Table~\ref{varindices} are overestimated due to pointing offset of 3\farcm3. 
\cite{goedhart05} suggested that the source was variable. The source is discarded from further analysis.

\noindent
{\bf G73.06+1.80}. The features at $-$2.59 and 5.93\,\kms\ showed variations with a period of 160\,d (\citealt{szymczak15}). 
Synchronous periodic changes of a faint emission at 6.95\,\kms\, were seen only in certain cycles.

\noindent
{\bf G75.782+0.343}. Weak features near $-$10.93 and $-$9.48\,\kms\, showed periodic (120\,d) variations (\citealt{szymczak15}). 
The feature at $-$1.98\,\kms\, showed significant variability. The other features exhibit correlated variations with moderate or 
small amplitude. The values of variability indices are probably overestimated because of the pointing offset of 1\farcm3.
This source is discarded from further analysis.

\noindent
{\bf G78.122+3.633}. This source was highly variable during our monitoring (Section~\ref{sec:complex-spectra}). Inspection of
earlier observations reveals a double peaked spectrum in the velocity range from $-$7.5 to $-$5.5\,\kms\ in 1995 April (\citealt{slysh99}),
but a single-peaked asymmetric spectrum at $-$6.1\,\kms\ in 1999 (\citealt{szymczak00}) with an amplitude of 40-60\,Jy. \cite{goedhart04}
suggested significant variations from 2000 January to 2003 February. Thus the source varied significantly on time-scales up
to 18\,years.

\noindent
{\bf G80.861+0.383}. The main feature at $-$4.06\,\kms\, exhibited two maxima near MJD 55540 and 55950 of duration of 60$-$80\,d 
superimposed on a slight decrease of the quiescent level from 5 to 3\,Jy. A new feature appeared at $-$11.12\,\kms\, reaching a maximum 
of 4\,Jy near MJD 55950 then dropped to 1.6\,Jy over a month. Note that the values of variability indices are probably overestimated 
due to the pointing offset of 3\farcm0. The source is discarded from further analysis.

\noindent
{\bf G81.871+0.781}. The features at 3.42 and 4.03\,\kms\ showed correlated variations of similar relative amplitude. The features 
at 4.56, 5.13 and 5.75\,\kms, did not vary. The features at 7.20 and 9.39\,\kms\, showed variability with different patterns.   

\noindent
{\bf G85.410+0.003}. The feature at $-$31.60\,\kms\, linearly increased from 6 to 9\,Jy. The feature at $-$29.54\,\kms\, of 65\,Jy 
remained constant before MJD 55200, then doubled in intensity over three months and attained a mean value of 105\,Jy after MJD 55290.
The feature at $-$28.53\,\kms\, exponentially decreased from 65 to 22\,Jy over the monitoring period.

\noindent
{\bf G90.921+1.487}. The two main features exhibited similar quasi-periodic (150$-$320\,d) behaviour with a prominent flare near MJD 55965, 
when the feature at $-$69.23\,\kms\, rose by a factor of 4.6 over a 80\,d period. Note that the source was observed with the 1\farcm5
pointing offset which probably causes a large scatter of flux density.  The source is discarded from further analysis.

\noindent
{\bf G94.602$-$1.796}. All the features in the source showed quasi-periodic ($\sim$195\,d) variations of different amplitudes.

\noindent
{\bf G108.184+5.519}. This source showed little variation. 

\noindent
{\bf G109.871+2.114}. Almost all features are highly variable. The flux density variations of the blue- and red-shifted features 
are synchronised and anti-correlated and interim high-amplitude periodic variations occurred for some features (\citealt{sugiyama08}; 
\citealt{szymczak14}).

\noindent
{\bf G111.256$-$0.770}. The source was highly variable. The feature at $-$41.18\,\kms\, peaked near MJD 55320 but was marginally detected 
after MJD 55610. The feature at $-$38.90\,\kms\, systematically decreased in intensity by a factor of 6.5 before MJD 56180 then 
increased to its initial level. The feature at $-$37.84\,\kms\, experienced an 85\,day burst starting around MJD 55920 (Section~\ref{sec:bursts})
and a further few bursts after MJD 56170. The feature at $-$36.79\,\kms\, increased from 1.5 to 4.5\,Jy before MJD 55320, then gradually decreased 
to 1.1\,Jy near MJD 56275. 

\noindent
{\bf G111.542+0.777}. The source showed moderate variability. The flux density of feature at -61.35\,\kms\, was $\sim$80\,Jy before 
MJD 55200, then dropped to a mean value of 70\,Jy and remained at this level until MJD 55865 and then increased to about 100\,Jy. 
The features at $-$60.73 and $-$57.31\,\kms\, did not show variation. The feature at $-$58.10\,\kms\, showed a sinusoidal-like variation 
of 0.5 in the relative amplitude from MJD 55280 to 56010. The emission at $-$56.03\,\kms\, linearly decreased by a factor of 1.5. 
The emission at velocities higher than $-$53.4\,\kms\, comes from the nearby sources G111.553+0.756 and G111.567+0.752 
(\citealt{pestalozzi06}) which showed moderate variation.

\noindent
{\bf G121.298+0.659}. Faint ($<$2\,Jy) emission in the velocity range of $-$29.7 to $-$24.6\,\kms\, was strongly variable. 
The feature at $-$24.16\,\kms\, exponentially increased from 0.7 to $\sim$4\,Jy. The feature at $-$23.41\,\kms\, showed quasi-periodic 
($\sim$330\,d) changes superimposed on an exponential decrease and a velocity drift of $-$0.16\,\kms\, (Section~\ref{sec:drifts}). 
The peak flux density of feature $-$22.53\,\kms\ was irregularly modulated by a factor of less than two.

\noindent
{\bf G123.066-6.309}. The features of this source showed low-amplitude (10$-$15\,per cent) quasi-periodic (50$-$90\,d) variations.

\noindent
{\bf G136.845+1.167}. The main feature at $-$45.14\,\kms\, showed sinusoidal-like variation with a decreasing amplitude. 
The source is discarded from further analysis because the pointing offset of 1\farcm4.

\noindent
{\bf G173.482+2.446}. The source showed little or no variation within the noise. 

\noindent
{\bf G183.348$-$0.575}. The features at $-$15.20 and $-$14.68\,\kms\, are variable. The feature at $-$4.89\,\kms\, 
decreased from 17 to 11\,Jy before MJD 55135, then remained stable almost to the end of the monitoring period.

\noindent
{\bf G188.794+1.031}. The features at $-$5.54 and $-$5.15\,\kms\, are significantly variable with the same pattern. 
The emission near $-$5.15\,\kms\, was marginally detected after MJD 55470. A faint ($\sim$1\,Jy) feature near 10.79\,\kms\, 
is a side-lobe response from G188.946+0.886 (\citealt{green12}).

\noindent
{\bf G188.946$-$0.886}. This is a known periodic (416\,d) source with low-amplitude sinusoidal variations in all the features 
(\citealt{goedhart04}). The quiescent state flux density of the feature at 11.53\,\kms\, decreased by a factor of two.

\noindent
{\bf G189.030+0.784}. The source was observed with a 1\farcm3 pointing offset and the variability indices are not reliable.
The source is discarded from further analysis.

\noindent
{\bf G192.600$-$0.048}. A few features with faint (1$-$2\,Jy) emission at velocities lower than 3.4\,\kms\, showed irregular 
flare behaviour (Section~\ref{sec:bursts}). The major features are relatively stable with the exception of low ($<$20\,per cent) 
amplitude, short ($\sim$30\,d) flares near MJD 55935 and 56240.

\noindent
{\bf G196.454$-$1.677}. The source was highly variable and all the features showed the same variability pattern. The three 
major features increased by a factor of 3$-$13; quasi-periodic ($\sim$70\,d) modulations of increasing amplitude occurred 
between MJD 55530 and 55930 then the flux density of the features 14.69 and 15.13\,\kms\, decreased to its pre-flare levels.
\cite{goedhart04} reported a low amplitude periodic ($\sim$668\,d) variation which has been confirmed for longer time-series. 
Faint (1$-$2\,Jy) emission appeared near MJD 55930 near 13.8 and 19.6\,\kms. 

\noindent
{\bf G232.620+0.996}. This was a moderately variable source. All the features showed a linear increase by a factor 1.8.
  
\begin{figure*}   
\centering
\includegraphics[width=\textwidth,height=9in,trim=-0.15cm -0.15cm -0.15cm -0.15cm,clip]{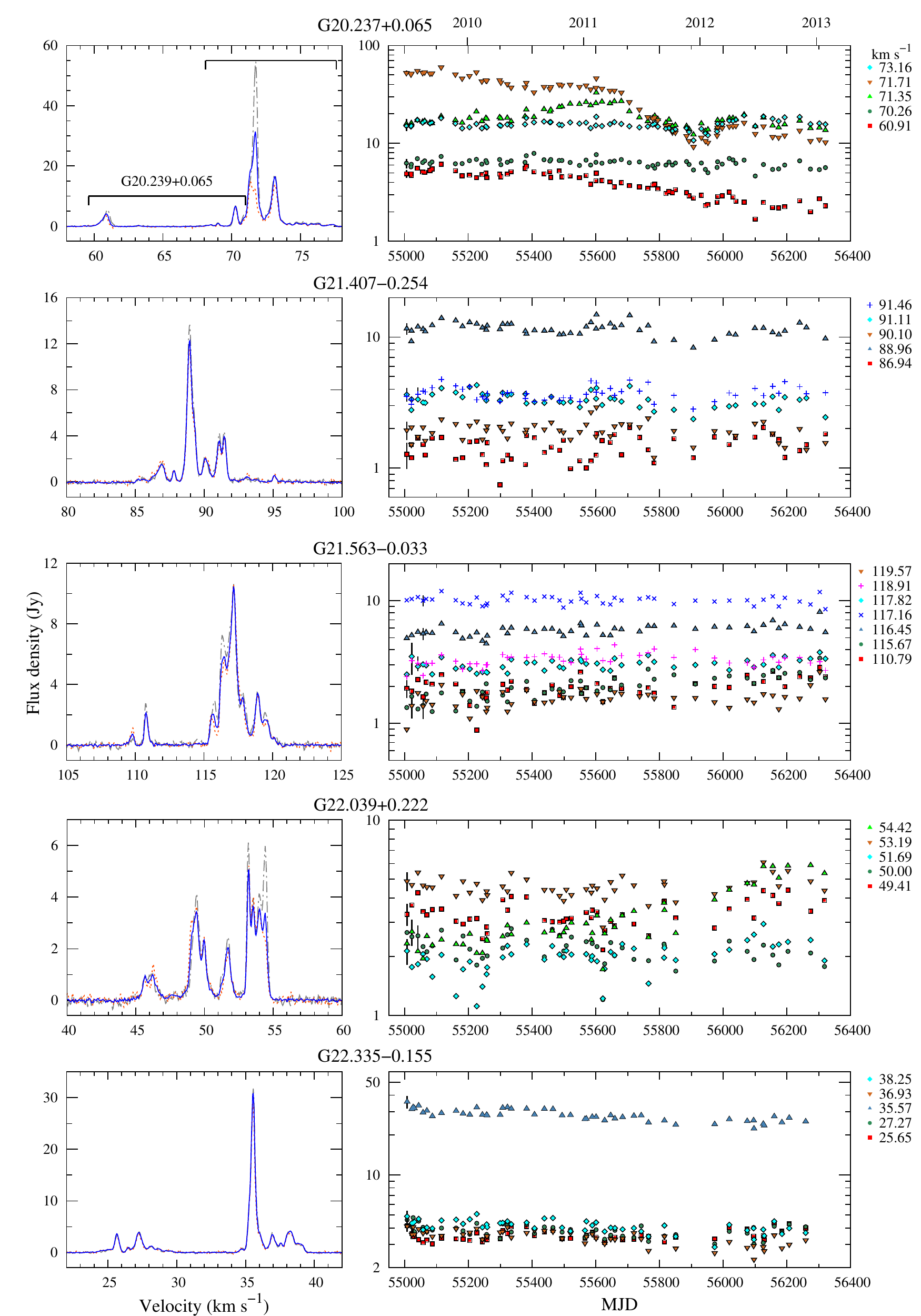}
\caption{Left panels: Spectra of 6.7\,GHz methanol masers showing the average (solid), high (dashed) and low (dotted) emission levels.
Right panels: Light curves of selected 6.7\,GHz methanol maser features. Typical measurement uncertainty is shown by the bar for one
of the first data points or is comparable to the symbol size. (Only a portion of this figure is shown here to demonstrate its form 
and content. The complete figure is available online.)
\label{spectra-lightcurves}}
\end{figure*}

\subsection{Variability statistics}
 \label{sec:var-stat}
\begin{figure}   
\resizebox{\hsize}{!}{\includegraphics[angle=0]{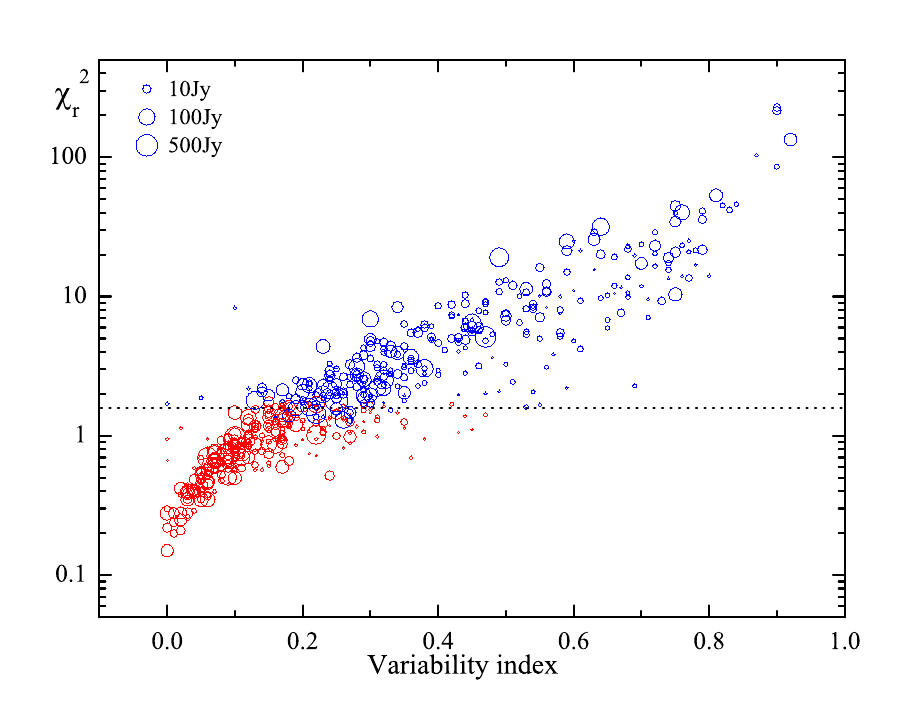}}
\caption{Reduced $\chi^2$ value versus the variability index for all maser features in the sample.
The dotted line marks the average threshold for variability detection. Non-variable and variable features are shown
with red and blue circles, respectively. The size of the symbol is proportional to the logarithm of the peak flux 
density of feature.   
\label{chi2_vi}}
\end{figure}

There are nine features with $VI\ge$0.35 which sit below the variability detection threshold (Fig.~\ref{chi2_vi}).
Close inspection of these features reveals that all but one are faint ($\la$2\,Jy) and their light curves include
a few points with extreme values that cannot be considered as outliers. For instance the feature 41.00\,\kms\, in G31.158+0.046 
has $VI=0.42$ but its fluctuation index is only 0.08 (Table~\ref{varindices}). All nine features have $FI\le$0.15. 
Figure~\ref{chi2_vi} also shows three features with $VI\le$0.10 above the variability detection threshold. 
These features are weak ($\le$3\,Jy) and one of them experiences a long ($\sim$500\,d) flare with the relative 
amplitude exceeding 2. The other two features show weak (2-4\,Jy) increasing and decreasing trends over the entire 
monitoring period. The fluctuation index of these three features ranges from 0.06 to 0.55. 
We conclude that the variability index does not always provide a reliable measure of variability for faint features
and it is insensitive for long lasting, low amplitude changes in the flux density. The fluctuation index is more 
useful in these cases.

\begin{figure}   
\resizebox{\hsize}{!}{\includegraphics[angle=0]{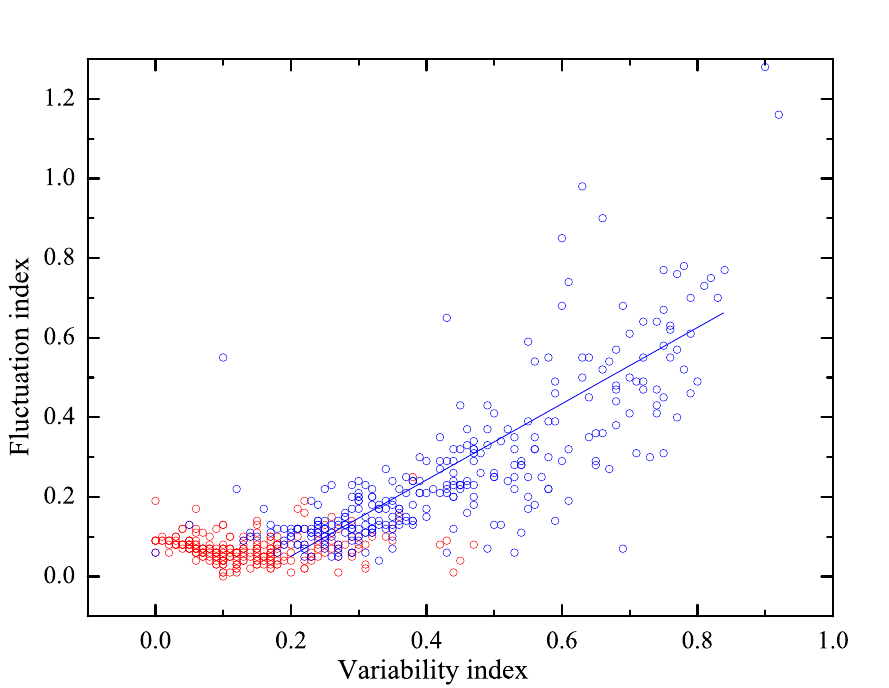}}
\caption{Comparison between the variability index and the fluctuation index for all features in the sample. 
$\chi^2$ detects variations in 303 features (blue circles) and does not detect in 263 features (red circles).
The line denotes a least-squares fit to the data after excluding non-variable features.
The circles correspond to 303 features where $\chi^2$ detects variations (blue) and 263 features where $\chi^2$ does not detect variations.
\label{vi_fi}}
\end{figure}

Figure~\ref{vi_fi} shows the fluctuation index versus variability index. It is clear that for $0.0<VI<0.35$ 
the fluctuation index of features not detected as variable with $\chi^2$-test is constant and its mean value
is $0.070\pm0.002$. For variable features with $VI>$0.20 and $FI>$0.05 these two indices are correlated with 
a correlation coefficient of 0.78 and for these limits $VI = FI - 0.17$ is obeyed. A few features with low $VI$ 
and high $FI$ have light curves with low S/N (signal to noise ratio) and experience low amplitude flares (e.g. the feature at 105.87\,\kms, 
in G33.093$-$0.073, Table~\ref{varindices}). Features with high $VI$ and low $FI$ are usually faint ($\la$2\,Jy)
and exhibit variability with low relative amplitudes. We confirm that all three measures of variability agree well
but their significance should be verified by considering the S/N of the feature.  

\begin{figure}   
\resizebox{\hsize}{!}{\includegraphics[angle=0]{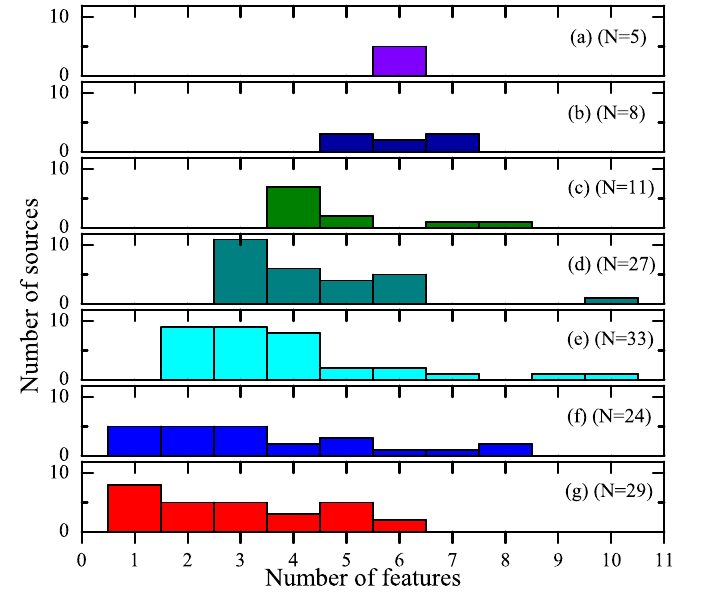}}
\caption{Histograms comparing the number of sources and variable features. Shown are the number of sources versus the number of features in the spectrum.
Panels from the top to the bottom represent the sources with decreasing number of variable features; (a) six features are variables, (f) one feature is variable.
Panel (g) represent the sources which all features are not variable using the $\chi_\mathrm{r}^2$ criterion. The total number of sources in each group is given in
the parentheses. Note that two sources with 7 and 8 variable features are not shown.  
\label{h_nonvar-var}}
\end{figure}

There are 137 sources in the sample for which variability parameters were unambiguously determined. 21.2~per cent (29/137) 
of them have spectra with all the features non-variable (Fig.~\ref{h_nonvar-var}). 41.6~per cent (57/137) of the sources have 1-10 
features in the spectra of which one or two features are variable. 27.7~per cent (38/137) of the sources have 3-4 variable
features in the spectra composed of 3-10 features. 29.9~per cent (41/137) of  sources showed variability for all their features.  

\subsection{Periodic variations}  
We find that 9 out of 137 masers with well determined variability measures are periodic sources, thus representing $\sim$6.5~per cent
of our sample. Spectral features with cyclic behaviour were detected in the following sources: G22.357+0.066 (179\,d), 
G24.148$-$0.009 (183\,d), G25.411+0.105 (245\,d), G30.400$-$0.296 (222\,d), G37.554+0.201 (237\,d), G45.473+0.134 (195\,d),
G73.06+1.80 (160\,d), G75.782+0.343 (120\,d) and G188.946+0.886 (416\,d). It is interesting that a few of the periodic sources have 
a period close to 180\,d. Our further observations after 2013, not reported here, ruled out possible seasonal effects for these sources.
A thorough analysis of the periodicity of five of these was reported elsewhere (\citealt{szymczak14}) and  a forthcoming paper will discuss the remainder.  
G196.454$-$1.677 has been noted as a sinusoidal-like variable with a period of $\sim$668\,d (\citealt{goedhart04}) but our monitoring
is probably too short to confirm this behaviour. On the other hand we have not detected any object with a period shorter than
100\,d probably due to the lower cadence for the majority of the sources. This may explain why our detection rate of periodic sources
is about half of that postulated by \cite{maswanganye15}.

\subsection{Complex variations}
 \label{sec:complex-spectra}
Several of the sources exhibit complex changes in the shape and intensity of the spectra which are not immediately visible 
in the light curves (Fig.~\ref{spectra-lightcurves}). In order to visualize the bulk variability of 6.7\,GHz maser emission 
we created dynamic spectra (Fig.~\ref{dyn-spectra}) using linear interpolation of the flux densities between consecutive 
32\,m telescope spectra taken at roughly monthly cadence or shorter. 

For G20.237+0.065 the shoulder near velocity 71.35\,\kms\, experienced a flare with a maximum of $\sim$27\,Jy near MJD 55620 
and the feature at 71.71\,\kms\, shows a periodically ($\sim$480\,d) modulated decline from 56 to 11\,Jy (Fig.~\ref{dyn-spectra}).

G23.707$-$0.198 showed complex variations in the entire spectrum (Fig.~\ref{dyn-spectra}). The emission near 78.21\,\kms\,  
became the strongest around MJD 55575 and then drifted to 77.81\,\kms\, at the end of the monitoring. The velocity drift of 
$\sim-$0.4\,\kms\,yr$^{-1}$ in this object, which has a systemic velocity of 68.9\,\kms\, may suggest an inflow motion with a
radial acceleration of 0.2\,\kms\,yr$^{-1}$, but it could be simply an effect of differing variability of blended features. The flux density 
in almost all spectral channels varied on time-scales from 5\,months to 3.6 years with markedly different patterns. 
The features from 74.2 to 77.0\,\kms\, exhibited flares lasting from 5 to 14 months.

A spectacular and complex burst occurred in G24.329+0.144 (Fig.~\ref{dyn-spectra}). Its properties are described in Section~\ref{sec:bursts}.

For 24.634$-$0.324, the emission at velocities higher than 43.2\,\kms\, shows complex variations (Fig.~\ref{dyn-spectra}).
The emission of 43.91\,\kms\, feature is a superposition of a linear increase from 1.5 to 4\,Jy and short (2$-$3 months) flares 
which profiles cannot be recovered due to sparse sampling. Weak ($\sim$2\,Jy) emission near 45.8\,\kms\, disappeared, whereas 
a new feature near 46.4\,\kms\, appeared about MJD 55420 and then monotonically increased. 

Weak emission of G25.826$-$0.178 at velocity higher than 94.5\,\kms\, shows complex changes (Fig.~\ref{dyn-spectra}).
Faint ($\sim$3-4\,Jy) flares at 99.53 and 95.36\,\kms\, occurred around MJD 55230. Its persistent features show irregular variations
in amplitude.
 
The emission in the velocity range of 61.8$-$66.3\,\kms\, of G49.599$-$0.249 is strongly blended and significantly variable with 
the relative amplitude up to 3.5 (Fig.~\ref{dyn-spectra}). The variations of individual features are not correlated which resulted 
in significant changes in the spectral shape over the monitoring period.

The spectrum of G78.122+3.633 experiences complex changes (Fig.~\ref{dyn-spectra}). The strongest feature at $-$6.14\,\kms\, showed 
moderate variability; its flux density of $\sim$33\,Jy was stable before MJD 55750, then slightly increased to $\sim$40\,Jy. 
The other features show correlated and strong variations; a shallow minimum occurred near MJD 55300, a large drop by a factor of 
4$-$15 occurred from MJD 55620 to 55880 and a modulated increase started after MJD 56430.

\subsection{Synchronous and anti-correlated variations}
 \label{sec:synchro} 
Synchronised and anti-correlated variations of maser features were first detected in G109.871+2.114 (\citealt{sugiyama08}).  
We have found further examples of similar variations in three sources.

\begin{figure}
\resizebox{\hsize}{!}{\includegraphics[angle=0]{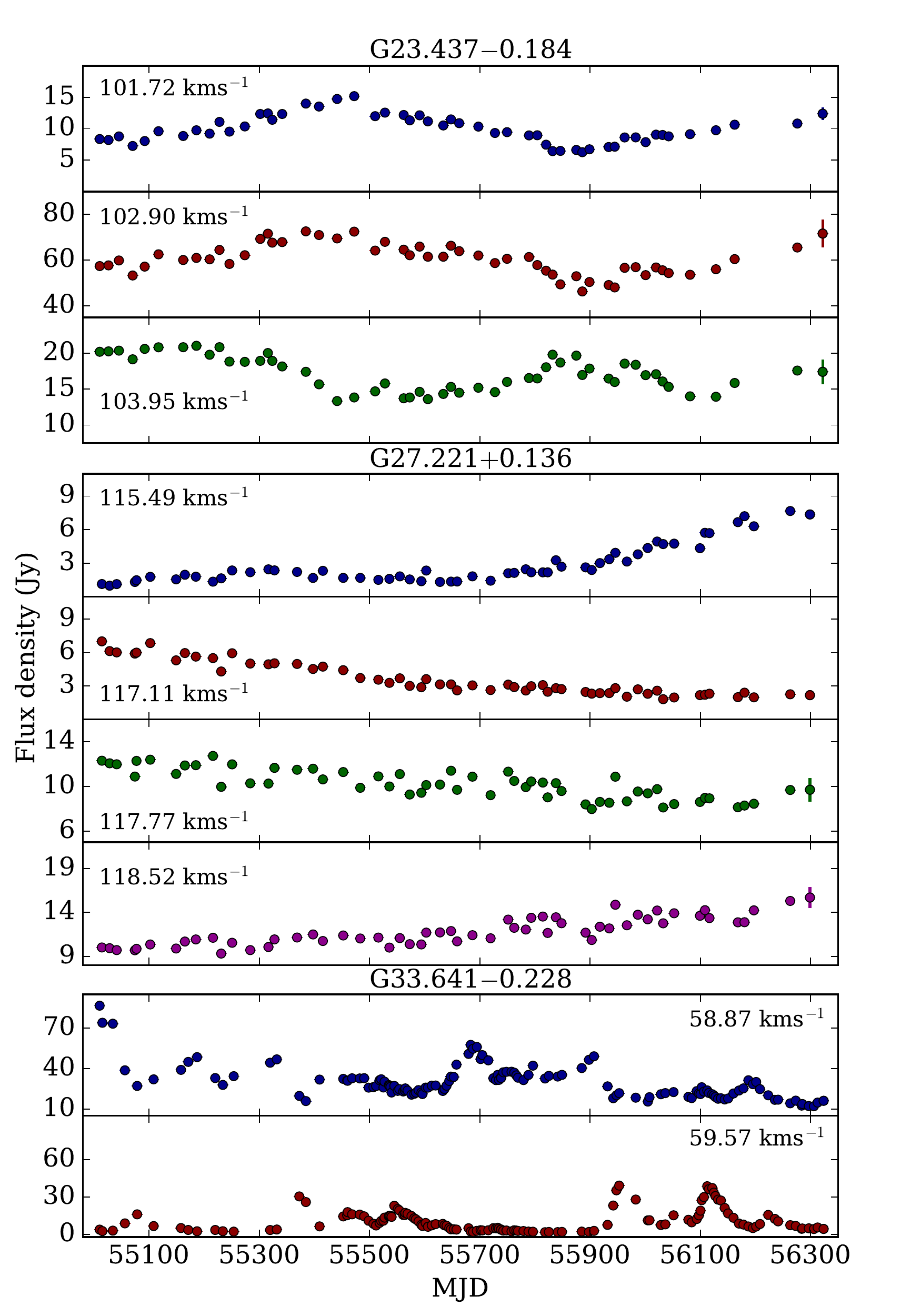}}
\caption{Examples of synchronous variations of the methanol maser features in the three sources. Each panel is labelled with
the velocity of feature. Typical measurement uncertainty is shown by the bar for the last data point or
is smaller or comparable to the symbol size.
\label{synchro}}
\end{figure}

For G23.437$-$0.184 the flux density in the range 101.5 to 102.9\,\kms\, varies with a sinusoidal like pattern, while 
synchronous and anti-correlated variations occur for the emission of 103.3$-$104.5\,\kms.  The light curves of features at 
101.72 and 103.95\,\kms\, are shown in Figure~\ref{synchro} and the correlation coefficient is $-0.39$.
The variability pattern changes characteristics around 103.12\,\kms.  
The interferometric maps imply that the emission comes from a warped linear structure with a clear velocity gradient 
(\citealt{hu16}) which seems to be a strong indication of a circumstellar disc seen nearly edge-on.

Features peaked at 117.11 and 117.77\,\kms\, in G27.221+00.136 show a slight decrease over the entire
observing period, while that at 118.52\,\kms\, increased (Fig.~\ref{synchro}). Correlation coefficients for the variations of 
the feature pairs 115.49/117.11\,\kms\, and 117.77/118.52\,\kms\, are $-0.59$ and $-0.57$, respectively.
VLBI maps revealed a core-halo structure of the emission with size of 100\,mas and very complex velocity structure (\citealt{bartkiewicz09}).
The components with anti-correlated variations coincide to within less than 2\,mas.

Features at 58.87 and 59.57\,\kms\, in G33.641$-$0.228 show anti-correlated and synchronised variability with the correlation 
coefficient of $-0.47$ over the entire data range (Fig.~\ref{synchro}). The components corresponding to these features coincide 
to within 10\,mas (\citealt{hu16}) and are located in the south-west edge of an arched structure (\citealt{bartkiewicz09}).

In agreement with previous studies (e.g. \citealt{cesaroni90}), we suggest that the detection of synchronised and 
anti-correlated variability in the sources showing disc-like morphologies supports a model of maser amplification 
with switching between radial and tangential modes (\citealt{szymczak14}). 

\subsection{Burst activity}
 \label{sec:bursts}
\subsubsection{Short lived bursts} 
\begin{figure}   
\resizebox{\hsize}{!}{\includegraphics[angle=0]{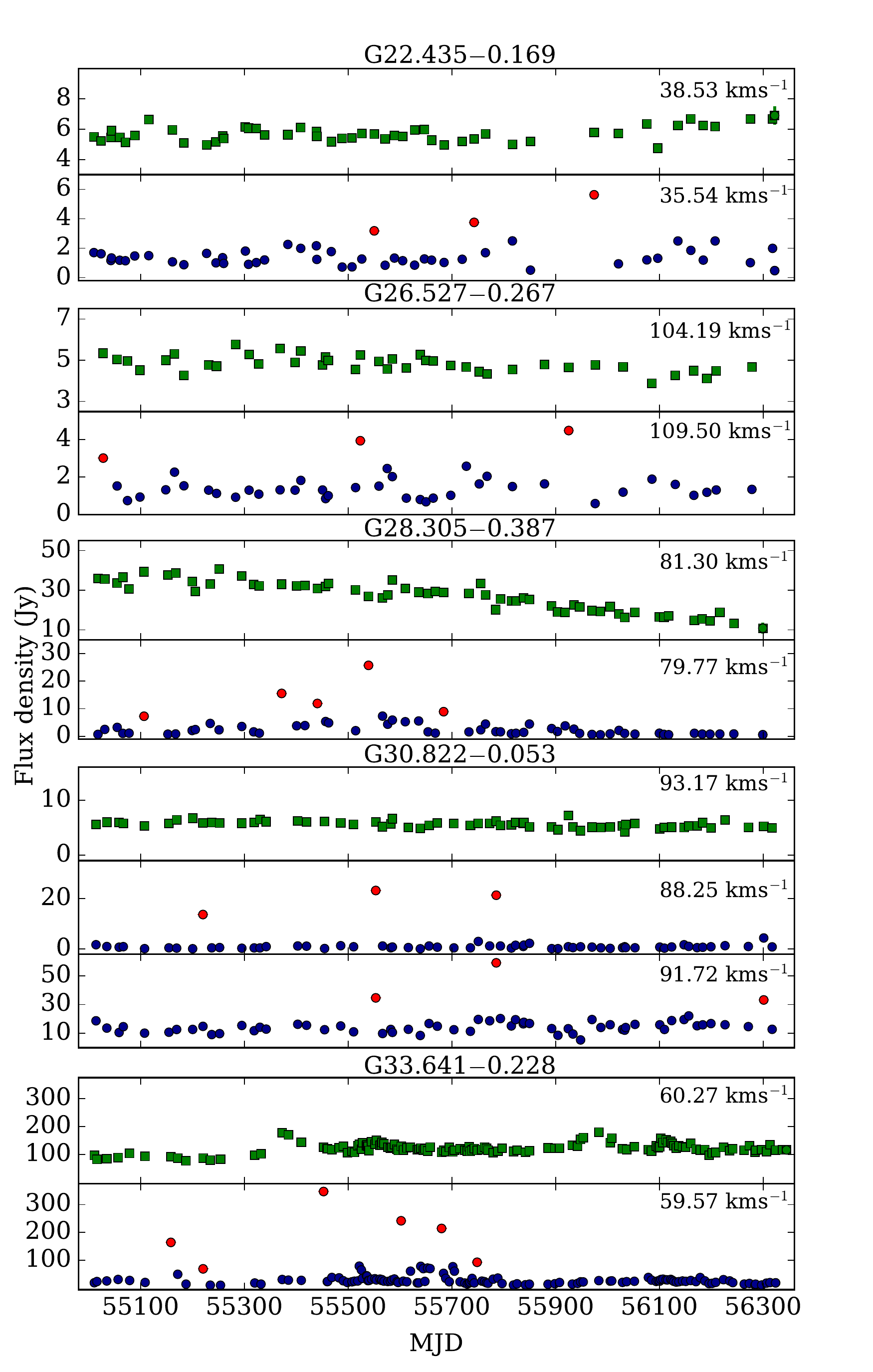}}
\caption{Examples of short flares in the methanol maser line. Light curves of flaring and comparison 
features are represented by circles and squares, respectively. Red circles show the 'flickering' events.
Typical measurement uncertainty is smaller or comparable to the symbol size.
\label{flicker}}
\end{figure}

Some maser features exhibit large and rapid flux density enhancements relative to their stable states.  In most cases these events 
are represented by outliers in the light curves and mimic a chain of corrupted spectral channels in the dynamic spectrum. 
Figure~\ref{flicker} displays the light curves of `flickering' features in the five sources and
Table~\ref{short-bursts} lists their parameters. Each burst meets the criterion of having a relative amplitude higher than two.

\begin{table}
 \caption{List of short lived bursts. The velocity of bursting feature $V_\mathrm{p}$, time of flare peak MJD, 
          peak flux density $S_\mathrm{p}$ and relative amplitude $R$ 
          are given.
\label{short-bursts}}
\begin{tabular}{l l r c c}
\hline
Name            & $V_\mathrm{p}$   & MJD    &   $S_\mathrm{p}$   & $R$ \\
                & (km\,s$^{-1}$)   &        &       (Jy)         &     \\
\hline
G22.435$-$0.169 & 35.54    & 55550 &   3.2  &  3.3  \\
                &          & 55743 &   3.8  &  2.6  \\
                &          & 55974 &   5.6  & 10.0  \\   
G26.527$-$0.267 & 109.50   & 55028 &   3.0  &  3.1  \\
                &          & 55523 &   4.0  &  2.9  \\
                &          & 55925 &   4.5  &  6.7  \\   
G28.305$-$0.387 &  79.77   & 55106 &   7.3  &  8.6  \\
                &          & 55371 &  15.5  &  13.1  \\
                &          & 55440 &  11.9  &  2.1  \\
                &          & 55539 &  25.7  & 11.7  \\
                &          & 55684 &  8.9   &  7.0  \\           
G30.822$-$0.053 &  88.25   & 55220 &  13.7  & 92.0  \\
                &          & 55553 &  23.2  & 44.5  \\
                &          & 55785 &  21.3  & 53.2  \\
                &  91.72   & 55553 &  34.6  &  2.5  \\
                &          & 55785 &  60.0  &  2.9  \\
                &          & 56301 &  33.2  &  2.2  \\           
G33.641$-$0.228 &  59.57   & 55158 &  164.5 & 10.1  \\
                &          & 55220 &   69.7 &  5.4  \\
                &          & 55452 &  346.8 & 13.5  \\
                &          & 55602 &  241.9 & 10.6  \\
                &          & 55680 &  214.5 &  5.0  \\
                &          & 55748 &   93.3 &  3.9  \\       
\hline              
\end{tabular}
\end{table}

G22.435$-$0.169 experienced at least three rapid flares at 35.54\,\kms. We consider the events as real 
because they were seen in the two polarisations and were absent in the adjacent feature at 38.53\,\kms. 
The source was mapped with high angular resolution at two different occasions (\citealt{bartkiewicz09}; 
\citealt{hu16}) and no emission was detected at 35.54\,\kms\, suggesting that the relative amplitude 
during flares is higher than 100.

Three `flickering' events occurred in G26.527$-$0.267 at 109.50\,\kms\, with relative amplitudes of 3-7.  
The feature at 79.77\,\kms\, in G28.305$-$0.387 flickered five times (Fig.~\ref{flicker}). In this source 
the emission of the feature comes from a region located $\sim$160\,mas away from the strongest emission
at 80.8 to 82.5\,\kms\, (\citealt{hu16}).  This may suggest that the flickering feature arises from 
a region of weak and diffuse maser emission located far from the central core structure. One can speculate 
that in those environments the maser is likely to be unsaturated and even a tiny contrast in density or 
maser path length may produce significant changes in the output flux density.

Synchronised and rapid flares occurred at MJD 55553 and 55785 for features 88.25 and 91.72\,\kms\, of G30.822$-$0.053 (Fig.~\ref{flicker}).
The first feature flickered also at MJD 55220 while the last feature flared also at MJD 56301. The emission at 88.25\,\kms\,
declined down to the level of detectability that implies the relative amplitude of flickers higher than 90. 

High amplitude `flickering' occurs in G33.641$-$0.228 at 59.57\,\kms\, (Fig.~\ref{flicker}). The flux density of this feature 
never dropped below the detection limit. High resolution data (\citealt{bartkiewicz09}; \citealt{hu16}) indicate that this emission 
comes from the S-W edge of an arched structure $\sim$10\,mas away from the region of anti-correlated and synchronous emission at 
58.87 and 59.26\,\kms\, discussed in Section~\ref{sec:synchro}.
High cadence observations over 294\,d in 2009-2012 reveal five flares of the 59.57\,\kms\, feature (\citealt{fujisawa14c}).
One of these flares is evident in our data; they reported the 506\,Jy peak at MJD 55449 while we found 347\,Jy peak three days 
later. This event was rarely sampled and one can determine only the lower limit of the relative amplitude of 13.5. \cite{fujisawa14c} 
derived the typical rise and fall times of one and five days, respectively. They suggest that such short lived bursts are powered 
by energy released in magnetic reconnection. Our data for G33.641$-$0.228 indicates that the width at half-maximum of the flickering 
feature is 0.21\,\kms\, in a quiescent state and drops to 0.20\,\kms\, in the flares. This precludes a line narrowing effect for 
the possibly unsaturated amplification.

The present survey indicates that short lived bursts of the maser emission are not unique to G33.641$-$0.228. These bursts are
observed for one feature of the spectrum or synchronously occurred for more features in restricted velocity range. The relative 
amplitude of the flickers ranges from 2 to 92 with the median value of 6.7.

\subsubsection{Long lived bursts}
\begin{figure}  
\centering
\includegraphics[angle=0,scale=0.40]{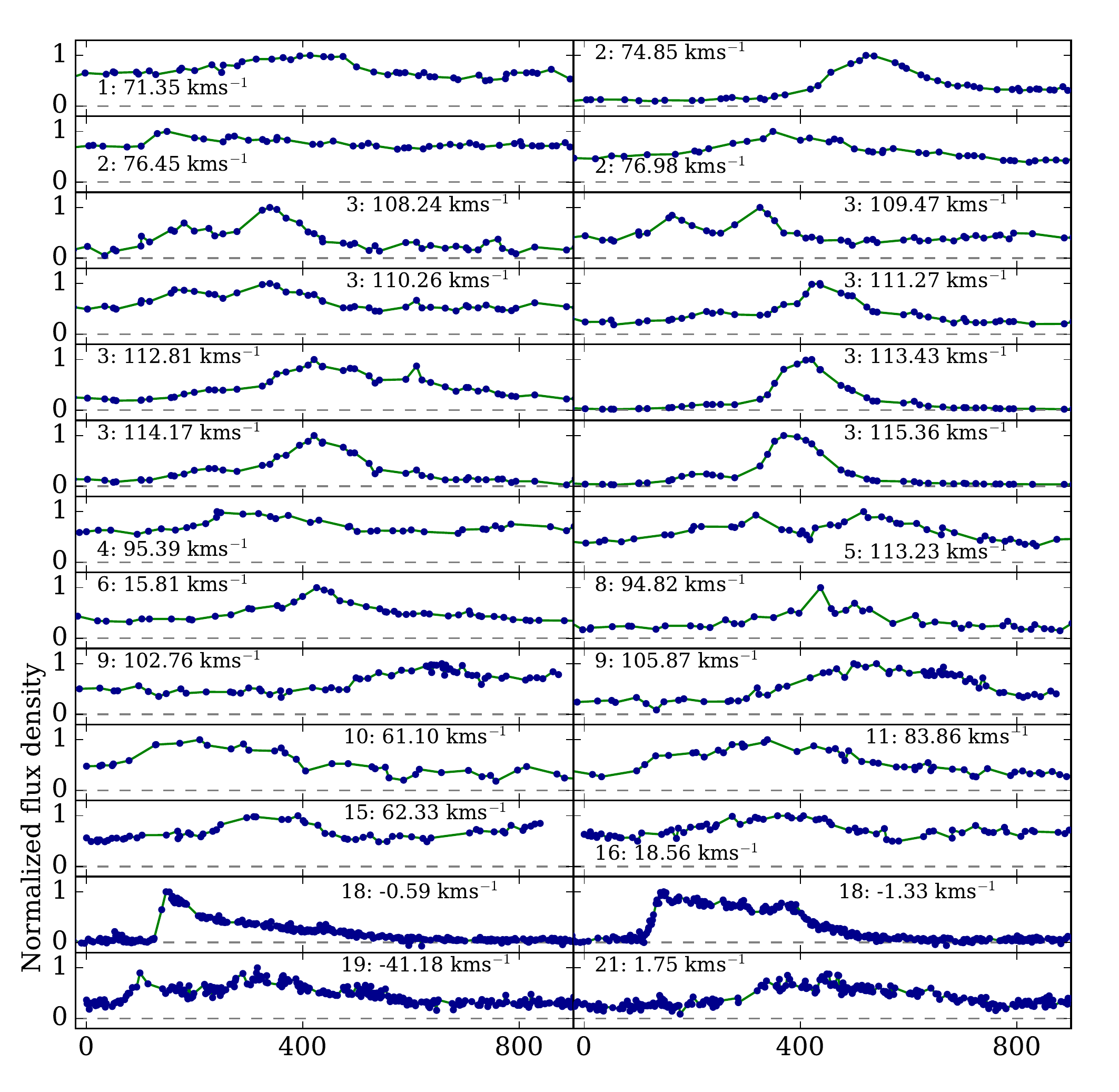}
\includegraphics[angle=0,scale=0.40, trim= 0 0 0 18]{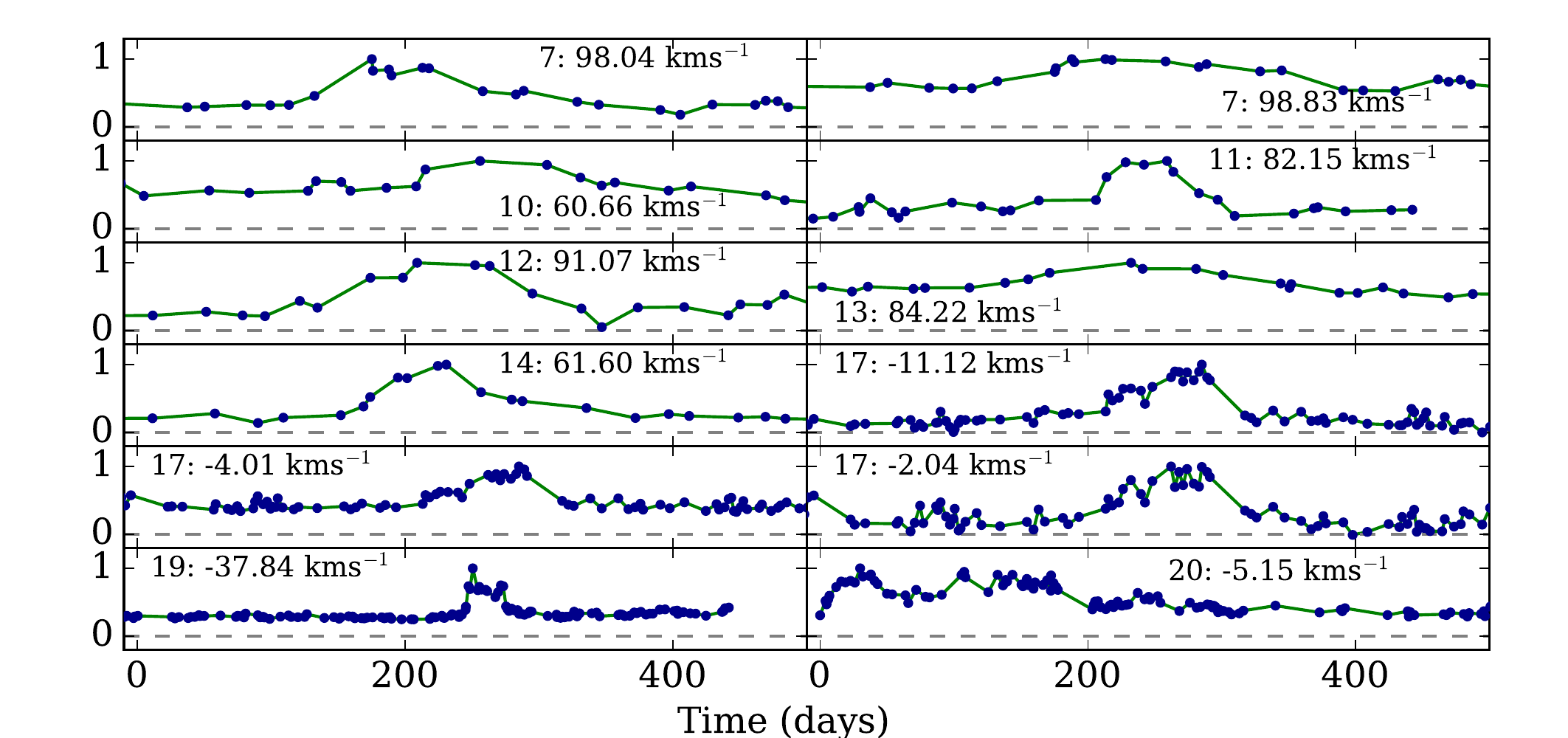}
\caption{Selection of flare profiles of the methanol maser features. Each profile is labelled with the number of source as
given in Table~\ref{long-bursts} and the radial velocity of the feature. Note different time-scale in the upper and lower panels.
\label{flares-all}}
\end{figure}

Some maser features show a single bursting episode on all time-scales that could be probed effectively in our monitoring.  
Generally the flares are recognised as events for which the post- and pre-flare flux densities are equal but there is evidence 
for bursts of features showing a gradual rise or fall over the entire monitoring period. Our main criterion for the recognition 
of a burst is a relative amplitude higher than 0.4. The most typical flare profiles are shown in Figure~\ref{flares-all} and 
their parameters are listed in Table~\ref{long-bursts}.

\begin{table*}
 \caption{List of long lived bursts. The velocity of bursting feature  $V_\mathrm{p}$, time of flare peak MJD, rise $t_\mathrm{r}$ 
   and decay $t_\mathrm{d}$ times, maximum flux density $S_\mathrm{max}$  and relative amplitude $R$ are given. The last two columns 
   list the slope fitted to the relationship between the line-width $\Delta$v and flux density, together with its correlation coefficient.
\label{long-bursts}}
\begin{tabular}{l l l r c c c c c c}
\hline
No &Name            & $V_\mathrm{p}$   & MJD    & $t_\mathrm{r}$ & $t_\mathrm{d}$ & $S_\mathrm{max}$   & $R$ & ln($\Delta$v)/ln($S$)  &  r \\
   &              & (km\,s$^{-1}$)   &        &       (d)      &    (d)         & (Jy)         &     &  & \\
\hline
1  & G20.237+0.065   &  71.35          & 55620  &  283         &   143         &   27.4        & 0.57 & -0.29 & -0.54 \\ 
2  & G23.707$-$0.198 &  74.85          & 55832  &  149         &   211         &    9.9        & 2.96 &  0.33 &  0.67 \\
   &                 &  76.45          & 55460  &   48         &   426         &   15.4        & 0.45 &       &       \\
   &                 &  76.98          & 55645  &  118         &   185         &   18.2        & 0.45 &       &       \\
3  & G24.329+0.144   & 108.24          & 55592  &   80         &    57         &    1.7        & 2.53 &  0.05 &  0.09 \\
   &                 &                 & 55750  &   60         &    98         &    2.5        & 3.24 &       &       \\
   &                 & 109.47          & 55574  &   61         &    75         &    4.0        & 1.00 &  0.03 &  0.11 \\
   &                 &                 & 55736  &   72         &    85         &    4.7        & 2.05 &       &       \\
   &                 & 110.26          & 55574  &  108         &    90         &   13.0        & 0.88 &  0.05 &  0.31 \\
   &                 &                 & 55750  &   86         &   149         &   14.1        & 1.25 &       &       \\
   &                 & 111.27          & 55847  &   96         &   248         &    6.2        & 9.33 & -0.19 & -0.88 \\
   &                 & 112.81          & 55832  &  266         &   372         &   14.0        & 2.78 &  0.20 &  0.81 \\
   &                 & 113.43          & 55638  &   70         &    52         &    5.7        & 1.56 & -0.22 & -0.96 \\
   &                 &                 & 55832  &  142         &   224         &   50.0        &22.04 &       &       \\
   &                 & 114.17          & 55638  &  110         &    52         &    1.9        & 1.75 &       &       \\
   &                 &                 & 55832  &  142         &   243         &    5.0        & 8.43 &       &       \\
   &                 & 115.18          & 55805  &  115         &   148         &   65.7        &11.23 &       &       \\
   &                 & 115.36          & 55638  &  110         &    52         &   13.1        & 3.60 &       &       \\
   &                 & 116.54          & 55832  &  142         &   379         &    2.8        & 3.07 &  0.01 &  0.02 \\
   &                 & 119.83          & 55832  &  264         &   388         &    4.5        & 2.49 &  0.04 &  0.21 \\
4  & G29.955$-$0.016 &  95.39          & 55655  &  125         &   259         &   48.7        & 0.85 &  0.11 &  0.74 \\
5  & G30.225$-$0.180 & 113.23          & 55586  &  149         &   199         &   13.8        & 1.16 &  0.05 &  0.24 \\
   &                 & 113.23          & 55919  &  134         &   264         &   18.5        & 1.89 &       &       \\
6  & G31.059+0.093   &  15.81          & 55642  &  192         &   165         &   14.9        & 2.05 &  0.02 &  0.13 \\
   &                 &  16.46          & 55642  &  188         &   151         &   21.1        & 0.79 & -0.02 & -0.17 \\
7  & G31.581+0.077   &  95.54          & 55335  &  101         &   176         &    5.4        & 0.69 &  0.04 &  0.11 \\
   &                 &  98.04          & 55297  &   63         &   214         &    4.2        & 2.23 &       &       \\
   &                 &  98.83          & 55310  &   76         &   201         &    8.9        & 0.63 &  0.17 &  0.35 \\
8  & G32.045+0.059   &  94.12          & 55761  &  206         &   129         &   74.1        & 2.82 &       &       \\
   &                 &  94.82          & 55821  &  266         &    69         &   53.6        & 1.76 & -0.32 & -0.88 \\      
9  & G33.093$-$0.073 & 102.76          & 56107  &  160         &    63         &    7.5        & 0.68 & -0.09 & -0.16 \\
   &                 & 105.87          & 55971  &  221         &   298         &    3.3        & 3.35 & -0.26 & -0.74 \\
10 & G34.244+0.133   &  60.66          & 55687  &   48         &    90         &    2.8        & 0.44 &  0.15 &  0.19 \\
   &                 &  61.10          & 55280  &  168         &   156         &    4.6        & 0.92 &  0.20 &  0.48 \\
11 & G36.115+0.552   &  82.15          & 56130  &  100         &    68         &    9.4        & 1.35 &  0.21 &  0.87 \\
   &                 &  83.86          & 55645  &  242         &   270         &    9.7        & 1.41 & -0.15 & -0.48 \\
12 & G37.598+0.425   &  85.80          & 55384  &  129         &    83         &   13.8        & 2.77 & -0.01 & -0.06 \\
   &                 &                 & 55788  &  145         &   111         &   11.6        & 1.13 &       &       \\
   &                 &  87.12          & 55387  &  132         &    80         &   24.9        & 1.76 &  0.03 &  0.17 \\
   &                 &                 & 55710  &   67         &   174         &   20.6        & 0.53 &       &       \\
   &                 &  88.57          & 55329  &   74         &   165         &    3.8        & 1.18 &  0.23 &  0.56 \\
   &                 &  91.07          & 55373  &  118         &    79         &    5.3        & 2.35 & -0.28 & -0.87 \\
   &                 &  93.13          & 55384  &  143         &    83         &    3.5        & 6.00 & -0.08 & -0.32 \\
13 & G38.203$-$0.067 &  81.85          & 55403  &   94         &   111         &    4.0        & 0.74 &  0.23 &  0.58 \\
   &                 &  84.22          & 55403  &  121         &   118         &   11.1        & 0.82 &  0.06 &  0.27 \\
14 & G43.038$-$0.453 &  61.34          & 55590  &  122         &   238         &    2.1        & 4.89 & -0.20 & -0.42 \\
   &                 &  61.60          & 55590  &   79         &   141         &    2.3        & 3.89 & -0.20 & -0.29 \\
15 & G49.599$-$0.249 &  62.33          & 55324  &   80         &   142         &   12.7        & 0.39 & -0.56 & -0.95 \\
16 & G53.618+0.035   &  18.56          & 55387  &  263         &   191         &    5.2        & 0.76 & -0.01 & -0.05 \\
   &                 &  18.87          & 55385  &  266         &   210         &    8.4        & 0.90 &       &       \\
17 & G80.861+0.383   &$-$11.12         & 55930  &   56         &    49         &    4.2        & 1.80 & -0.07 & -0.29 \\
   &                 &$-$4.01          & 55946  &   72         &    37         &   11.1        & 1.83 & -0.01 & -0.04 \\
   &                 &$-$2.04          & 55946  &   72         &    33         &    1.7        & 1.83 & -0.09 & -0.22 \\
18 & G109.871+2.114  &$-$1.33          & 55252  &   41         &   419         &    6.1        &14.38 &       &       \\
   &                 &$-$0.59          & 55310  &   32         &   474         &    7.0        &16.50 &       &       \\
19 &G111.256$-$0.770 &$-$41.18         & 55298  &   33         &   121         &    3.3        & 1.00 & -0.25 & -0.46 \\
   &                 &$-$37.84         & 55908  &   11         &    38         &    7.0        & 5.67 & -0.10 & -0.51 \\
20 & G188.794+1.031  &$-$5.15          & 55057  &   30         &   283         &    3.9        & 8.75 & -0.08 & -0.33 \\
21 &G192.600$-$0.048 &1.75             & 55466  &   99         &   391         &    2.6        & 5.57 & -0.20 & -0.37 \\
\hline              
\end{tabular}
\end{table*}

There are 19 sources in the sample for which one or more features experienced a single burst. 
In G24.329+0.144 and G37.598+0.425 two or more features showed multiple flaring episodes. The first source shows
a remarkable global variability where all the features experience synchronised flares (Fig.~\ref{dyn-spectra}). 
For instance, from MJD 55466 to 55780 the emission at 115.36\,\kms\, rose by a factor of 57. This extreme flare 
was preceded by two weaker bursts peaked at MJD 55314 and 55793 with a relative amplitude of 2.7 and 12.7,
respectively and followed by two faint bursts around MJD 56013 and 56122 with relative amplitudes of 3.3 and 0.9, 
respectively. Here the relative amplitude is with regard to the initial flux density of 1.2\,Jy. The intervals 
between repeated flares were 312, 164, 222 and 105\,d, respectively. The feature profile changed during the brightest
flare showing a velocity drift of $-$0.1\,\kms. The feature peaking at 113.43\,\kms showed a similar variability 
pattern with a phase lag of $\sim$20\,d. If this phase lag is due to the difference in light travel time then 
the two components are separated by $\sim$3400\,au. The angular separation of the components is $\sim$0\farcs3
(\citealt{bartkiewicz16}; \citealt{hu16}), which corresponds to $\sim$2800\,au at a distance of 9.4\,kpc.  
It is striking that the two estimates agree well within the error budget.  Thus, the flares are likely to be caused 
by variations in the pump rate triggered by significant changes in the output of a central powering source repeated 
on time-scales of 100-300\,d.

\begin{figure}   
\resizebox{\hsize}{!}{\includegraphics[angle=0]{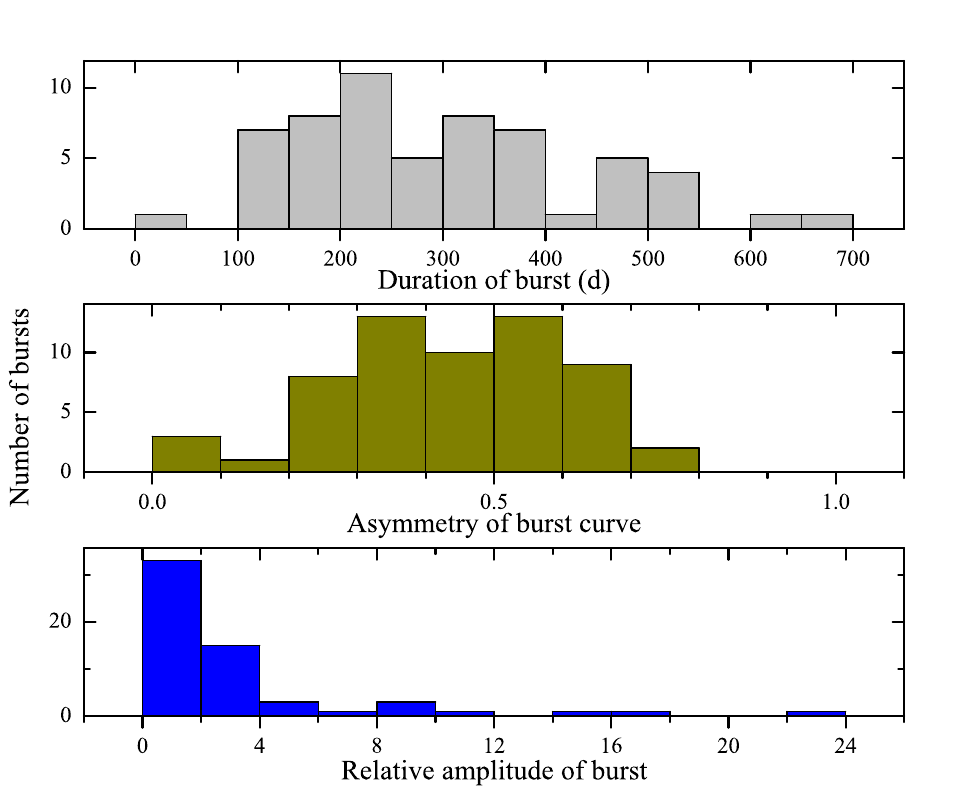}}
\caption{Histograms of long lived methanol maser bursts. The duration of bursts, the asymmetry of flare profile and the relative
amplitude of bursts are shown.
\label{long-hist}}
\end{figure}

The duration of the bursts ranges from 49 to 652\,d with average and median values of $296\pm18$ and 277\,d, respectively
(Fig.~\ref{long-hist}). The asymmetry of the flare curves, defined as the ratio of the duration of the rising branch 
to the total duration of the burst, ranges from 0.06 to 0.79. The median value of 0.43 implies a symmetric flare profile 
for the majority of the burst episodes. The relative amplitude of bursts varies from 0.4 to 22.0 with a median of 1.8 
(Fig.~\ref{long-hist}).

There are several flaring features which appear to be well separated in velocity and we attempted to fit Gaussian 
functions to their profiles in order to examine the relation between the line-width at half-maximum ($\Delta$v) 
and the peak flux density ($S_{\mathrm{p}}$). 

\begin{figure} 
\resizebox{\hsize}{!}{\includegraphics[angle=0]{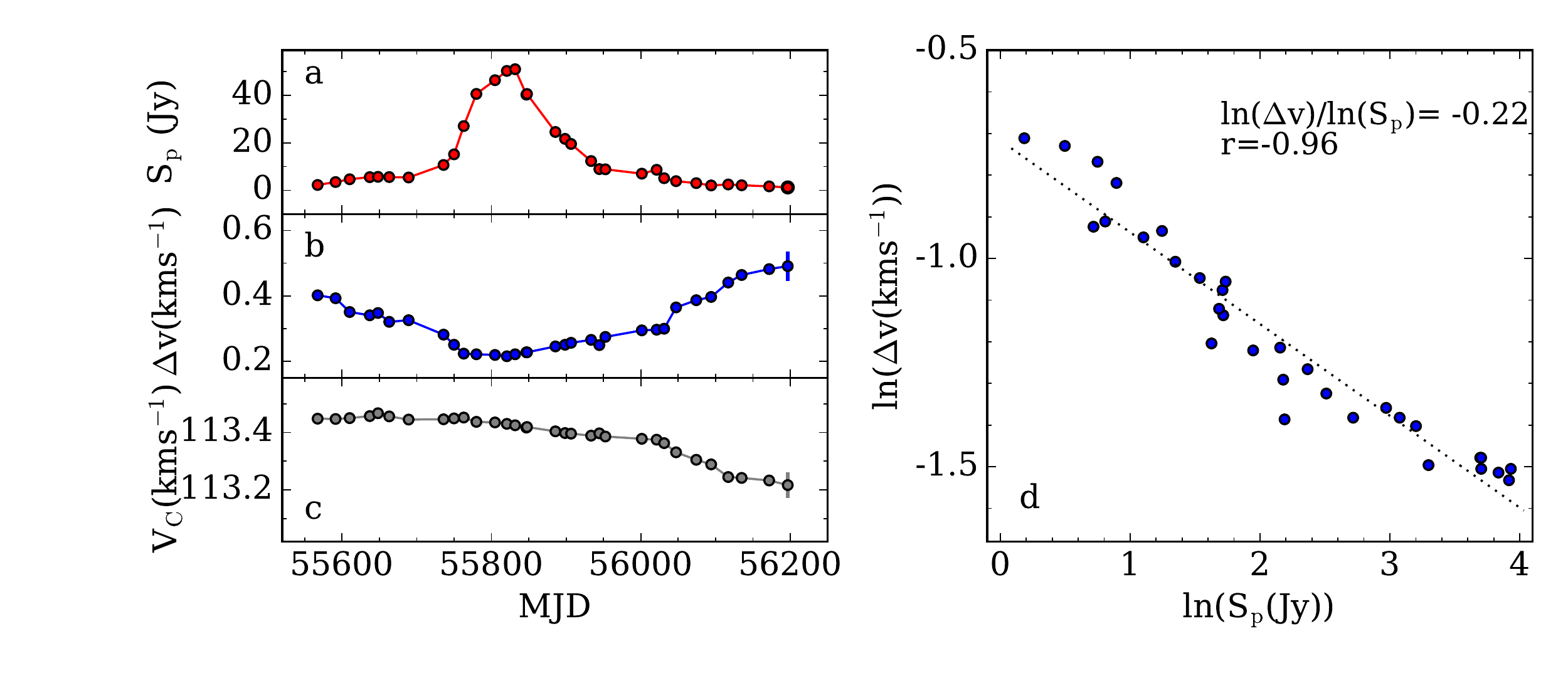}}
\caption{Line-width and flux density relationship for the 113.43\,\kms\, feature in G24.329+0.144. 
The line parameters obtained from Gaussian fitting (a) the flux density at feature peak $S_\mathrm{p}$, 
(b) the line-width at half maximum $\Delta$v, (c) the velocity of the profile centre $V_\mathrm{c}$ 
are shown as a function of time. Typical measurement uncertainty is shown by the bar for the last data point
either is smaller or comparable to the symbol size.
(d) Line-width versus peak flux density is shown, the dotted line 
is the fitted relation to the data points and its slope and correlation coefficient are given.
\label{fig:dvds_24}}
\end{figure}

Columns 9 and 10 of Table~\ref{long-bursts} give the slope fitted to the ratio ln($\Delta$v)/ln($S_{\mathrm{p}}$) and its 
corresponding correlation coefficient. The highest correlation coefficients of $r< -0.8$ are only for 5 features of which 
the slopes of ln($\Delta$v)/ln($S_{\mathrm{p}}$) ranges from $-0.19$ to $-0.56$. Figure ~\ref{fig:dvds_24} shows the plots 
of the flux density, line-width at half maximum and velocity of the profile centre versus time and the line-width versus 
peak flux density for the feature 113.43\,\kms\, of G24.329+0.144. We conclude that only a few features may obey 
the line-width $-$ flux density relation predicted for unsaturated amplification (\citealt{goldreich74}). Line narrowing 
is not obvious for the majority of features, probably due to the transformation of spectral features during the flares 
and to spatial blending.

\subsection{Velocity drifts}
 \label{sec:drifts}
Seven sources in the sample have individual spectral features showing velocity drifts of 0.15$-$0.26\,\kms\, over the entire observing period.
We fitted Gaussian curves to the observed line profiles of features with average peak velocities of 82.81, 82.89, 74.55 and $-$23.41\,\kms\ 
in G23.207$-$0.378, G31.047+0.356, G36.115+0.552 and G121.298+0.659, respectively. The dynamic spectra of these features are shown in Fig.~\ref{dyn-drift}.

The feature at 82.81\,\kms\, from G23.207$-$0.378 drifts at rate of $-$0.035\,\kms\,yr$^{-1}$, which may suggests an inflow motion, since
the systemic velocity of the source is 79.2\,\kms. Inspection of high angular resolution maps taken at MJD 53320 (\citealt{bartkiewicz09})
revealed a single maser cloud at 82.84\,\kms, which is also seen in contemporaneous VLA data (\citealt{hu16}). This indicates the persistence 
of the maser cloud on a time-scale of $\sim$7\,yr but its structure has evolved considerably over the period of monitoring.

The feature at 82.89\,\kms\, from G31.047+0.356 has a Gaussian profile and drifts at the rate of $-$0.036\,\kms\,yr$^{-1}$. This linear drift is
still preserved when the VLBI data taken at MJD 54173 (\citealt{bartkiewicz09}) are considered. The emission of the feature comes from a single maser 
cloud separated from a nearby cloud by 25\,mas. The systemic velocity of the source is 77.6\,\kms\, therefore the observed velocity drift can indicate inflow motion.

The feature at 74.55\,\kms\, from G36.115+0.552 shows a drift at a rate of +0.073\,\kms\,yr$^{-1}$. Its profile is well fitted with a Gaussian
function but there are weak features in adjacent spectral channels which may affect the behaviour of the feature. Indeed, the VLBI data
(\citealt{bartkiewicz09}) show two weak maser clouds which coincide with the 74.55\,\kms\, cloud within 4\,mas.

The feature at $-$23.41\,\kms\, from G121.298+0.659 shows a velocity drift of $-$0.035\,\kms\,yr$^{-1}$. Contemporaneous VLA data (\citealt{hu16}) 
confirm that the feature profile is Gaussian. The systemic velocity of the object is probably $-$25.1\,\kms\, thus the drift may indicate inflow motion.

We conclude that the above-mentioned targets are good candidates for future VLBI studies to verify if the radial velocity acceleration of features really traces 
the accreting gas, as was demonstrated for AFGL5142 (\citealt{goddi11}).

There is also evidence for velocity drifts of the following features: at 60.91\,\kms\, from G20.239+0.065; at 23.88\,\kms\, from G26.598$-$0.024 and
at 114.97\,\kms\, from G26.601$-$0.221. The profiles of these features cannot be fitted with Gaussian curves and different variability patterns 
of a few blended features are the most natural explanation for the observed drifts.

\section{Discussion}
\subsection{Comparison with previous studies}
Monitoring the variability of methanol masers at 6.7\,GHz has been subject of a number of studies. 
Table~\ref{comp-goedhart04} summarises a comparison of our variability statistics with that reported by  \cite{goedhart04}
for a sample of 54 sources. They considered a feature as variable when the variability index, 
defined by their equation 1, is higher than 0.5, whereas in our analysis this criterion is met if a feature 
has the reduced $\chi^2$ higher than the corresponding value for a 99.9~per cent significance level of variability.
We infer that the percentage of variable features in Goedhart et al.'s study is nearly the same as ours (Table~\ref{comp-goedhart04}). 
Assuming that a source is non-variable if all features in its spectrum are non-variable we found that 19\,per~cent
of sources in Goedhart et al.'s sample were non-variable, compared to 21\,per~cent in the present sample.
We conclude that the quantitative characteristics of maser variability on time-scales of one month to 4\,yr are
very similar in the two studies. 

Eleven sources in our sample are common to  Goedhart et al.'s sample. Sources G35.200$-$1.736, G52.663$-$1.092 and G196.454$-$1.677
were highly variable at the two monitoring periods lasting 3.7 and 4.2\,yr. Their spectra were highly transformed whilst
individual features decreased by a factor of 3$-$5 after $\sim$10\,yr. In G59.78+0.06 the feature at 27.02\,\kms\, 
increased from 10 to 35\,Jy when observed by Goedhart et al. but it reached 42\,Jy after 10\,yr and remained stable
during our monitoring. However, features at velocities lower than 26.5\,\kms\, strongly transformed after 10\,yr and showed 
significant variability.
G23.010$-$0.411 and G23.437$-$0.184 showed synchronised and low or moderate amplitude variations on the shorter time-scales
with only small changes in the shapes of spectra over $\sim$10\,yr.
The spectral shape of G188.946$-$0.886 and its periodic variations were preserved.
G45.071+0.132 was non-variable during the two monitoring periods. The shape and intensity of its spectrum remained the same after 10\,yr.
Comparison of the variability characteristics of G78.122+3.633, G81.871+0.781 and G189.030+0.784 is not warranted because
the first two objects were observed by Goedhart et al. at low elevation, whereas our data for the last source are contaminated 
by confusion. Therefore, the above comparison indicates that for four out of eight sources the spectra changed significantly after 10 years.

\begin{table}
 \caption{Comparison with the study by \citealt{goedhart04}.
\label{comp-goedhart04}}
\begin{tabular}{@{}l @{}c @{\hspace{2.5mm}}c}
\hline
           & \citealt{goedhart04}  & This work \\

\hline
Duration of monitoring (yr) &      4.2  &    3.7 \\
Average observation cadence (month$^{-1}$) & 2-4 &     1 \\
Number of sources analysed &       54   &   137  \\
Percentage of variable features  &    55.0  &   53.5 \\
Percentage of non-variable sources  &  18.9  &  20.9 \\
Percentage of periodic sources   &    13.0  &   6.5 \\             
\hline              
\end{tabular}
\end{table}

In our sample there are 17 and 19 sources for which the maser features showed a steady increase and decrease, respectively. 
Here, we assume that a trend of steady increase or decrease is real if the initial and final flux densities differ by more 
than 20~per cent. In a further 8 sources the numbers of features showing steady increases and decreases in flux density 
are nearly equal. For a total number of 70 features with a monotonic variability the pattern of steady increase occurs in 32
features. We conclude that the numbers of sources and maser features with long ($\ge$3.7\,yr) steady growth and decay of 
flux density are similar. This is in contrast with finding by \cite{goedhart04} who reported that 12 out of 15 sources have 
features with steadily increasing flux density, which may hint that the growth phase of masers may be much longer than decay 
phase. A sensitivity of their monitoring was similar to ours and we suggest that the asymmetry between the number of 
increasing v. decaying features could be related to the different criteria for pattern recognition used by the two studies.

\subsection{Variability measures and luminosity relations}
Here, we investigate the relationships between variability properties and the luminosities of maser features and 
their powering stars. The distances were derived from trigonometric parallaxes, mainly from the compilation by \cite{reid14},
or else calculated by applying the recipe from \cite{reid09} to the systemic velocities listed in Table~\ref{sample}. 
Trigonometric distances are available for 43 maser sites in the sample. To resolve the kinematic distance ambiguity 
(KDA) we relied upon the data published in \cite{green11}. They used respectively the presence or absence of 
self-absorption in HI spectra in the proximity of the systemic velocity to decide whether the source is at the near or 
far kinematic distance. We found that the KDA could be resolved for 68 maser sites in Table~\ref{sample}.
For targets with known distances we have calculated the luminosity of each maser feature with uncorrupted variability
indices given in Table~\ref{varindices} and assuming the emission is isotropic. In several cases the maser features 
are blended and fitting a Gaussian function to the profiles was risky when high angular resolution data are not available. 
Therefore the integrated flux density of features was estimated from the peak flux density assuming the mean line-width 
of 0.4\,\kms\, (\citealt{bartkiewicz16}). 

\begin{figure*}   
\resizebox{\hsize}{!}{\includegraphics[angle=0]{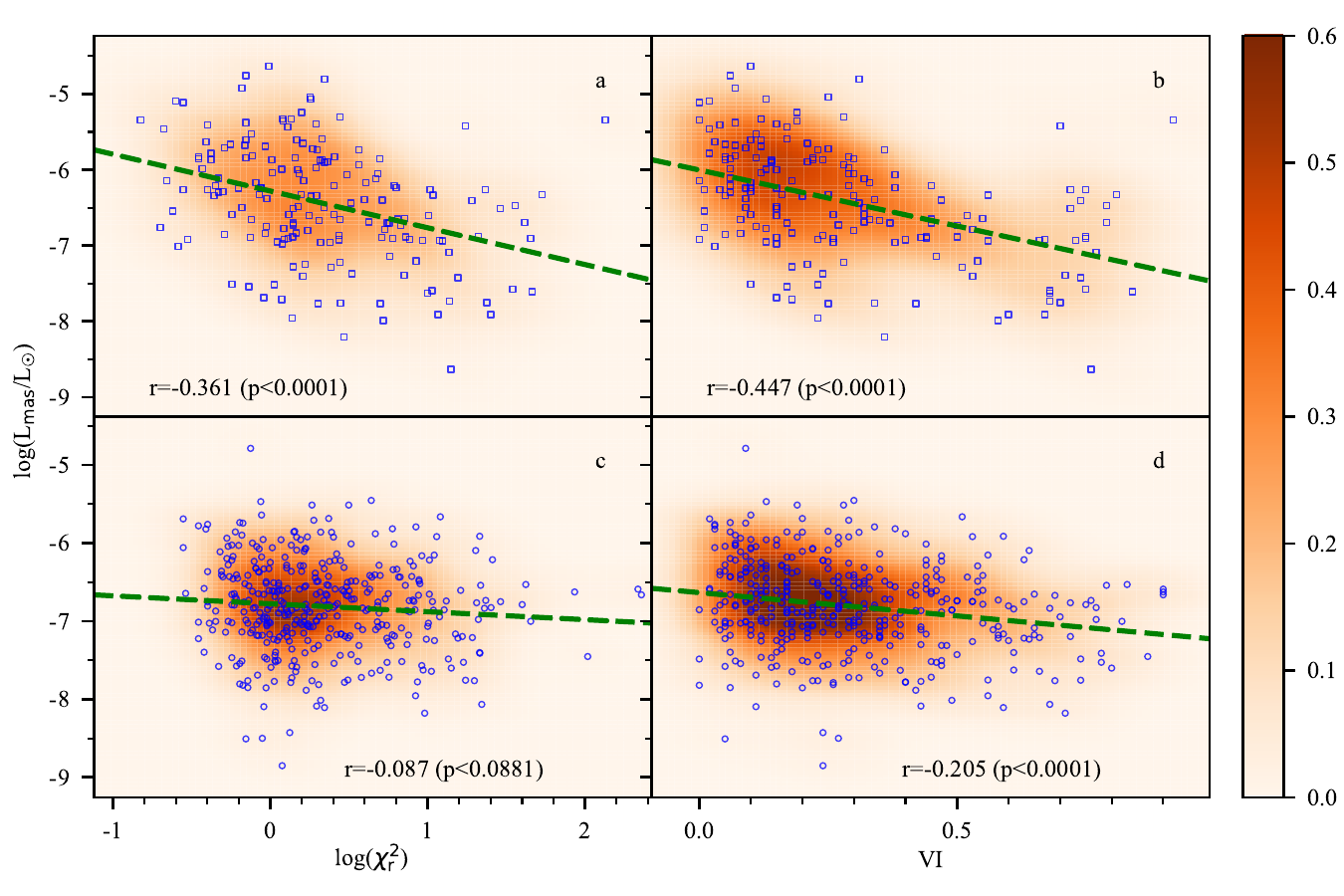}}
\caption{Luminosity of maser features and variability measures. The luminosity is shown versus reduced $\chi^2$ (panels a, c)
and variability index $VI$ (panels b, d). The squares and circles represent features of sources with trigonometric and kinematic 
distances, respectively. The dashed lines represent the best-fitting results. The colour wedge marks the number of features per 
$\Delta\chi_{\mathrm{r}}^2 \Delta{L}_{\mathrm{mas}}$ and $\Delta{VI} \Delta{L}_{\mathrm{mas}}$ bins.
\label{vi-lmas-correl}}
\end{figure*}

In the upper panels of Figure~\ref{vi-lmas-correl} we plot the maser feature luminosity as a function of reduced $\chi^2$ 
and variability index for the sources with known trigonometric distances. There is a moderate anti-correlation between 
the feature luminosity and variability index ($r=-$0.45 and $p$ value $<<0.01$) and a weaker anti-correlation between
the feature luminosity and reduced $\chi^2$ ($r=-$0.36 and $p$ value $<<0.01$). The same parameters  for the sources
where the feature luminosity was derived from the kinematic distances are plotted in the lower panels of Figure~\ref{vi-lmas-correl}.
The correlation coefficient for the feature luminosity $-$ variability index relation drops by a factor of two and no correlation is seen
between the feature luminosity ($L_{\mathrm{mas}}$) and $\chi_{\mathrm{r}}^2$. From this we conclude that maser features of low luminosity tend 
to be more variable than those of high luminosity. This anti-correlation disappears when the sources with less accurate distances are considered. 

The anti-correlation between the maser feature luminosity and variability measures does not allow us to distinguish between 
maser variations caused by changes in the background radiation or by changes in the optical depth of the transition.
Comparing the variability measures and bolometric luminosity of exciting star and other characteristics of the maser sites has 
the potential to resolve this ambiguity. In order to derive the bolometric luminosity of exciting star, we used the Python package 
for the SED fitting tool by \cite{robitaille07} and SED models from \cite{robitaille06} to model the photometric data from 3.4$\mu$m 
to 1.1\,mm taken from GLIMPSE (\citealt{benjamin03}), MIPSGAL (\citealt{carey09}), WISE (\citealt{wright10}), Herschel (\citealt{molinari16}), 
ATLASGAL (\citealt{schuller09}) and BOLOCAM (\citealt{ginsburg13}). A full discussion of the photometric data used and the results 
for this sample is beyond the scope of this paper and will be discussed in a future publication.

\begin{figure*}
\resizebox{\hsize}{!}{\includegraphics[angle=0]{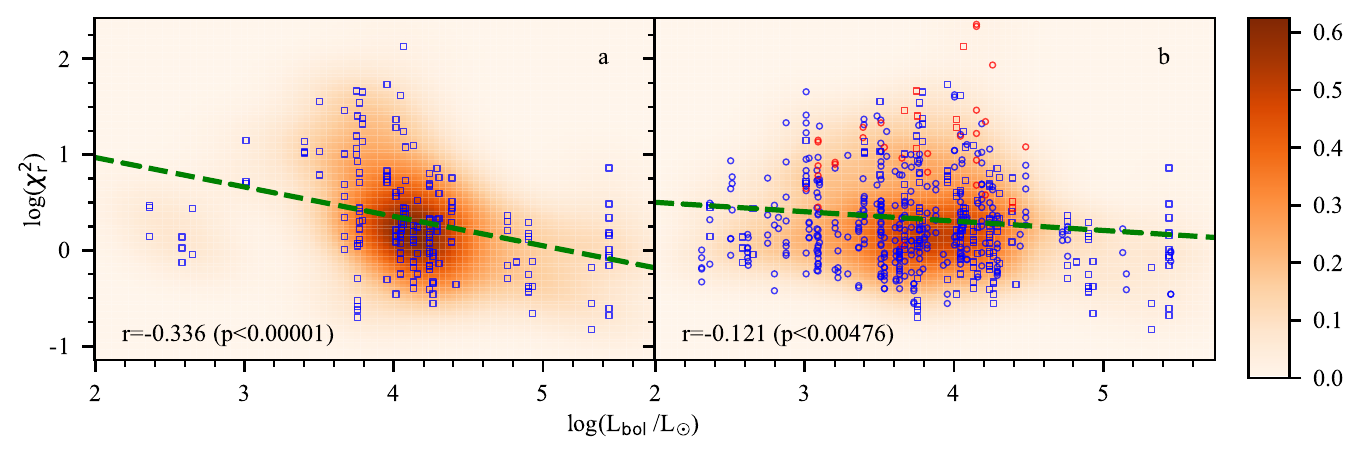}}
\caption{Reduced $\chi^2$ of maser features versus the bolometric luminosity. The meaning of squares and circles is the same as in Fig.~\ref{vi-lmas-correl}.
The features of sources with known trigonometric distances are shown in panel $a$ and the dashed line represents the least square fit to the data.
Panel $b$ shows all the features but the dashed line marks the best-fitting result when the flaring and periodic features (red symbols)
are excluded. The colour wedge marks the number of features per $\Delta\chi_{\mathrm{r}}^2 \Delta{L}$ bin.
\label{vi-lbol-correl}} 
\end{figure*}

Figure~\ref{vi-lbol-correl} shows the relationship between reduced $\chi^2$ and bolometric luminosity ($L_{\mathrm{bol}}$) for 125 maser 
sites (546 features). The associated correlation coefficient is $-$0.12 with a significance value of 0.005. There are 52 features showing 
flaring and periodic behaviour, the origin of which can be extrinsic to maser regions, such as variations of accretion rate in a binary system
(e.g. \citealt{araya10}; \citealt{parfenov14}). After excluding these 52 features the correlation coefficient increased to $-$0.16
($p<0.0004$), which implies weakly statistically significant correlation. When we confine the analysis to the sources with known trigonometric 
distances, the correlation coefficient for $L_{\mathrm{bol}} - \chi_{\mathrm{r}}^2$ relation rises to $-$0.26 ($p<0.002$). This implies that the
variability of maser features increases in the sources powered by less luminous MYSOs. 
We suppose that the statistical significance of $L_{\mathrm{bol}} - \chi_{\mathrm{r}}^2$ relation shown in Figure~\ref{vi-lbol-correl} may
be reduced because the bolometric luminosity we used is a measure of the total luminosity of the protocluster, whereas the maser intensity
more likely depends on the intrinsic luminosity of a single cluster member which supplies the far-infrared pumping photons (\citealt{cragg05}).

A recent survey of 6.7\,GHz continuum emission towards methanol masers (\citealt{hu16}) provides another constraint on the physical
properties of 113 of the targets in our sample. Continuum emission was detected within less than 3\arcsec\, from the maser position for 48
objects, so we assume that these maser and continuum sources are physically related. The Lyman continuum flux ($N_{\mathrm{i}}$) is
estimated using formula (\citealt*{carpenter90}): 

\begin{eqnarray}
\left(\frac{N_{\rm i}}{\mathrm{photons~s^{-1}}}\right)= 9\times10^{43}\left(\frac{S_{\nu}}{\mathrm{mJy}}\right)\left(\frac{D^2}{\mathrm{kpc}}\right)\left(\frac{{\nu}^{0.1}}{\mathrm{6.7GHz}}\right)
\label{Ni}
\end{eqnarray}

\begin{figure}
\resizebox{\hsize}{!}{\includegraphics[angle=0]{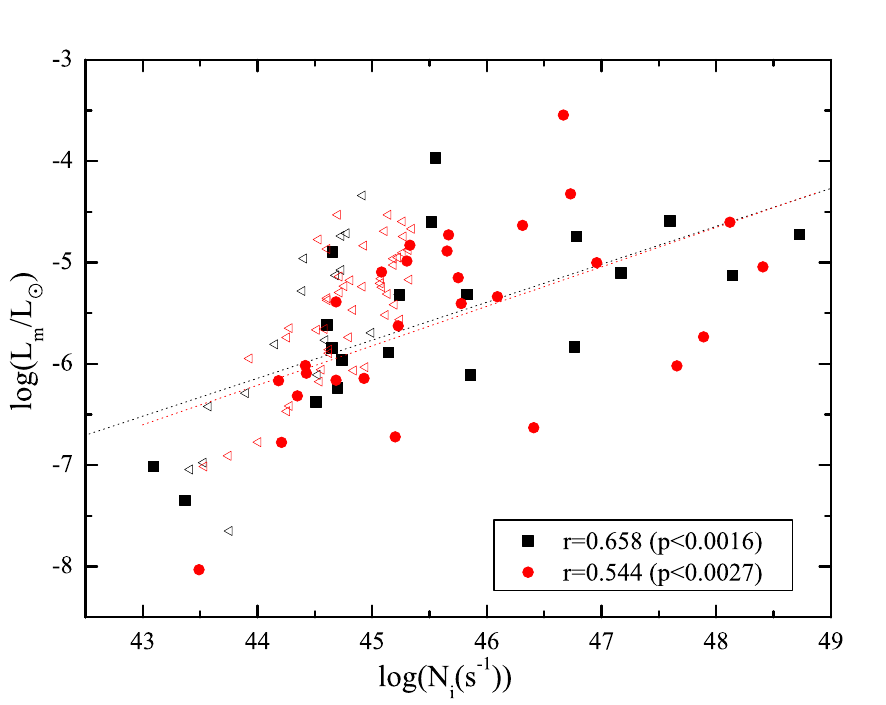}}
\caption{Isotropic maser luminosity versus the Lyman flux. The dotted black and red lines show the results of a power-law fit to
the sources with trigonometric (black squares) and the sources with kinematic (red circles) distances, respectively.
The triangles represent the upper limits of the Lyman flux.  
\label{Tot-mas-Lyman}}
\end{figure}

\noindent
where $S_{\nu}$ is the integrated radio flux density measured at frequency $\nu$ and $D$ is the distance to the source. 
Figure~\ref{Tot-mas-Lyman} shows the relation between the isotropic maser luminosity ($L_{\mathrm{m}}$), calculated from the
observed line integral, as a function of Lyman flux.  There is a significant correlation between these parameters. Again, we see that
the correlation coefficient is slightly higher for the subset of 20 sources with trigonometric distances than for that of 28 sources with
kinematic distances. We can state that the maser luminosity is related to the continuum photon flux of the most massive star in the cluster
which is able to produce an HII region detectable at cm wavelengths. More luminous massive stars would excite methanol molecules  
more efficiently over a larger region than less luminous stars; the pump rate would be higher and the path of maser amplification would be
longer. This conclusion is consistent with the results of previous studies (\citealt{urquhart13a, urquhart13b, urquhart15};
\citealt{devilliers14}) which have shown significant correlations between  methanol maser luminosity and the the mass and size of
maser-associated sub-millimetre continuum sources.

\begin{figure}
\resizebox{\hsize}{!}{\includegraphics[angle=0]{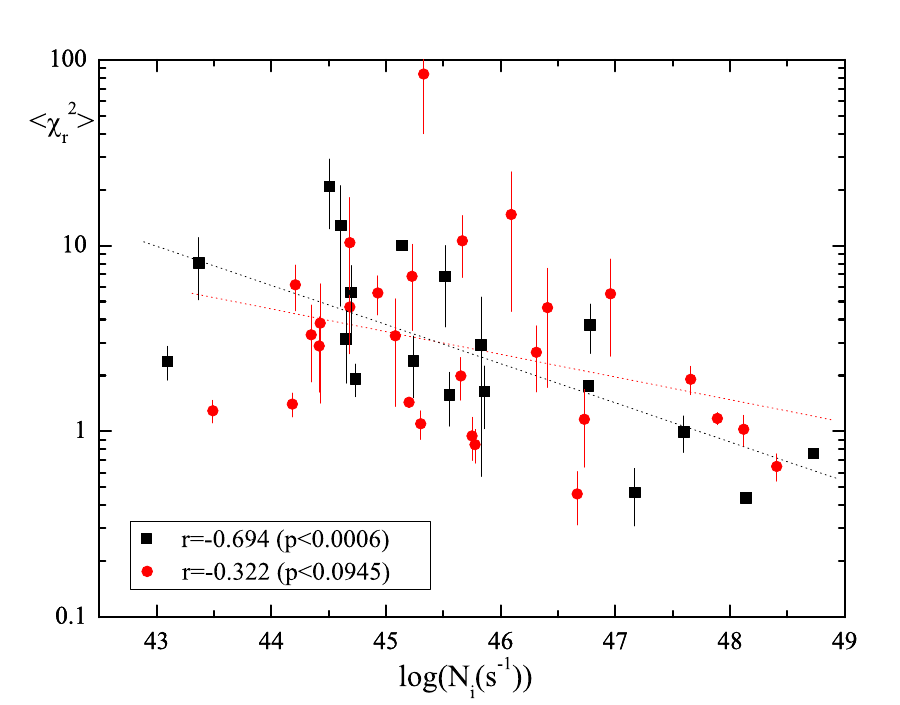}}
\caption{Mean reduced $\chi^2$ versus the Lyman flux. The dotted black and red lines represent the best-fitting results to the sources with trigonometric 
(black squares) and kinematic (red circles) distances, respectively.   
\label{chi2-Lyman}}
\end{figure}

Figure~\ref{chi2-Lyman} presents the average value of reduced $\chi^2$ versus the Lyman flux. Here, the average for each of 48 sources is
calculated from $\chi_{\mathrm{r}}^2$ of features given in Table~\ref{varindices}. The correlation coefficient is -0.694 for objects with trigonometric 
distances, demonstrating a significant anti-correlation between maser variability and Lyman photon flux. The correlation coefficient significantly decreases 
for the sources with the parameter determined from kinematic distances and no significant correlation exists. This analysis again suggests that 
maser emission excited by more luminous central stars is less variable than that that driven by fainter objects.

Summing up the above discussion we suggest that the variability of 6.7\,GHz maser emission decreases for objects where the bolometric luminosity 
of the exciting star is higher.

\subsection{Causes of variability}
Extraordinary outbursts in G24.329+0.144 appear to be particularly instructive to investigate for the causes of variations. 
The brightness temperature of its maser components ranged from ${\sim}10^7$ to $2\times10^9$ K during the VLBI observations
in June 2009 (\citealt{bartkiewicz16}). At this epoch the emission near 115.36\,\kms\, was marginally detected with 
a beam of $5.7\times4.7$ mas and had a brightness temperature $\le 2\times10^7$ K. The pre-flare flux density of this feature,
measured with the 32\,m dish, was $\sim$1.1\,Jy and it increased by a factor of $\sim$57 during the main flare. This corresponds 
to an increase in the optical depth of the maser radiation by more than 4 for unsaturated amplification, assuming  constant brightness
temperature of background radiation ($T_{\mathrm{bg}}$). The assumption that the maser operates in the unsaturated regime is supported 
by the fact that the flux density during the bursts always shows exponential growth and decline with a characteristic time of 45$-$105\,d. 
We have found a few features showing an anti-correlation between the intensity and line-width (Table~\ref{long-bursts}, Fig.~\ref{fig:dvds_24}),
which is theoretically foreseen for unsaturated amplification (\citealt{goldreich74}).

Using the method discussed in \cite{moscadelli17} we derive $T_{\mathrm{bg}}\le$30\,K, adopting the contributions from spontaneous
and dust emissions of $\le$2\,K and $\ll$20\,K, respectively. The 6.7\,GHz continuum emission of 0.26\,mJy detected in March 2012 with
the JVLA in C configuration (\citealt{hu16}) implies a contribution from ionized gas of $\le$1\,K.  Thus, the brightness temperature of
the maser component given by $T_{\mathrm{bm}} = T_{\mathrm{bg}}e^{|\tau|}$ corresponds to a maser line optical depth of $\tau$=13.4, 
in a quiescence phase. Therefore, $T_{\mathrm{bm}}$ would increase to $\sim$1.2$\times10^9$K at the flare maximum when $\tau$ increased 
to $\sim$17.4. This value of $T_{\mathrm{bm}}$ is still below the saturation threshold (\citealt{moscadelli17}) and, for a constant $\tau$, 
an increase of $T_{\mathrm{bg}}$ up to 1800\,K can account for the flux density observed at the flare maximum. However, no obvious process 
is known which could lead to the growth of $T_{\mathrm{bg}}$ by a factor of 60 on time-scales of 100$-$300\,d and we suggest it is improbable 
that that seed photon flux increase is a major contributor to the maser flare.

The second factor which can influence the flare curve is the maser optical depth. It can be expressed (\citealt{goldreich74}; \citealt{cragg02}) 
as $\tau\propto(\Delta P/\Gamma) (n/\Delta\Phi) l$, where $\Delta P$ is the net pump rate, $\Gamma$ is the collisional dumping rate of 
the maser levels, $n$ is the number density of the molecule, $\Delta\Phi$ is the line profile width and $l$ is the amplification path.
For the flaring feature in G24.329+0.144 a change of $\tau$ by about 30\,per~cent accounts for the observed flux density variations.
Nearly simultaneous bursts of several features in the source strongly indicate that they are excited radiatively. This supports the hypothesis
that the 6.7\,GHz maser transition is pumped by infrared photons (\citealt{cragg05}) and implies that the pump rate
plays a major role in triggering the burst activity. The same scenario has been proposed to explain a giant flare of the 6.7\,GHz maser
in S255 NIRS3 (G192.600$-$0.048) which strongly brightened in the infrared due to accretion events (\citealt{caratti17}; \citealt{moscadelli17}). 
A substantial increase in the infrared luminosity can enlarge the size of the excited region and provide a longer amplification path.

The present survey revealed many features showing uncorrelated variability but usually of low amplitude, on different time-scales. 
This behaviour may be induced by tiny fluctuations of the parameters affecting the optical depth or may arise due to correlations
between the velocity fields and the alignment of the structures along the line of sight in extended regions of maser formation (see \citealt{sobolev12} for review).

There is observational evidence that the maser emission comes from a cylindrical structure rather than from a spherical cloud. VLBI data
imply a typical transverse size of the maser components of 16\,au (\citealt{bartkiewicz16}), implying that a shock propagating
perpendicularly to a maser cylinder must have a velocity of 92\,\kms\, to travel the structure over 300\,d, disturbing the velocity coherence. 
The existence of such fast shocks is possible but it is very unlikely that the methanol maser would survive such shocks. 
Additionally, it is unlikely that the emission from different clouds located in maser regions of size 400-1000\,au could be excited nearly 
simultaneously (\citealt{bartkiewicz16}). We conclude that shocks are not an important factor in causing the maser variability at 6.7\,GHz. 
Calculations of spectral and spatial distributions of maser emission at 25\,GHz from a turbulent medium (\citealt*{sobolev98}) suggest that 
a velocity dispersion of turbulence of about 1\,\kms\, could account for the observed characteristics and may lead to  time variability by 
changes in the velocity coherence.

One of the unexpected findings of the present monitoring is the detection of short lived bursts from spectral features in five targets.
These events are usually confined to a single feature, suggesting the local enhancement of maser amplification, repeated in a random manner.
Detailed temporal characteristics of the events in G33.641$-$0.228 were reported by \cite{fujisawa14c}, who estimated that the size of the active 
region was much less than 70\,au. They proposed a rapid energy release in magnetic reconnection near the stellar surface or in 
a flaring disc as possible origin of this phenomenon. 

The present survey has shown a wide range of variability phenomena but no anti-flare sources, i.e. those characterized by a sudden dimming 
followed by a slow return to the original flux density were found. We notice that for the observed brightness temperature of $10^7$ to $10^9$\,K 
a rapid decrease in the maser optical depth by $\Delta\tau$ results in relatively small drop in $T_{\mathrm{bm}}$ for the case of unsaturated 
amplification ($T_{\mathrm{bm}} \propto e^{\tau}$). On the contrary an increase of $\tau$ by the same value would produce a relatively large 
increase in $T_{\mathrm{bm}}$. We suggest that a lack of anti-flare events is a natural consequence that the 6.7\,GHz methanol maser in 
the monitored sources largely operates below the saturation threshold.

\section{Summary}
We have presented the results of 3.7\,yr monitoring observations of large sample of HMYSOs in the 6.7\,GHz methanol maser line.  
We summarize our principal results as follows. 

\begin{enumerate}
\item 
Nearly 80\,per cent of the sources show variations in the flux density greater than 10\,per~cent in at least one spectral
features on all time-scales that could be sampled effectively in the study. 

\item
Diverse variability patterns appear; from a monotonic decrease/increase over the entire monitoring, independent
or correlated changes in feature peaks, bursts of individual features or their groups on different time-scales to very complex 
changes of the spectra with appearance and disappearance of maser features. 

\item
Cyclic variations are detected in nine sources with periods ranged from 120 to 416\,d. 

\item
Synchronised and anti-correlated variations of maser feature groups are identified in four sources. This specific behaviour 
may be caused by switching between radial and tangential modes of maser amplification in these sources if they have a disc-like morphology. 

\item
Several bursts of duration from less than a week to several months are observed in about 17\, per cent of the sources.
In only a few cases the line-width of the flaring feature decreases with increase in the flux density indicating the effect
of line narrowing in the unsaturated regime. 

\item
The radial velocity drifts detected for several features can be naturally explained mostly as due to different variability 
patterns of close and blended features, but there are a few candidate features suggesting inflow motion that deserve further studies. 

\item
A general trend is found that low luminosity features are more variables than high luminosity features. The maser luminosity is closely related
to the bolometric luminosity and Lyman flux of exciting HMYSO as we confirmed for the sub-sample of sources with trigonometric distances.
This strongly suggests that variation of infrared flux is a main cause of the burst activity of the maser influencing the pump rate. 
The variations of individual features in the spectra may be also due to small disturbances in velocity coherence and path of amplification
in the turbulent velocity field.

\end{enumerate}

\section*{Acknowledgements}
The authors thank the TfCA staff and the students for assistance with the observations. We are grateful to Anita Richards 
and Eric G\'erard for their valuable comments on earlier draft of the paper and the referee Simon Ellingsen for his detailed and constructive input. 
We also warmly thank FESTO Poland company for selfless crucial help with repairing the antenna control system breakdown.
This research has made use of the SIMBAD database, operated at CDS (Strasbourg, France) and NASA's Astrophysics Data System Bibliographic Services. 
The work was supported by the National Science Centre, Poland through grant 2016/21/B/ST9/01455.

\clearpage
\newpage
\begin{appendix}


\section{Dynamic spectra}
 \label{sec:dynamic-spectra} 
 
\begin{figure}  
\includegraphics[angle=0, scale=0.5]{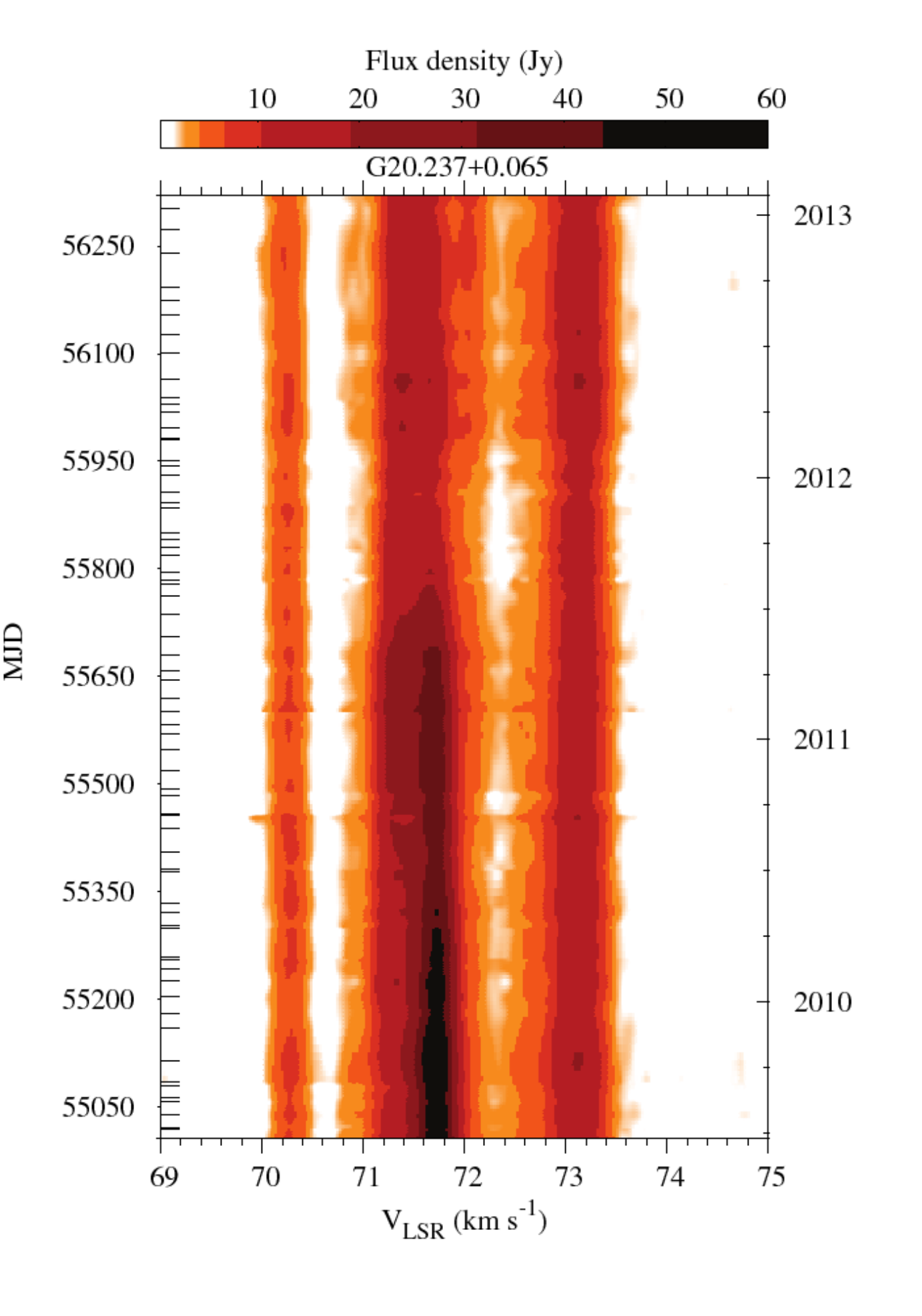}
\caption{Dynamic spectra for sources with complex variability. The velocity scale is relative to the local
standard of rest. The horizontal bars in the left coordinate correspond to the dates of the observed spectra.
(Only a portion of this figure is shown here to demonstrate its form and content. The complete figure is available online.)
\label{dyn-spectra}}
\end{figure}

\section{Dynamic spectra of spectral features with velocity drift}
\label{sec:dynamic-drift}
\begin{figure}   
\includegraphics[angle=0, scale=0.5]{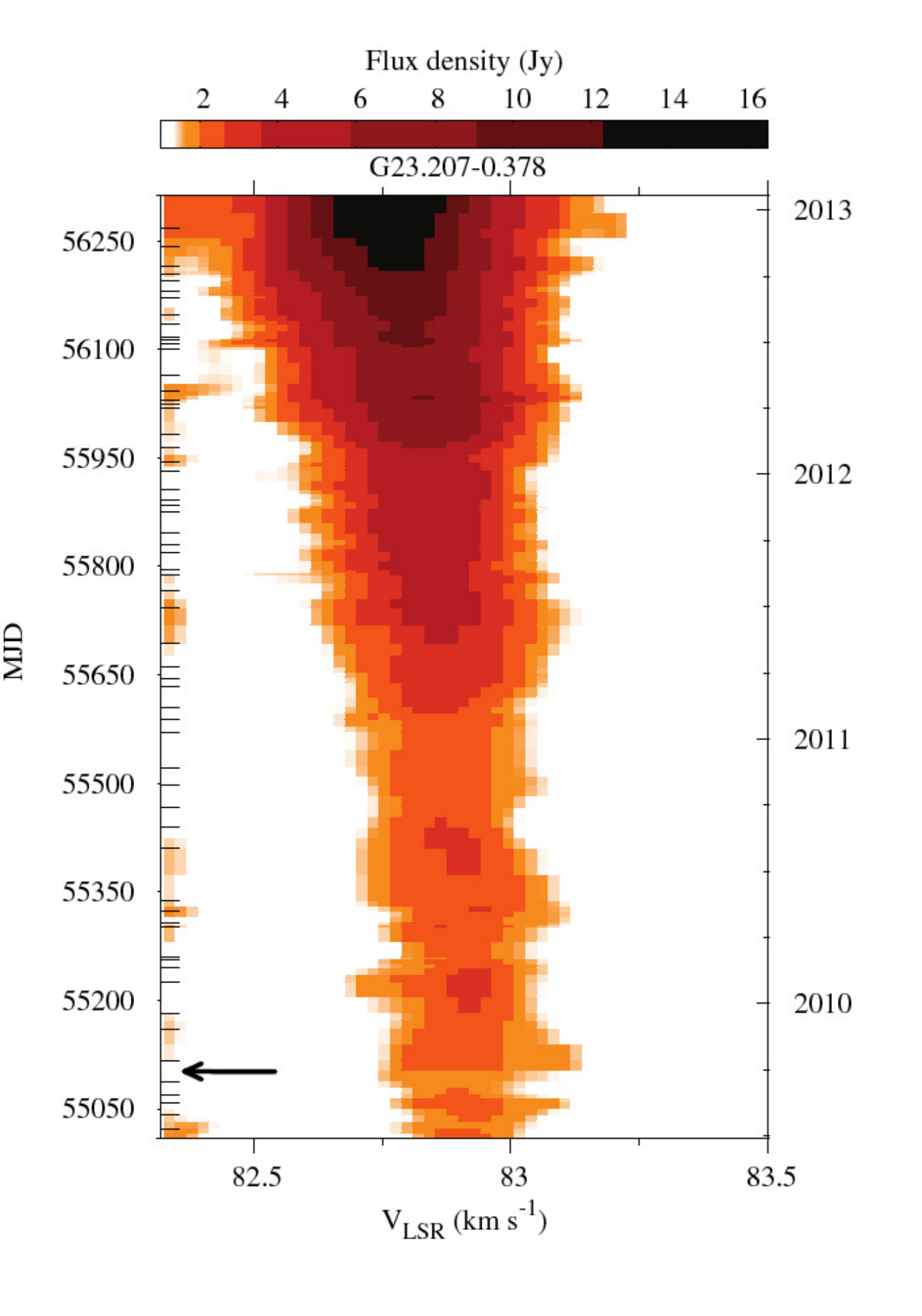}
\caption{Dynamic spectra for features with velocity drifts. The arrow indicates the systemic velocity for each source.
(Only a portion of this figure is shown here to demonstrate its form and content. The complete figure is available online.)
\label{dyn-drift}}
\end{figure}

\end{appendix}
\clearpage
\newpage
 {\bf ONLINE MATERIALS}

\addtocounter{figure}{1}
\begin{figure*}   
\centering
\includegraphics[width=\textwidth,height=9in,trim=-0.15cm -0.15cm -0.15cm -0.15cm,clip]{pl-eps-converted-to.pdf}
\caption{Left panels: Spectra of 6.7\,GHz methanol masers showing the average (solid), high (dashed) and low (dotted) emission levels.
Right panels: Light curves of selected 6.7\,GHz methanol maser features. Typical measurement uncertainty is shown by the bar for one
of the first data points or is comparable to the symbol size. 
\label{spectra-lightcurves}}
\end{figure*}

\begin{figure*}   
\resizebox{\hsize}{!}{\includegraphics[width=\textwidth,height=9in]{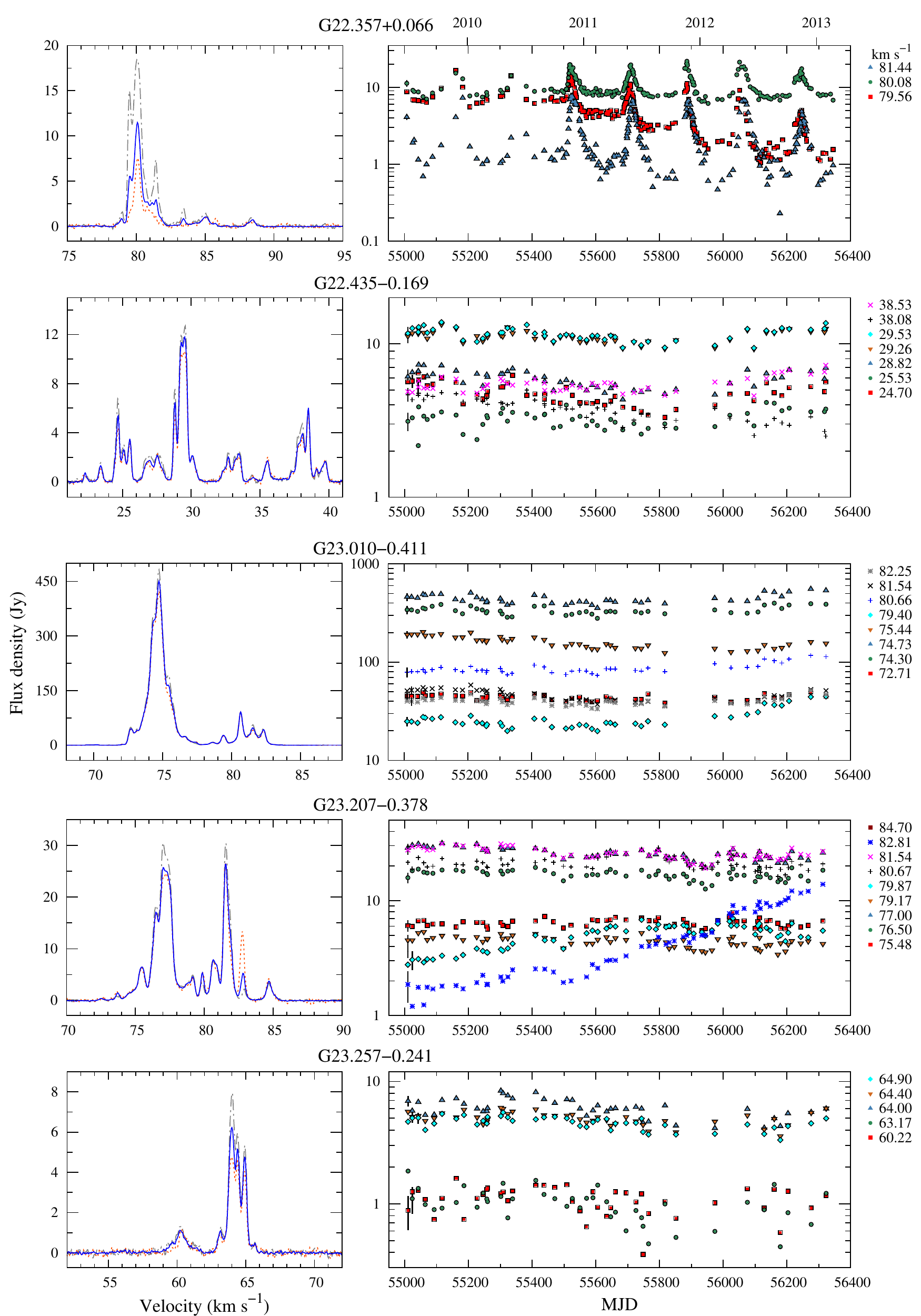}}
\contcaption{
\label{fig:continued}}
\end{figure*}

\begin{figure*}   
\resizebox{\hsize}{!}{\includegraphics[width=\textwidth,height=9in]{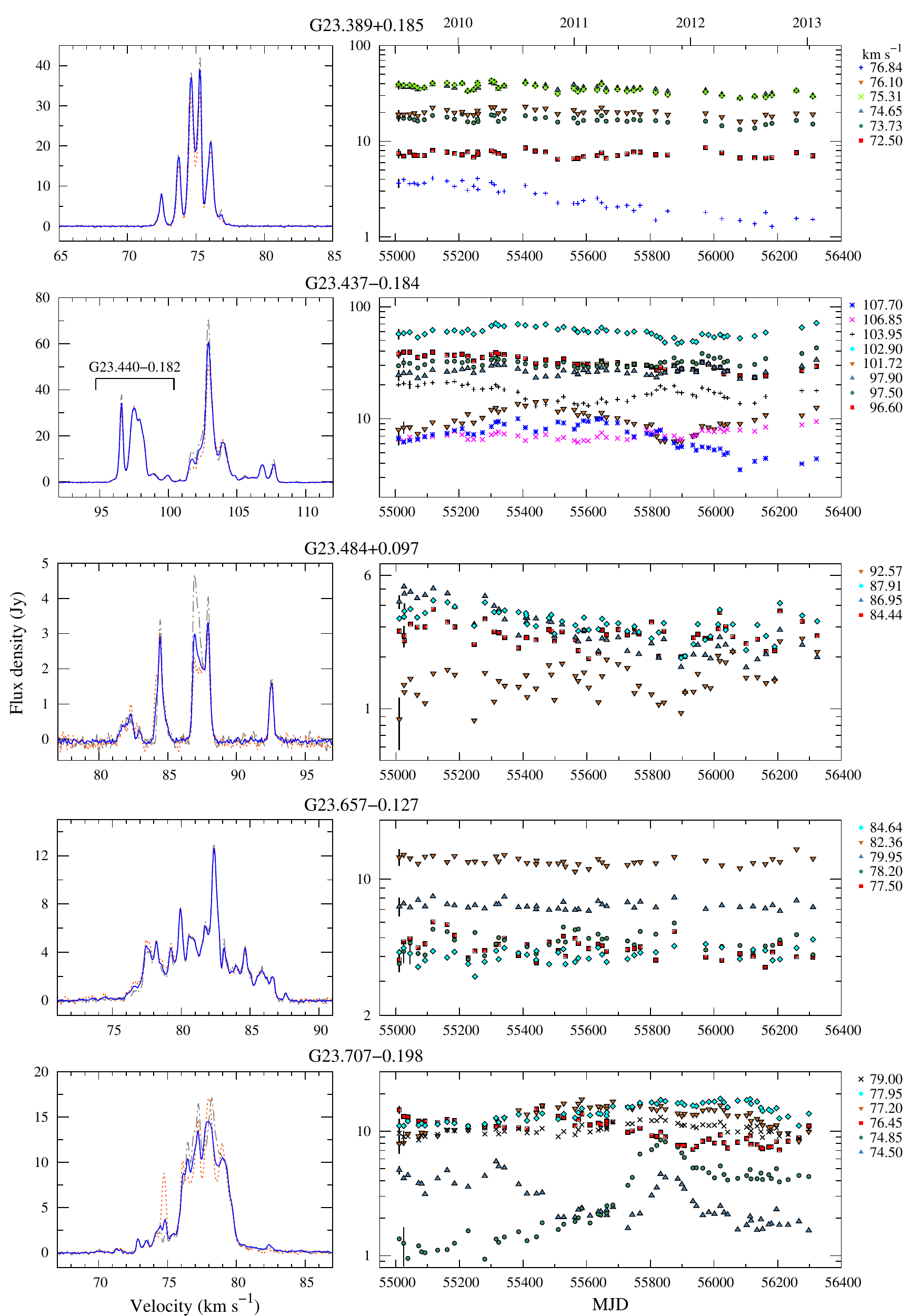}}
\contcaption{
\label{fig:continued}}
\end{figure*}

\begin{figure*}   
\resizebox{\hsize}{!}{\includegraphics[width=\textwidth,height=9in]{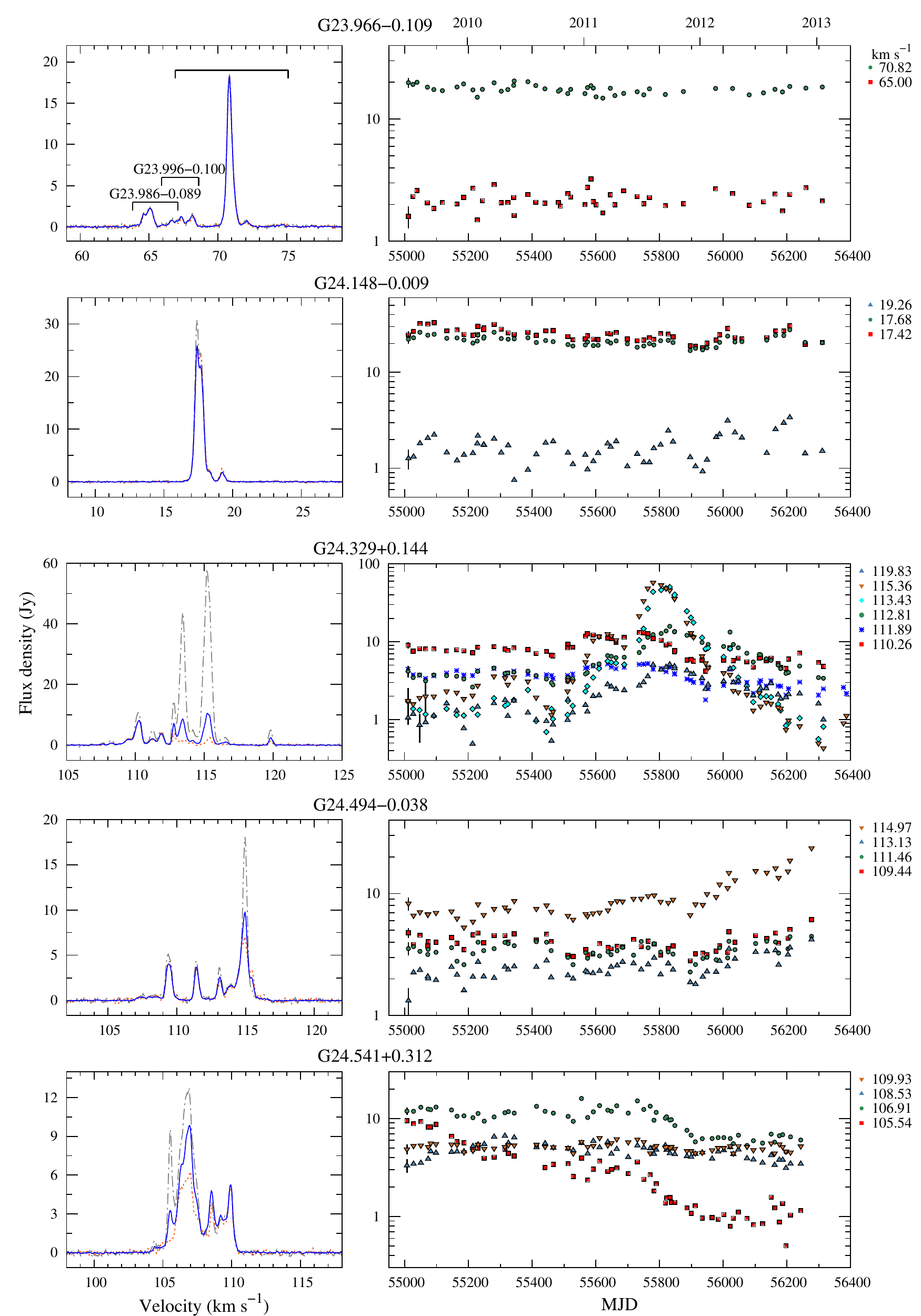}}
\contcaption{
\label{fig:continued}}
\end{figure*}

\begin{figure*}   
\resizebox{\hsize}{!}{\includegraphics[width=\textwidth,height=9in]{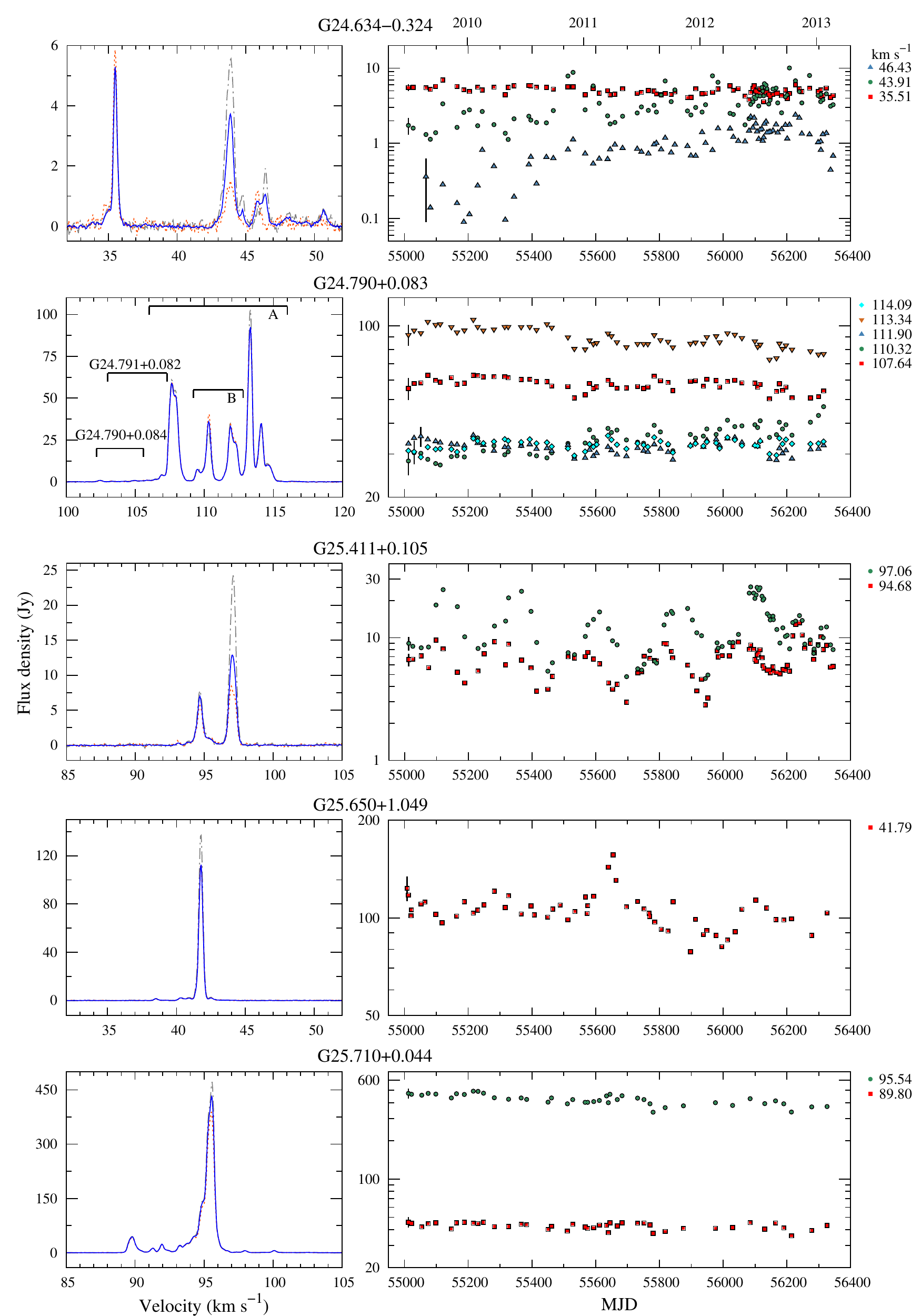}}
\contcaption{
\label{fig:continued}}
\end{figure*}

\begin{figure*}   
\resizebox{\hsize}{!}{\includegraphics[width=\textwidth,height=9in]{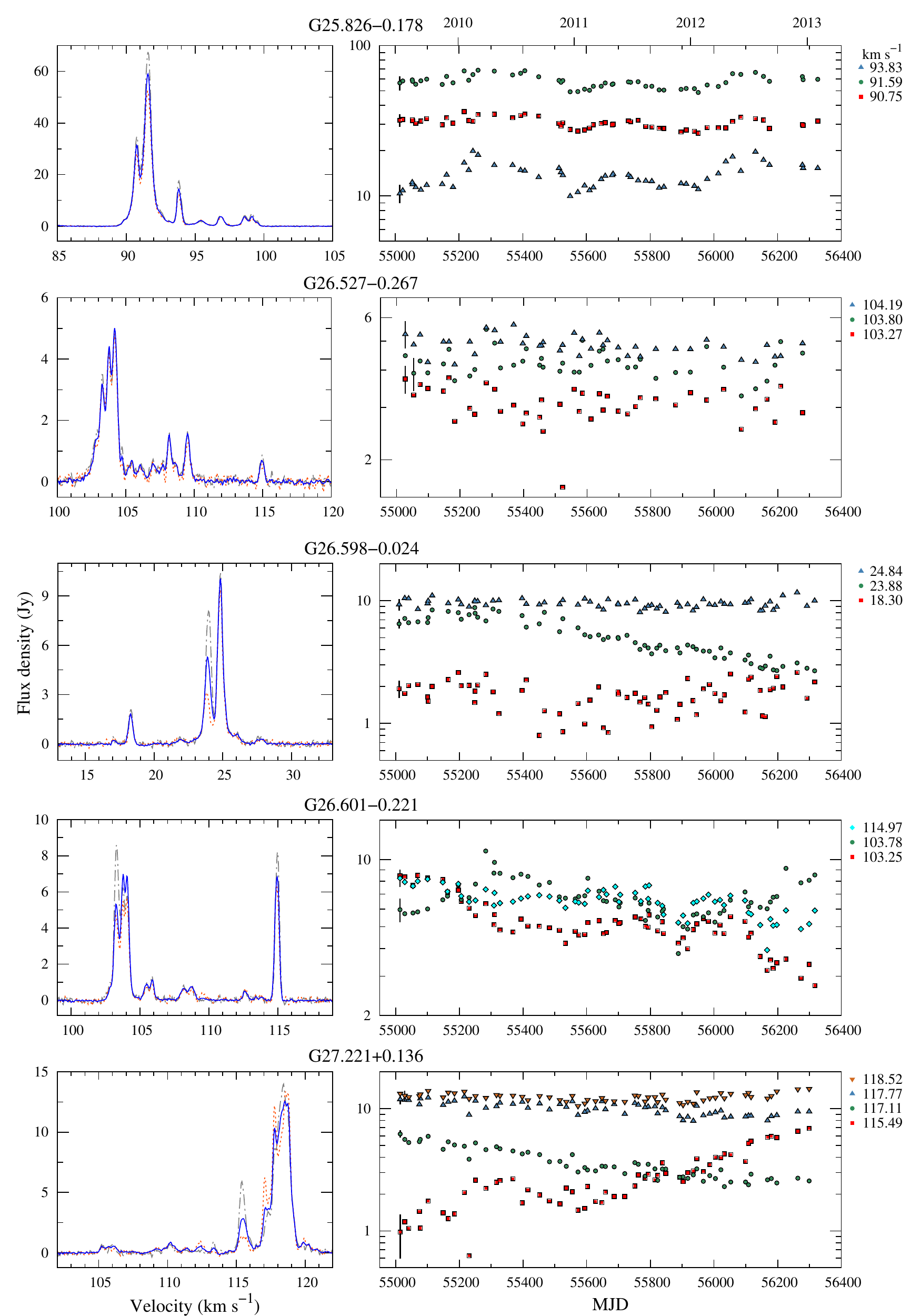}}
\contcaption{
\label{fig:continued}}
\end{figure*}

\begin{figure*}   
\resizebox{\hsize}{!}{\includegraphics[width=\textwidth,height=9in]{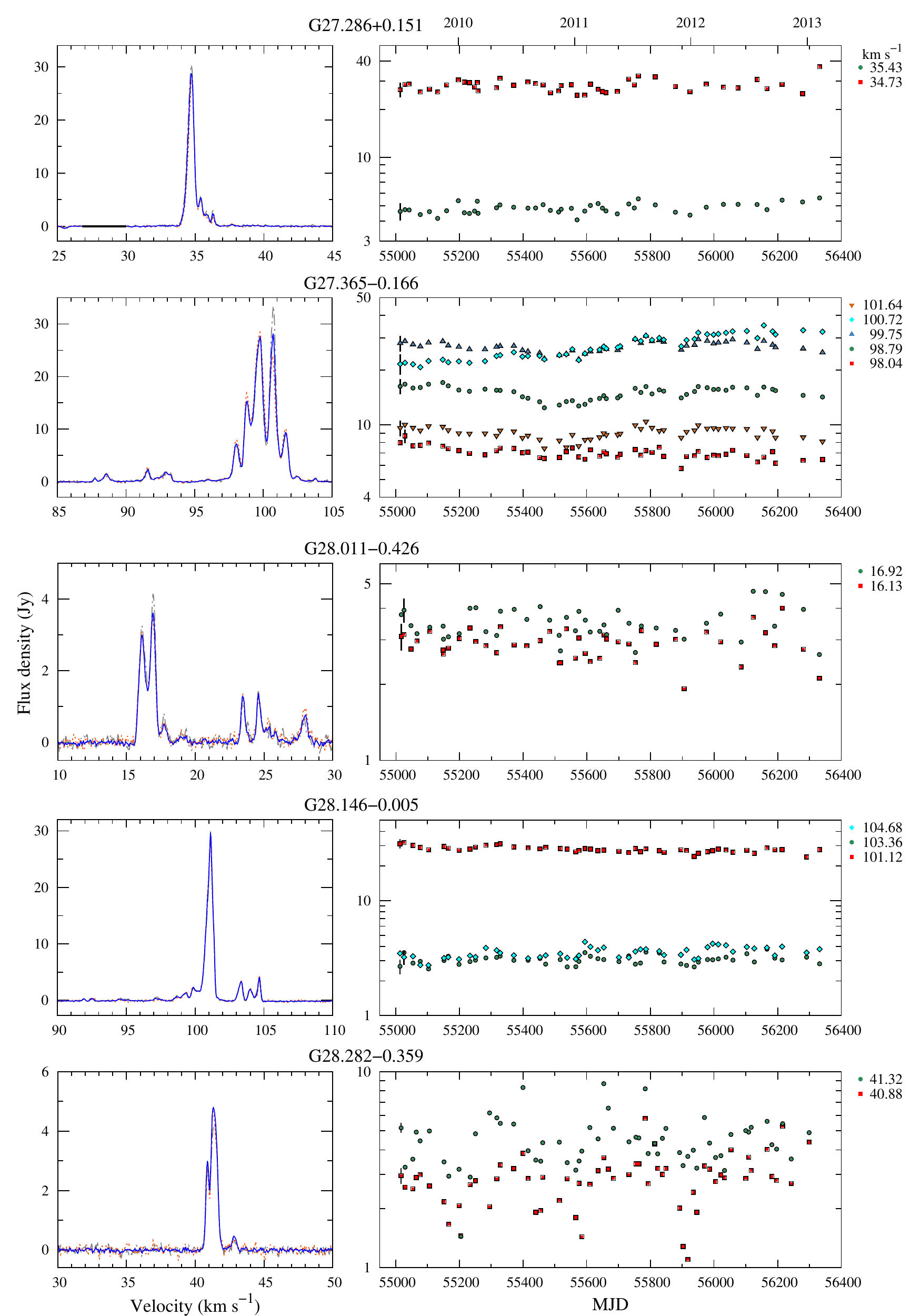}}
\contcaption{
\label{fig:continued}}
\end{figure*}

\begin{figure*}   
\resizebox{\hsize}{!}{\includegraphics[width=\textwidth,height=9in]{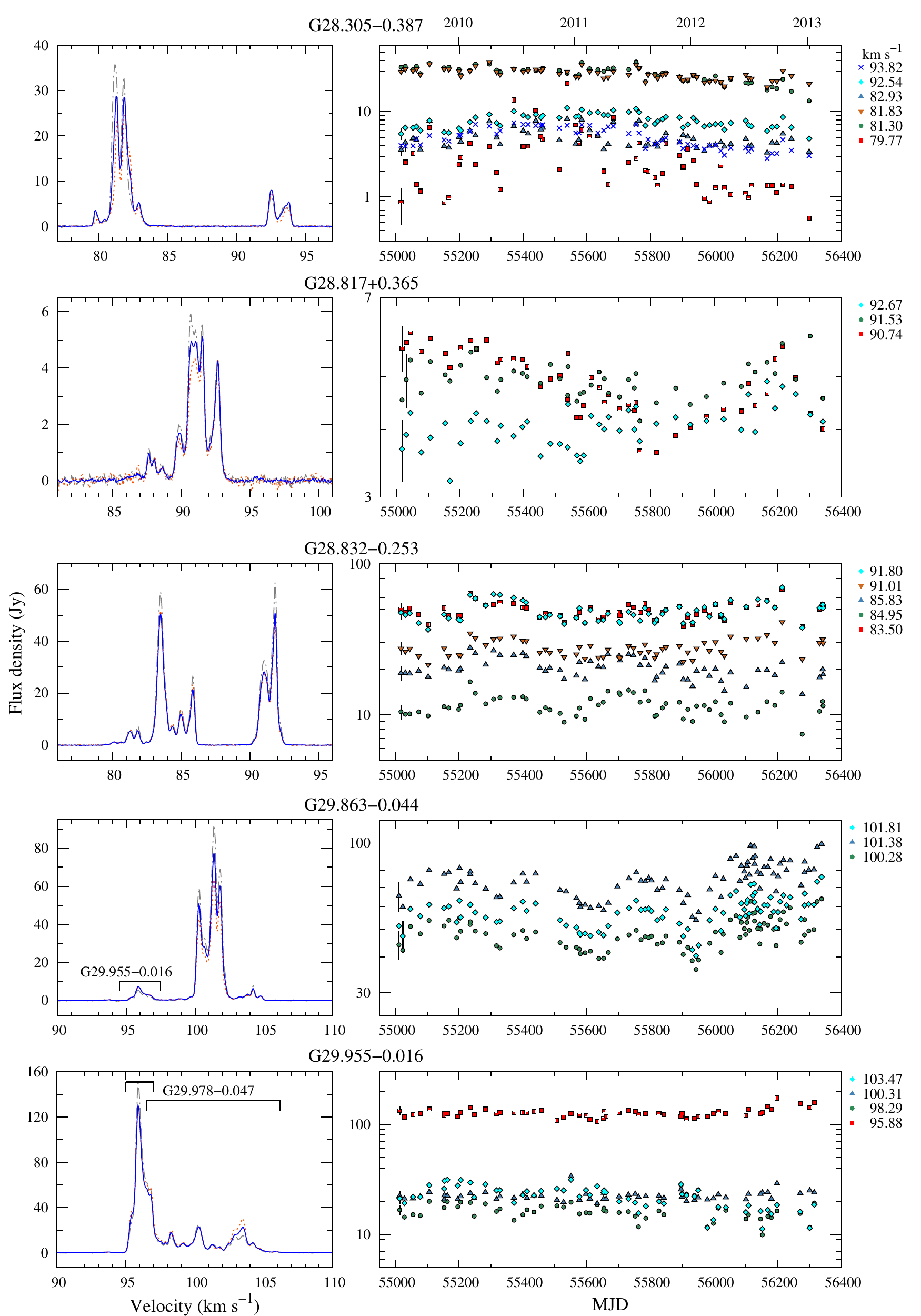}}
\contcaption{
\label{fig:continued}}
\end{figure*}

\begin{figure*}   
\resizebox{\hsize}{!}{\includegraphics[width=\textwidth,height=9in]{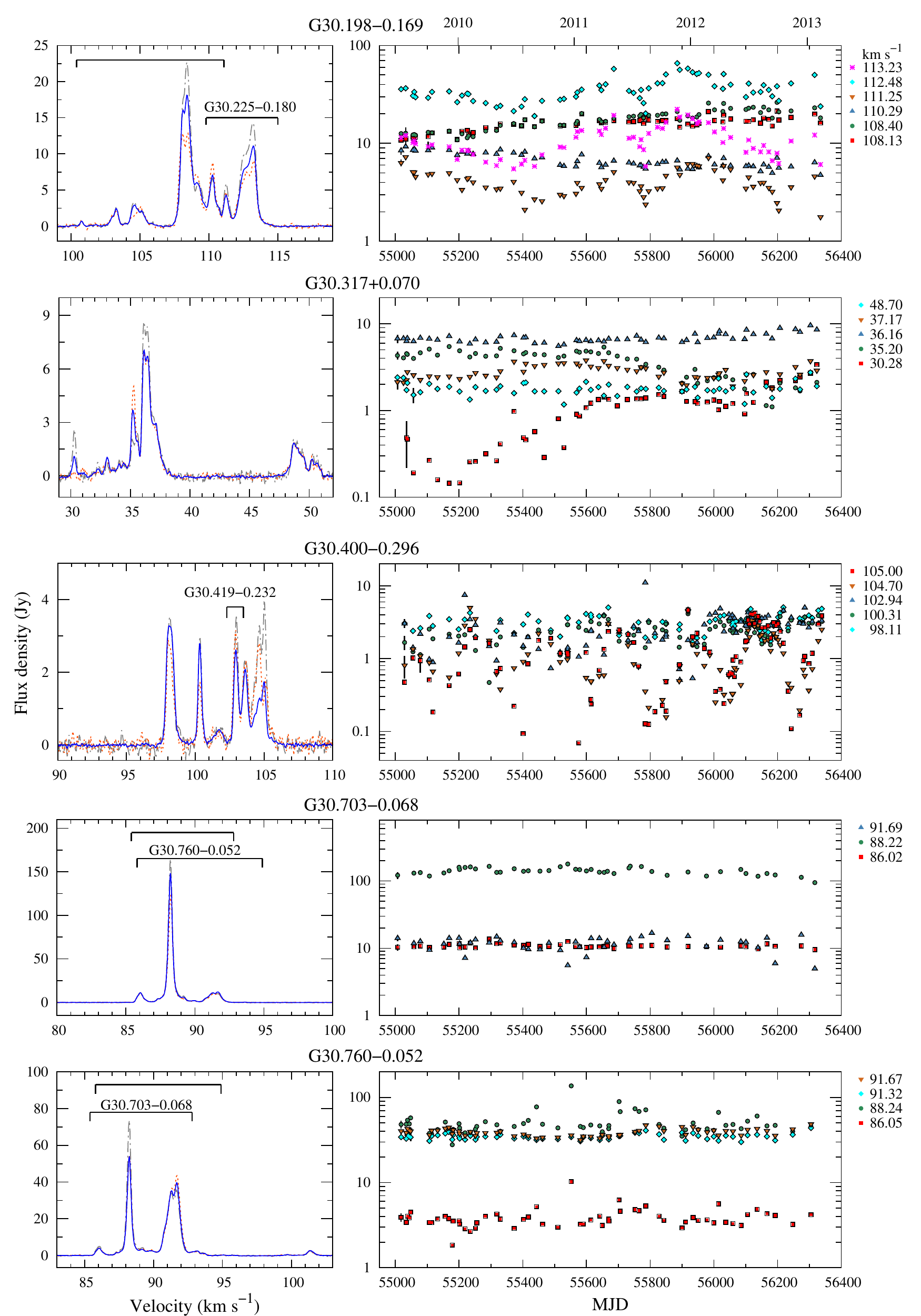}}
\contcaption{
\label{fig:continued}}
\end{figure*}

\begin{figure*}   
\resizebox{\hsize}{!}{\includegraphics[width=\textwidth,height=9in]{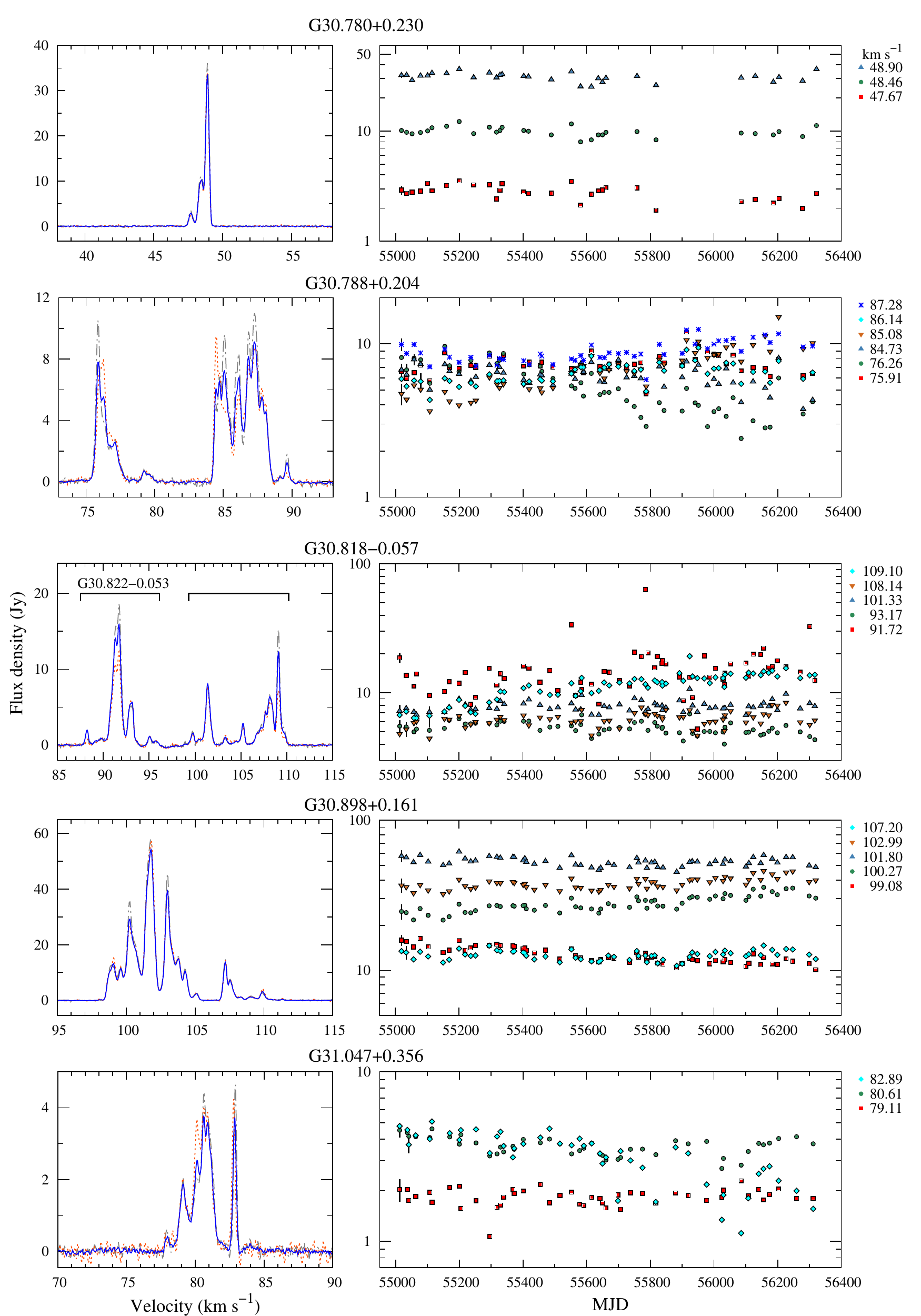}}
\contcaption{
\label{fig:continued}}
\end{figure*}

\begin{figure*}   
\resizebox{\hsize}{!}{\includegraphics[width=\textwidth,height=9in]{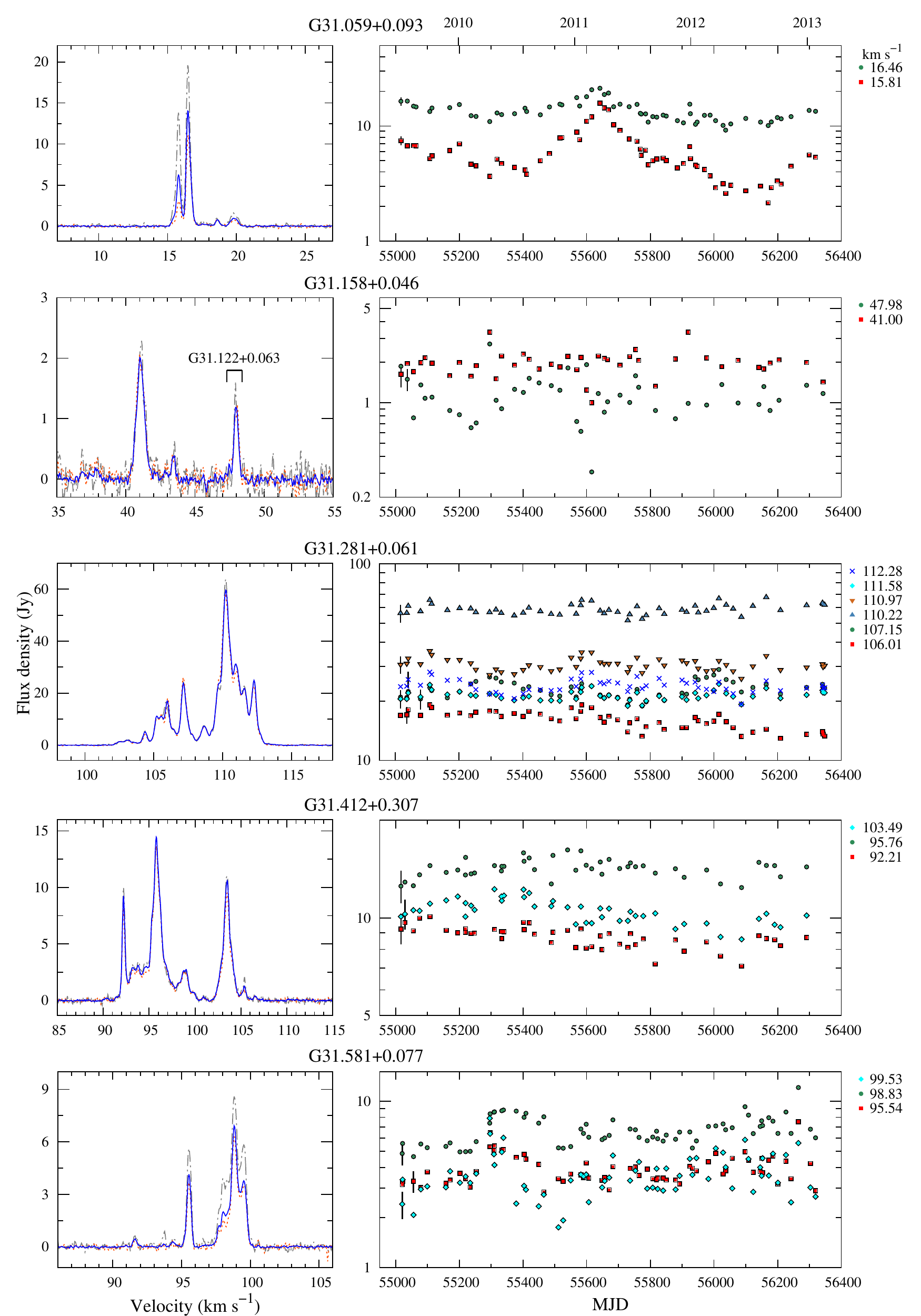}}
\contcaption{
\label{fig:continued}}
\end{figure*}

\begin{figure*}   
\resizebox{\hsize}{!}{\includegraphics[width=\textwidth,height=9in]{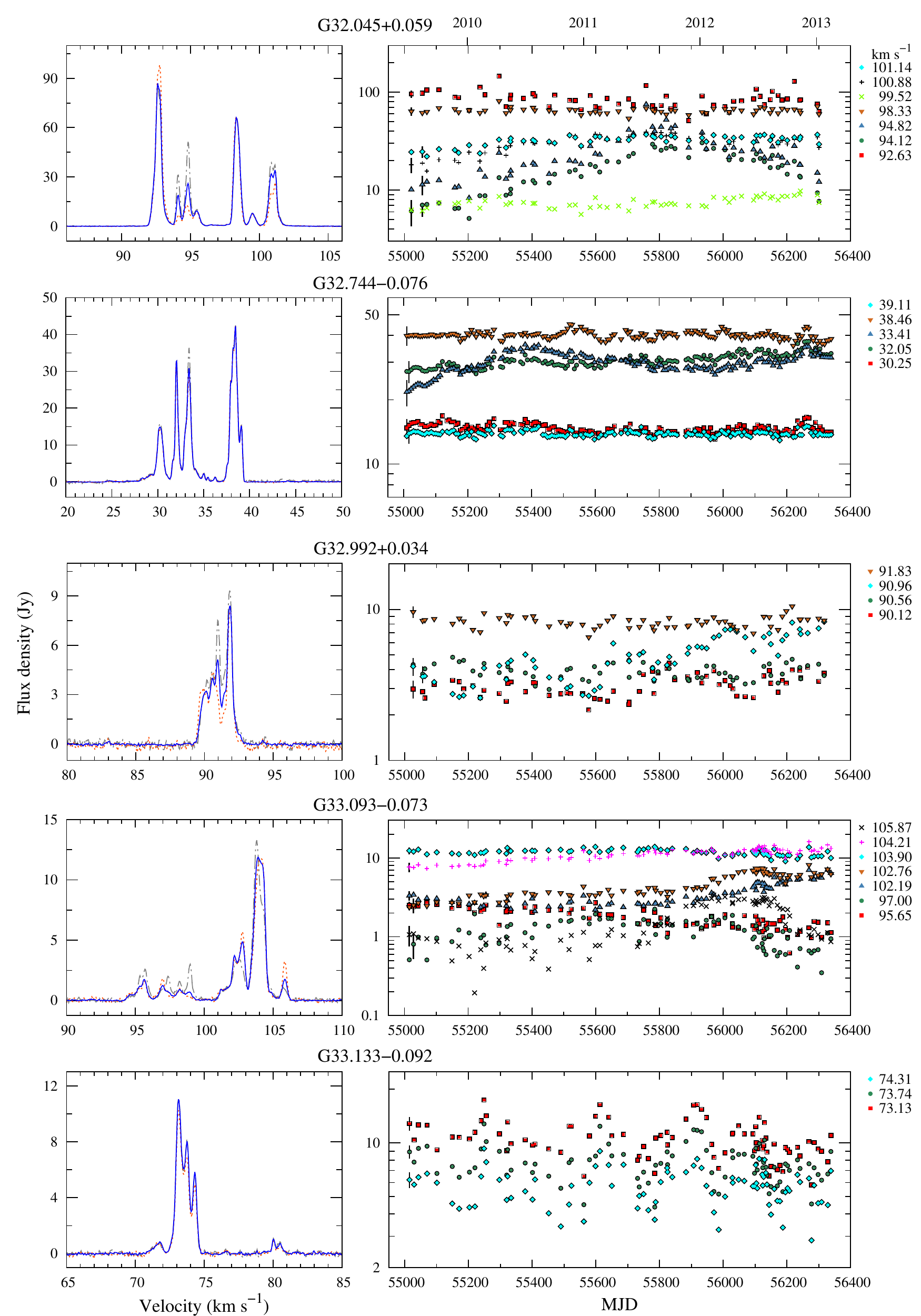}}
\contcaption{
\label{fig:continued}}
\end{figure*}

\begin{figure*}   
\resizebox{\hsize}{!}{\includegraphics[width=\textwidth,height=9in]{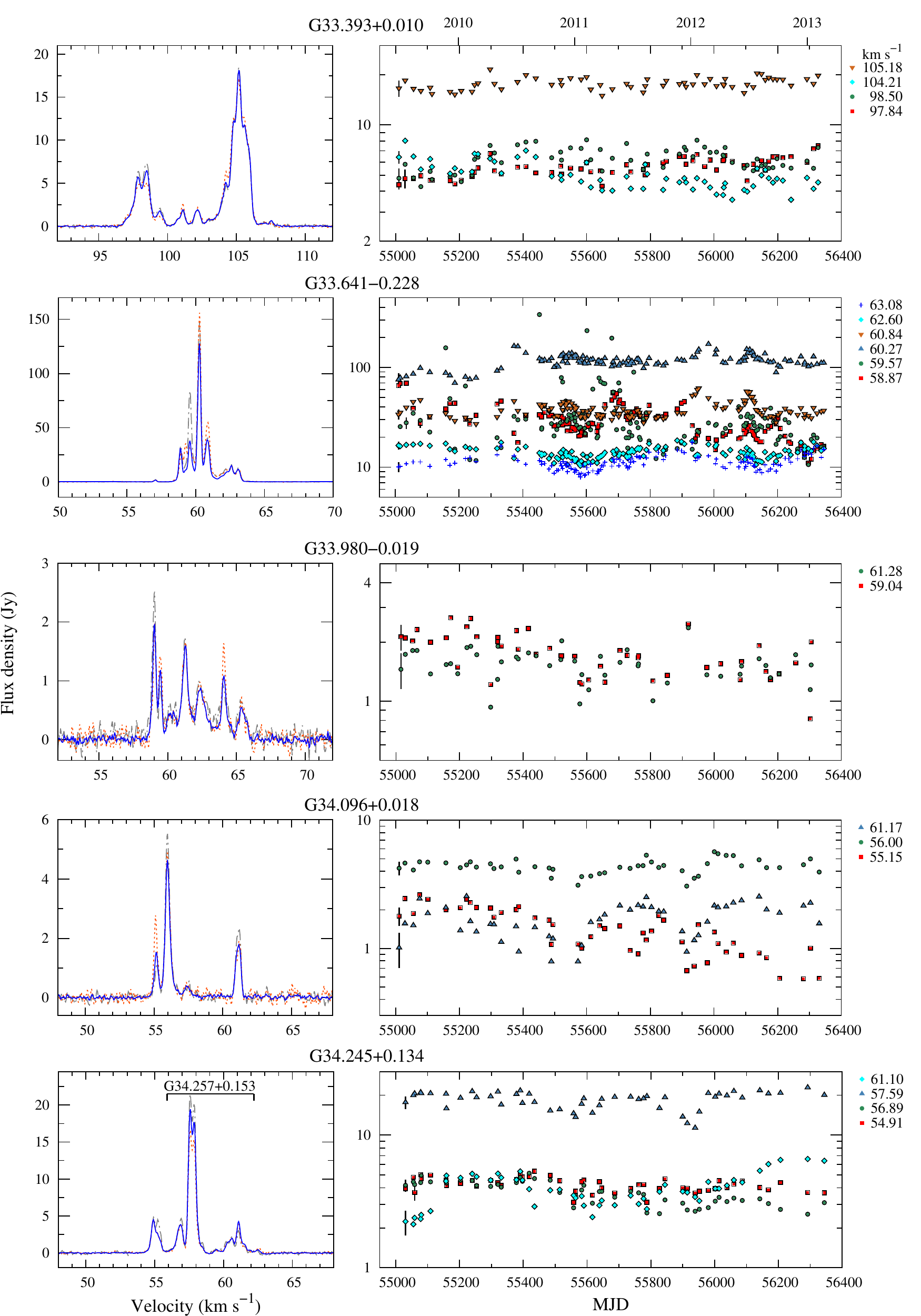}}
\contcaption{
\label{fig:continued}}
\end{figure*}

\newpage
\clearpage
\begin{figure*}   
\resizebox{\hsize}{!}{\includegraphics[width=\textwidth,height=9in]{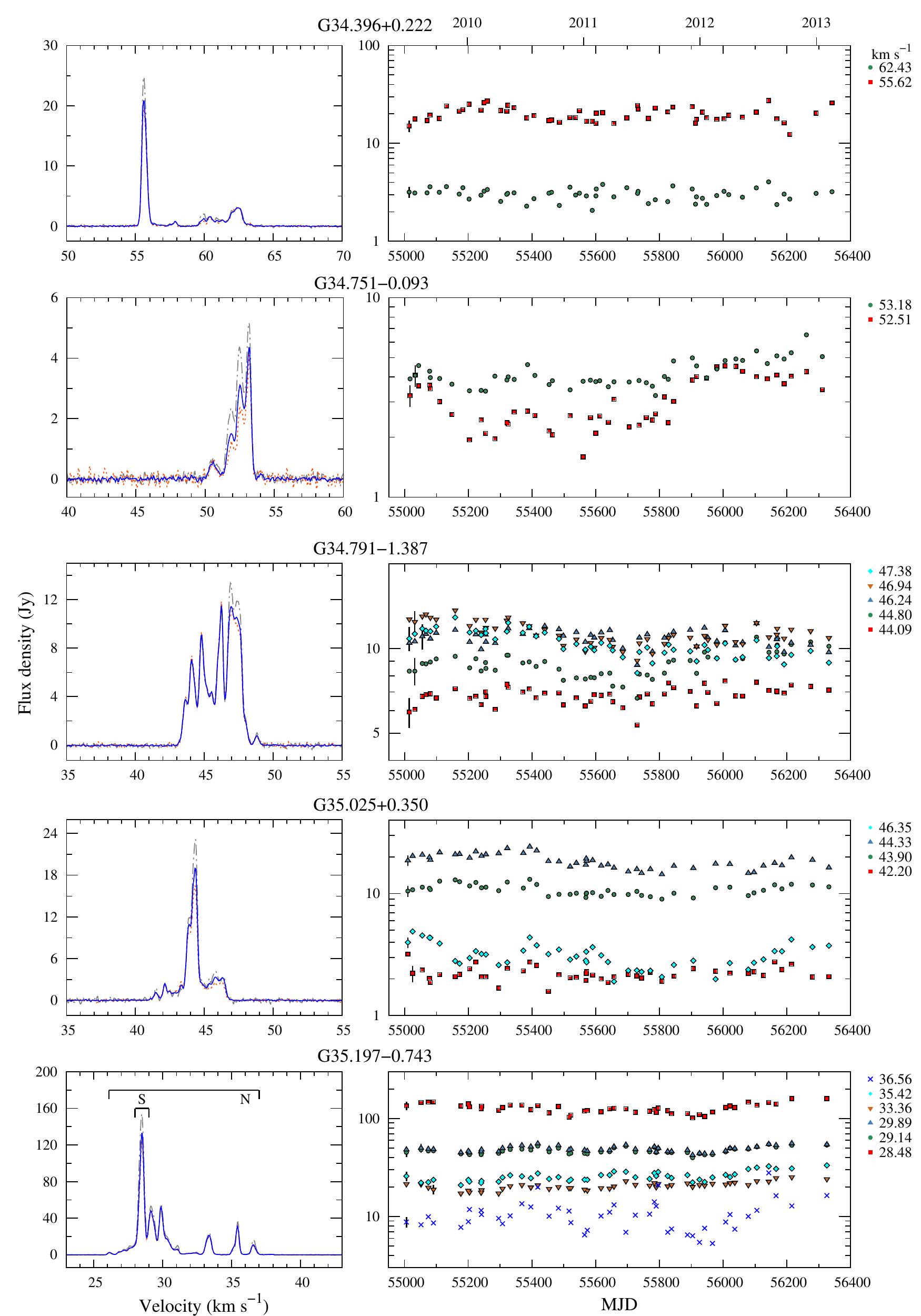}}
\contcaption{
\label{fig:continued}}
\end{figure*}

\begin{figure*}   
\resizebox{\hsize}{!}{\includegraphics[width=\textwidth,height=9in]{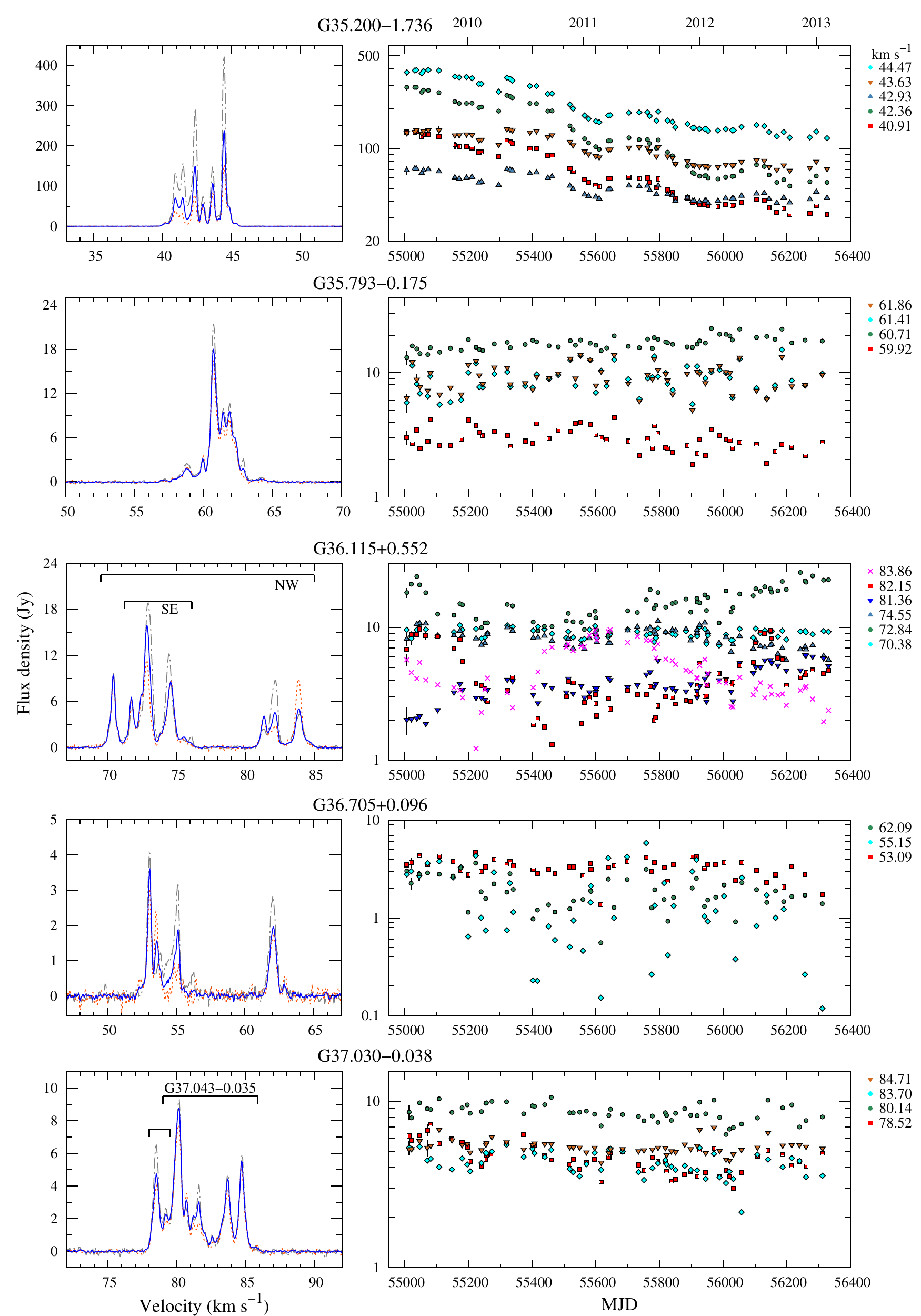}}
\contcaption{
\label{fig:continued}}
\end{figure*}

\begin{figure*}   
\resizebox{\hsize}{!}{\includegraphics[width=\textwidth,height=9in]{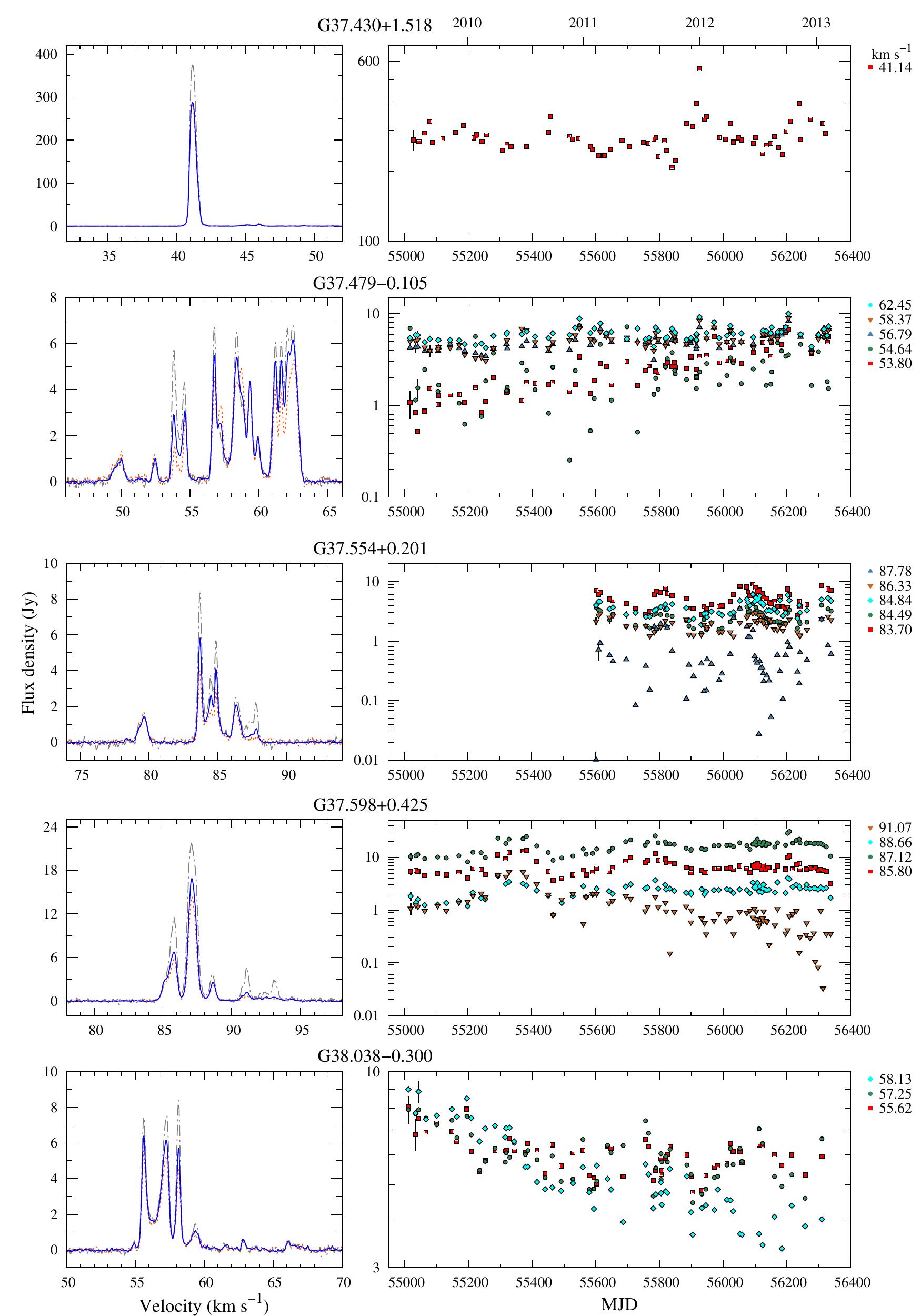}}
\contcaption{
\label{fig:continued}}
\end{figure*}

\begin{figure*}   
\resizebox{\hsize}{!}{\includegraphics[width=\textwidth,height=9in]{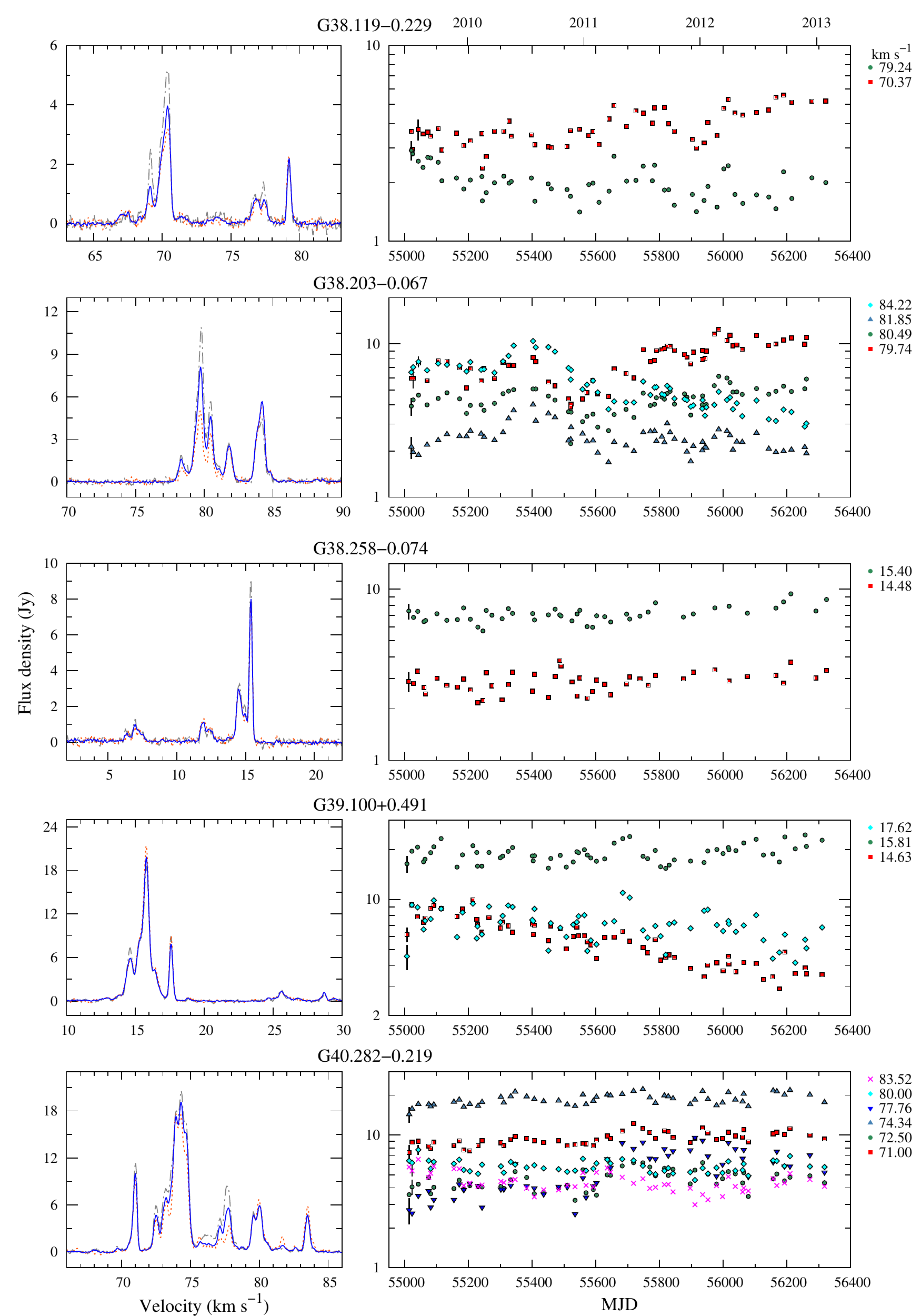}}
\contcaption{
\label{fig:continued}}
\end{figure*}

\clearpage
\begin{figure*}   
\resizebox{\hsize}{!}{\includegraphics[width=\textwidth,height=9in]{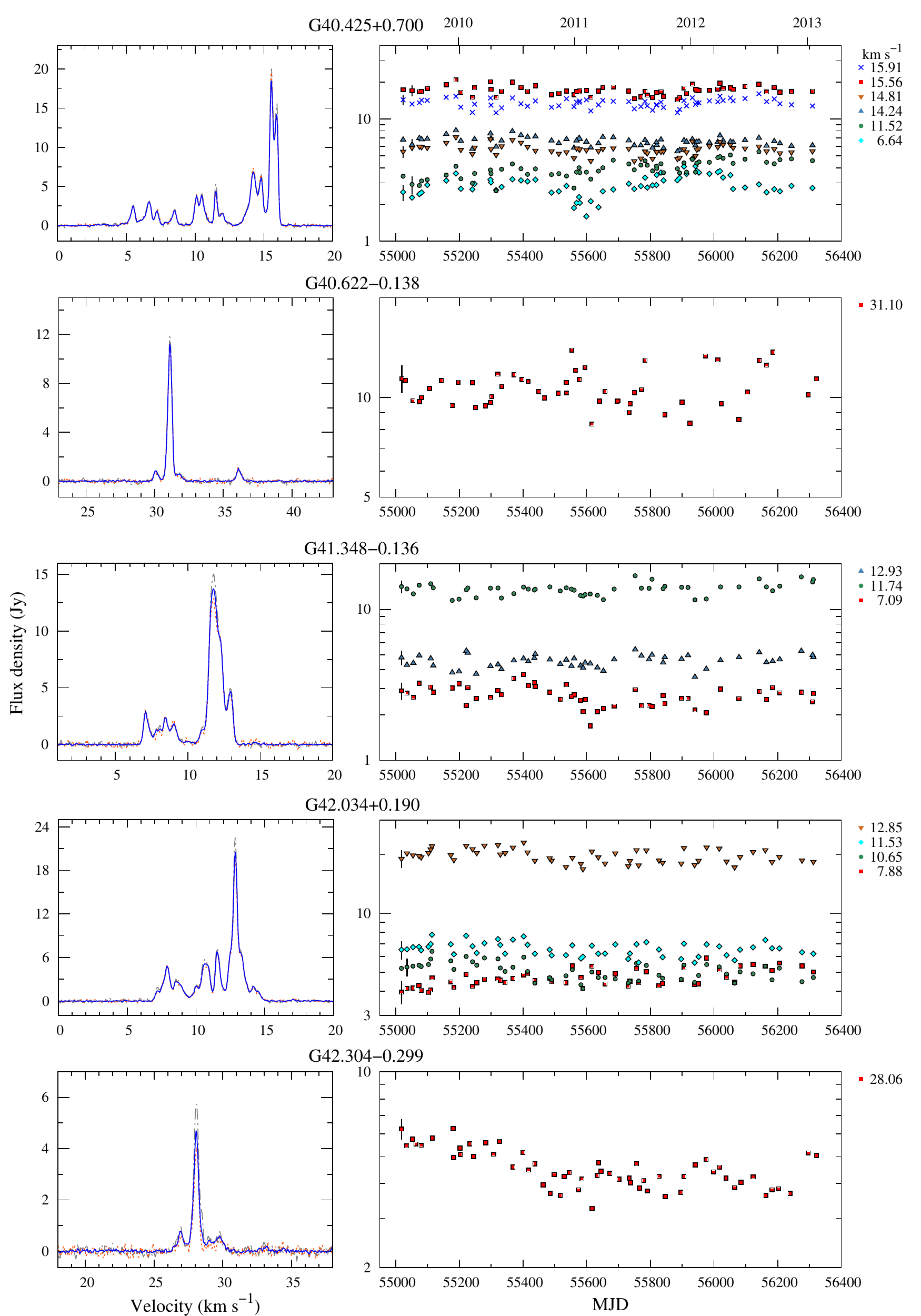}}
\contcaption{
\label{fig:continued}}
\end{figure*}

\begin{figure*}   
\resizebox{\hsize}{!}{\includegraphics[width=\textwidth,height=9in]{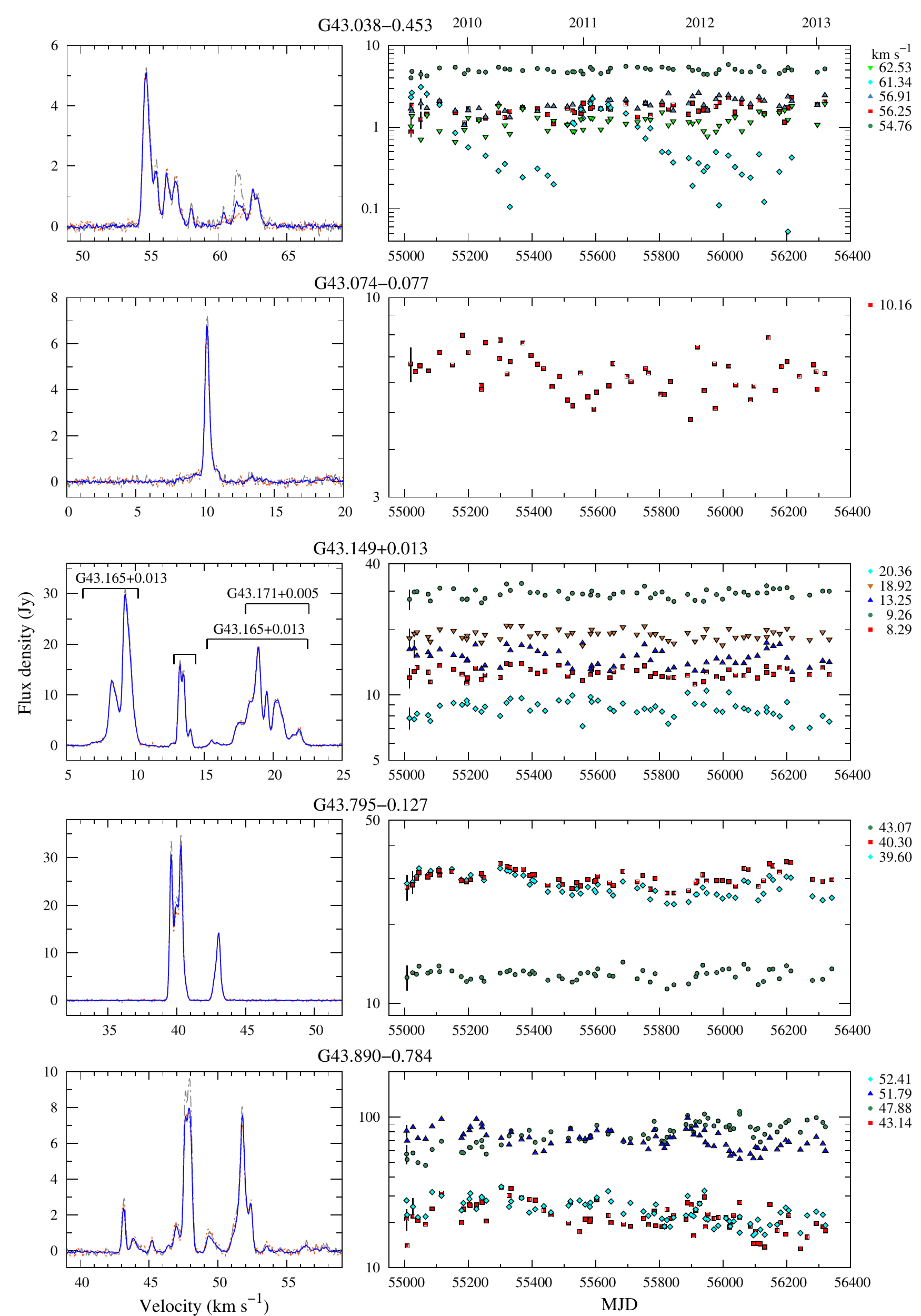}}
\contcaption{
\label{fig:continued}}
\end{figure*}

\begin{figure*}   
\resizebox{\hsize}{!}{\includegraphics[width=\textwidth,height=9in]{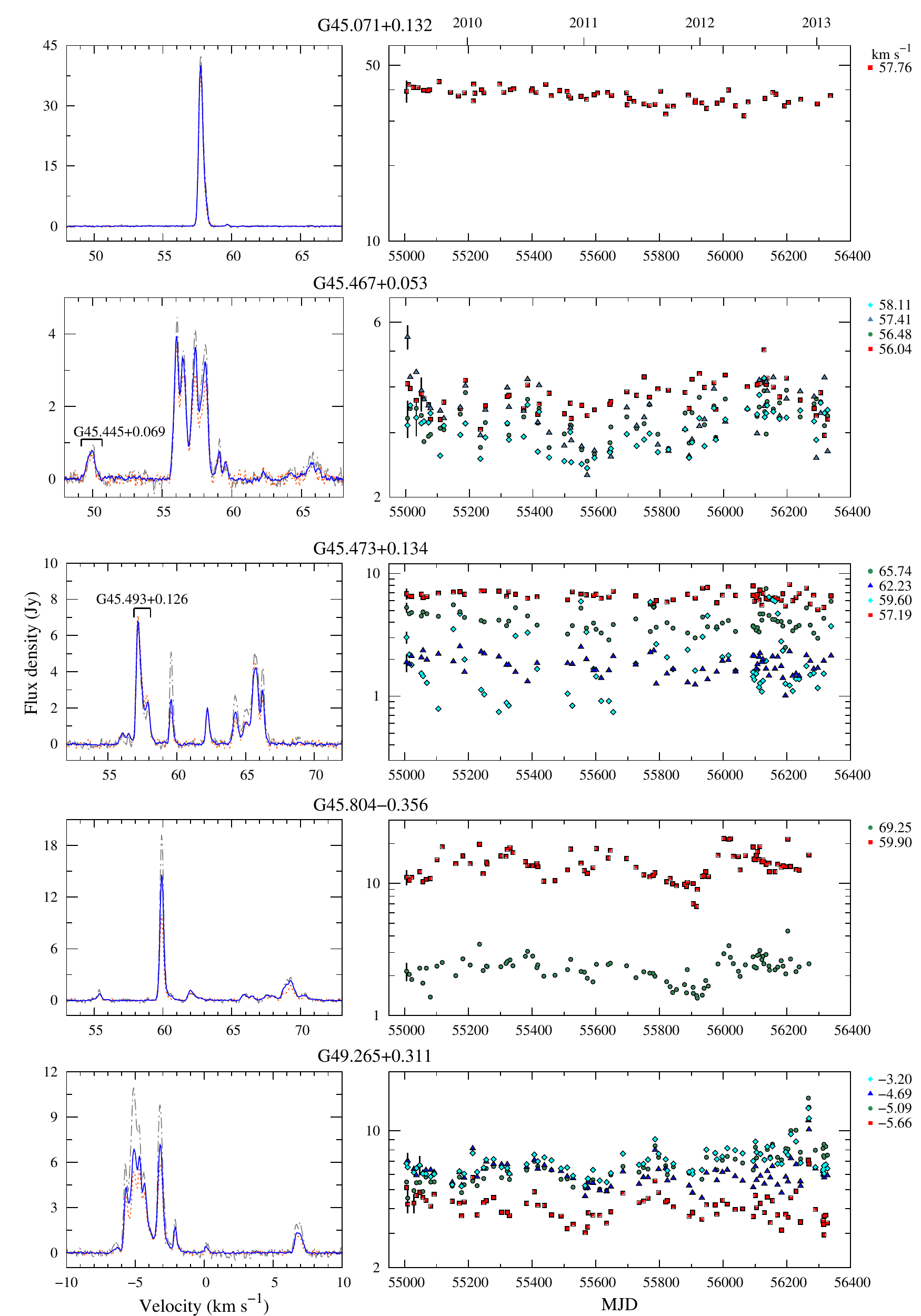}}
\contcaption{
\label{fig:continued}}
\end{figure*}

\begin{figure*}   
\resizebox{\hsize}{!}{\includegraphics[width=\textwidth,height=9in]{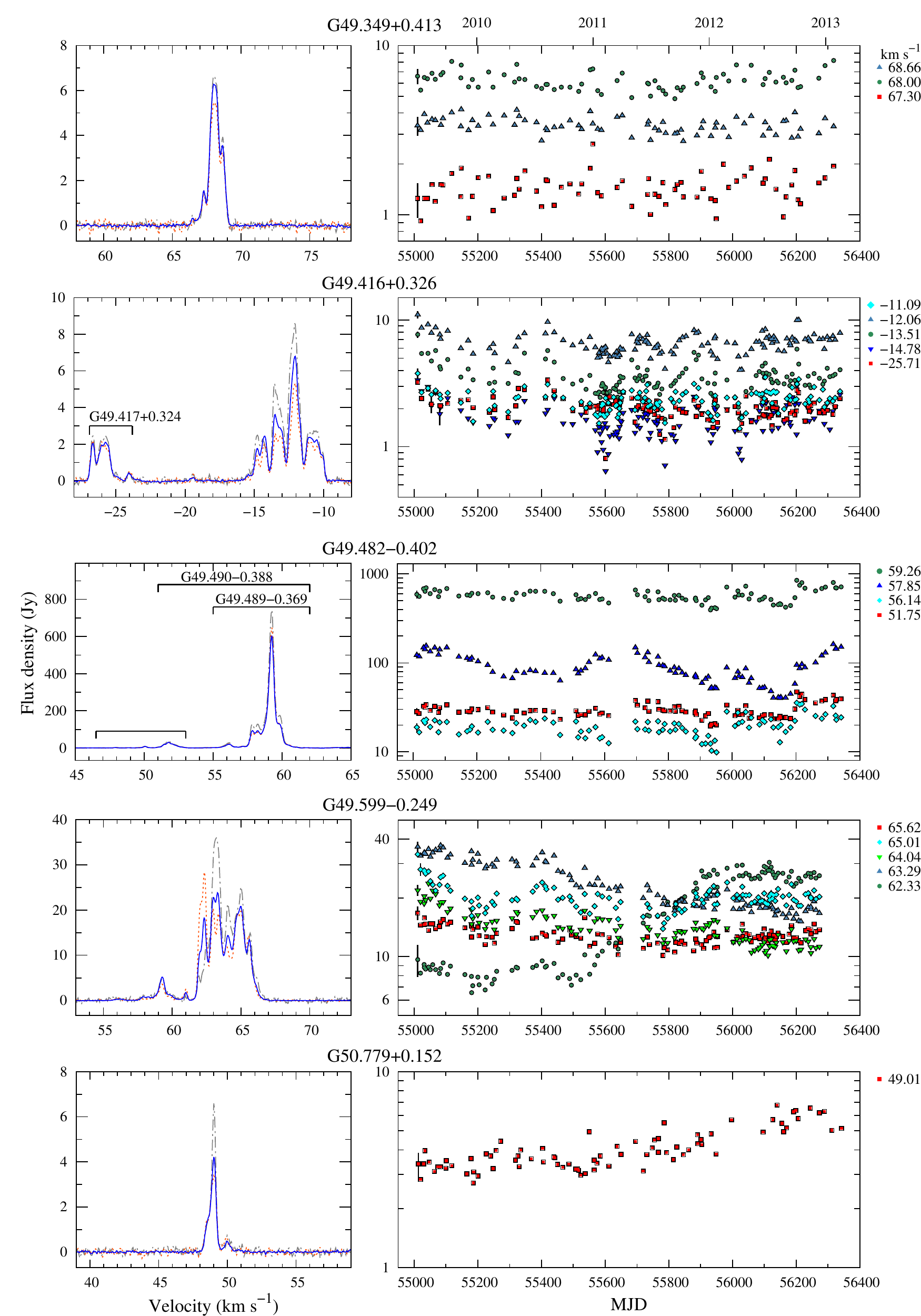}}
\contcaption{
\label{fig:continued}}
\end{figure*}

\begin{figure*}   
\resizebox{\hsize}{!}{\includegraphics[width=\textwidth,height=9in]{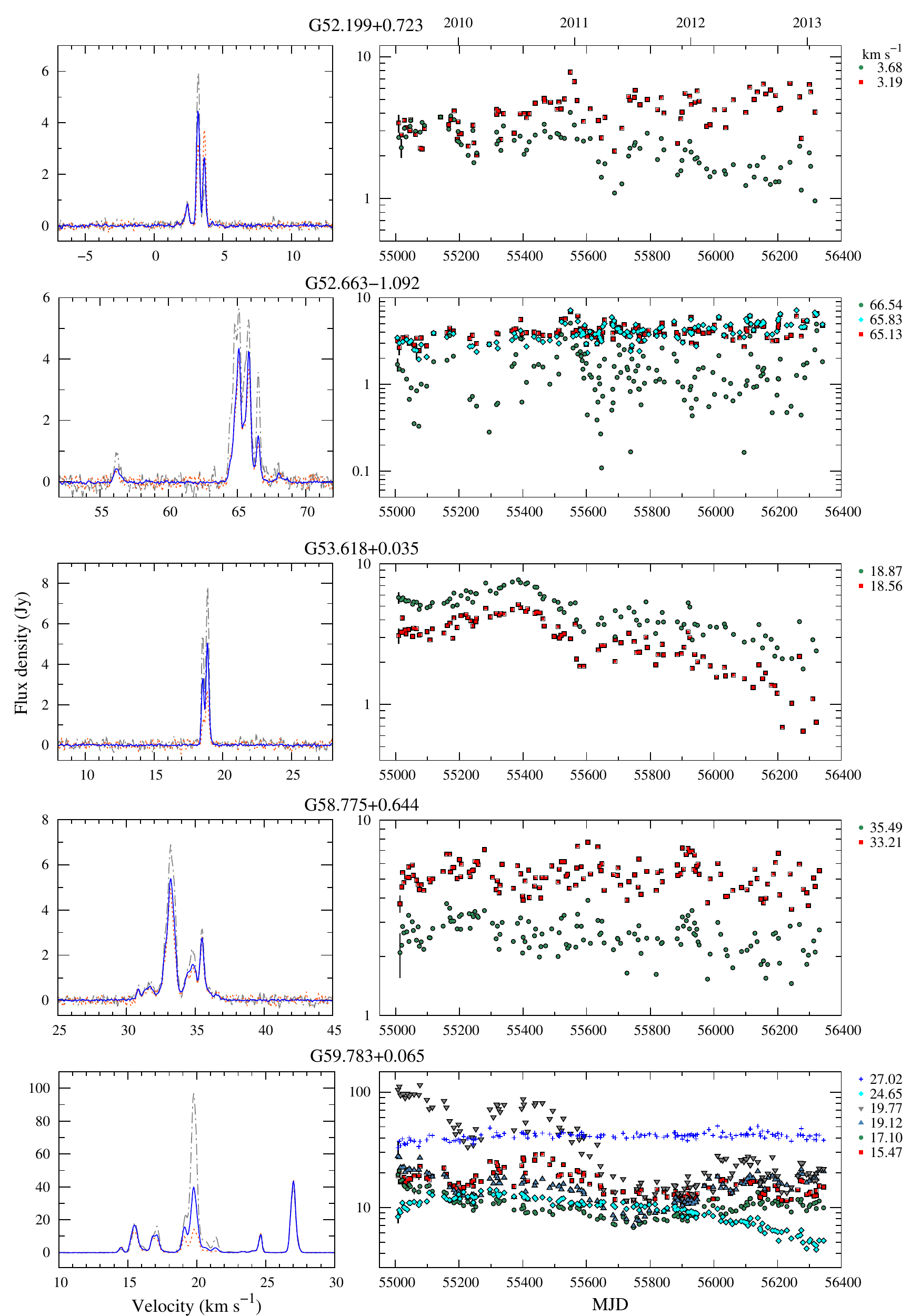}}
\contcaption{
\label{fig:continued}}
\end{figure*}

\begin{figure*}   
\resizebox{\hsize}{!}{\includegraphics[width=\textwidth,height=9in]{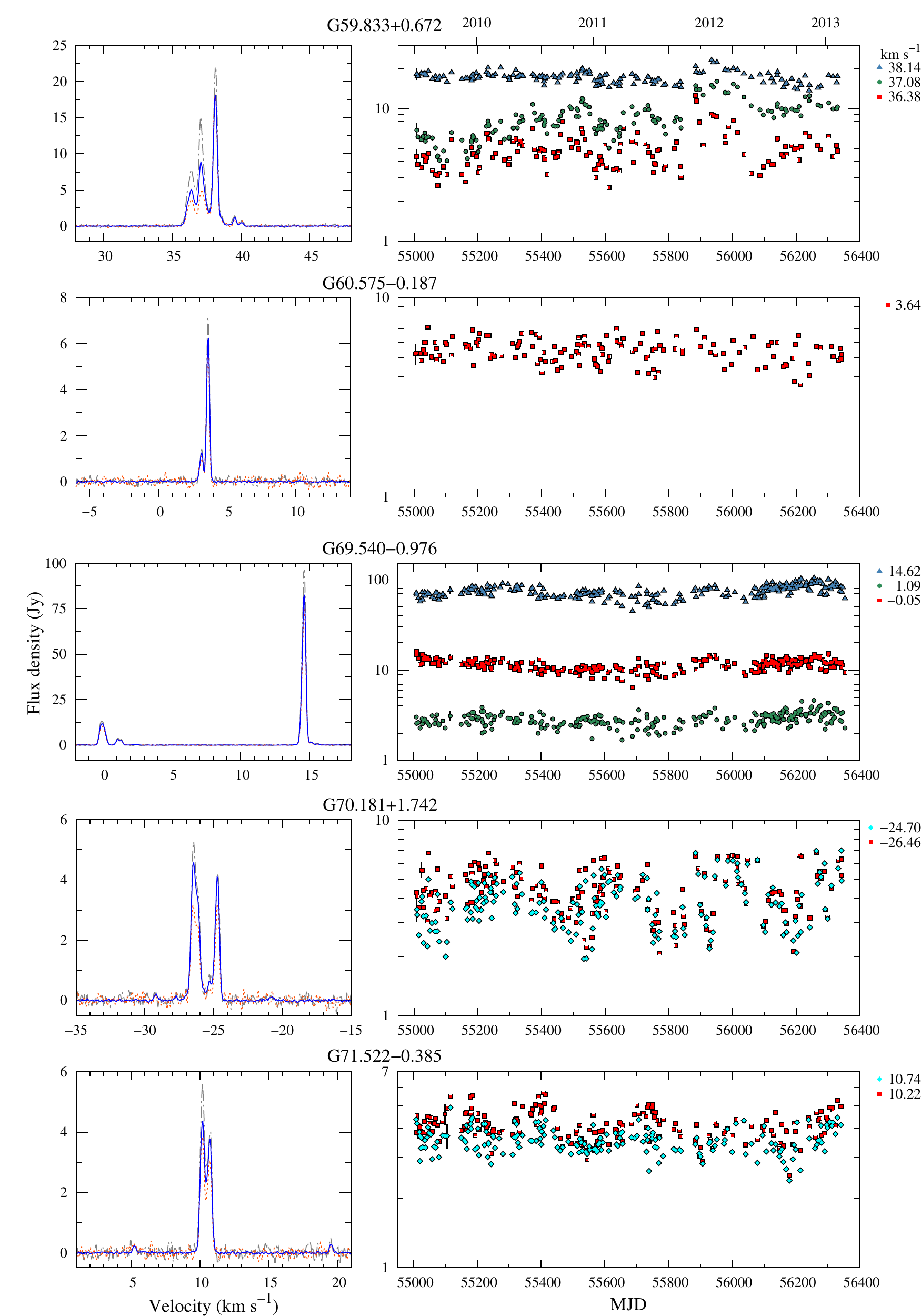}}
\contcaption{
\label{fig:continued}}
\end{figure*}

\begin{figure*}   
\resizebox{\hsize}{!}{\includegraphics[width=\textwidth,height=9in]{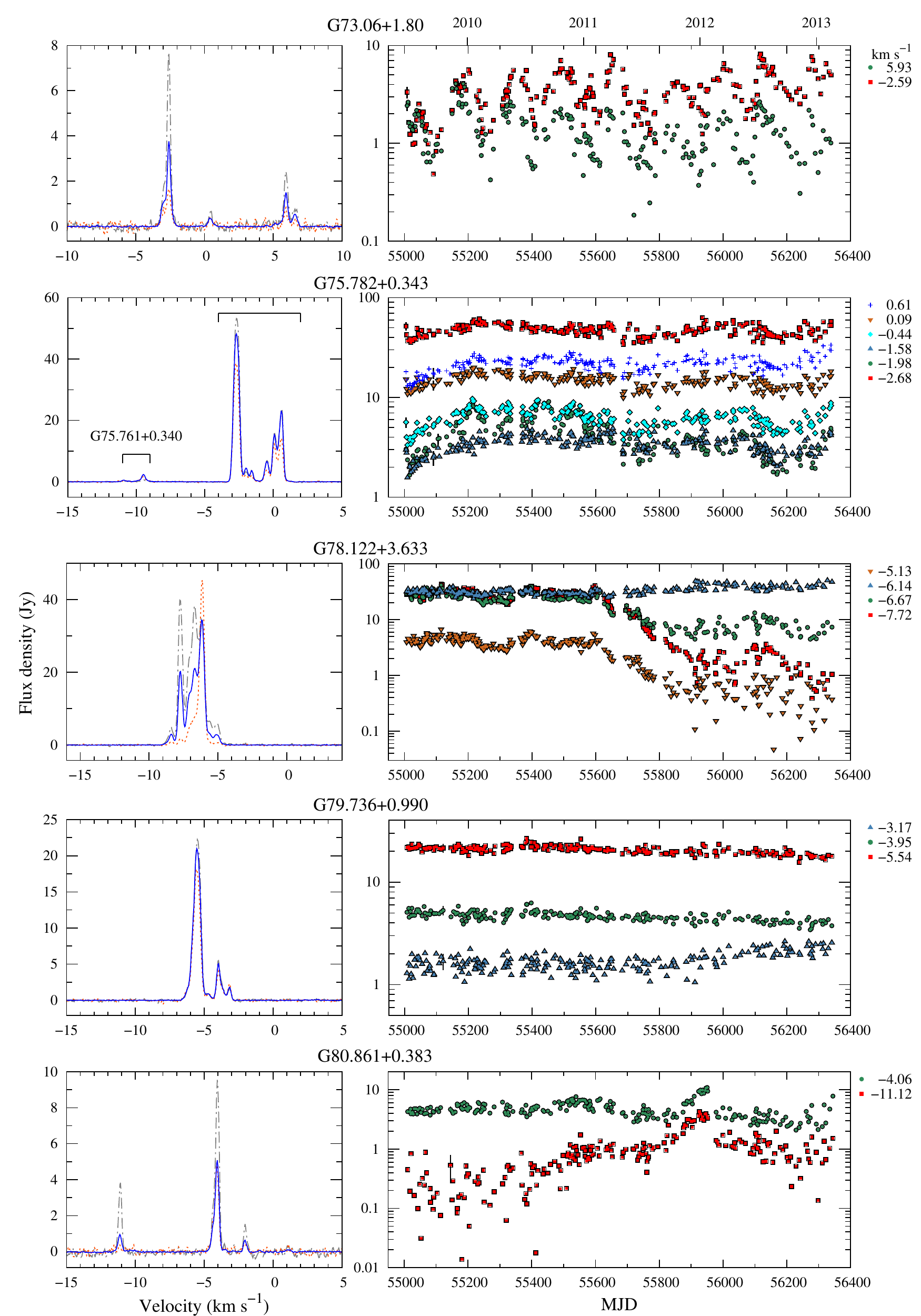}}
\contcaption{
\label{fig:continued}}
\end{figure*}

\begin{figure*}   
\resizebox{\hsize}{!}{\includegraphics[width=\textwidth,height=9in]{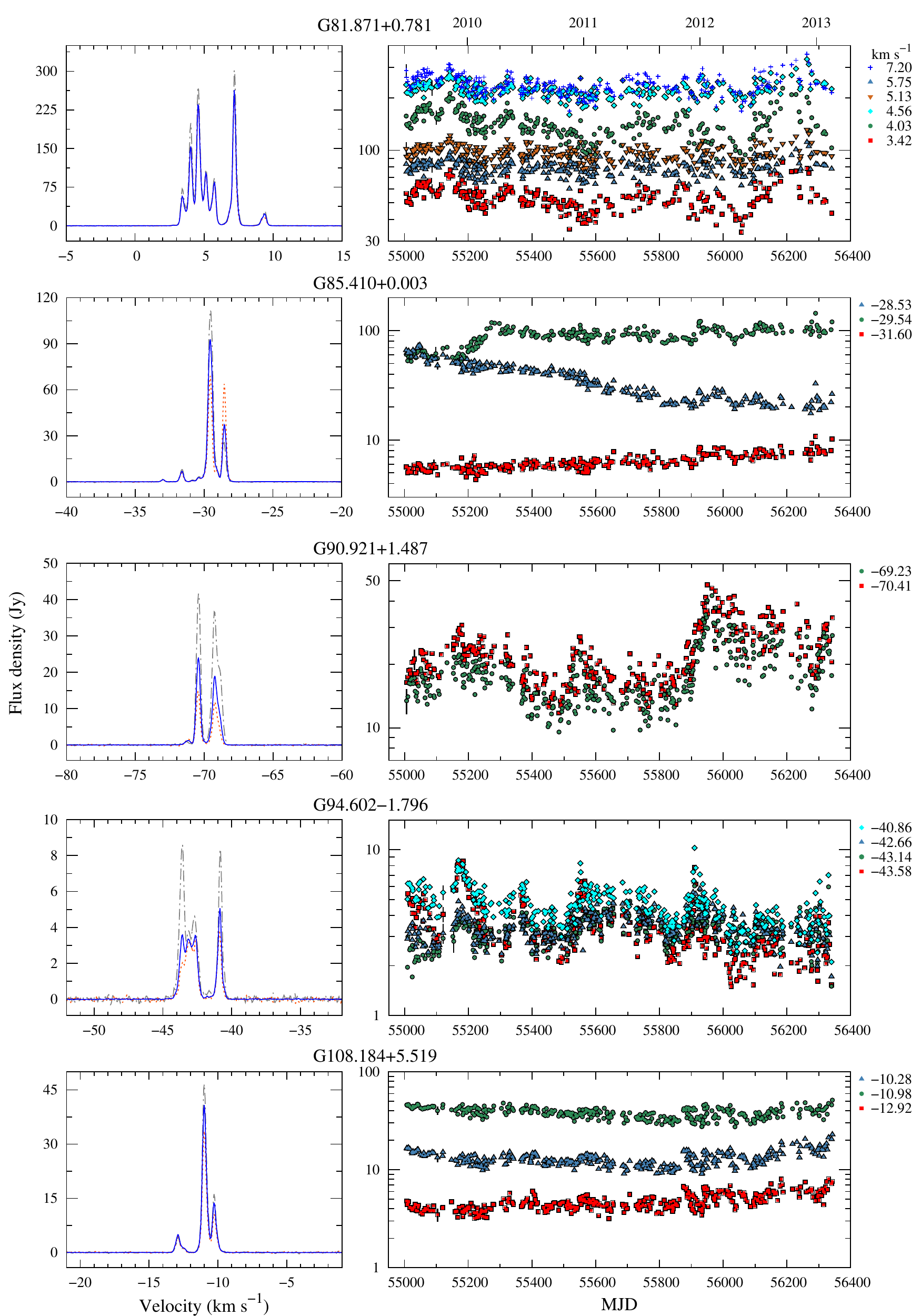}}
\contcaption{
\label{fig:continued}}
\end{figure*}
\newpage
\clearpage
\begin{figure*}   
\resizebox{\hsize}{!}{\includegraphics[width=\textwidth,height=9in]{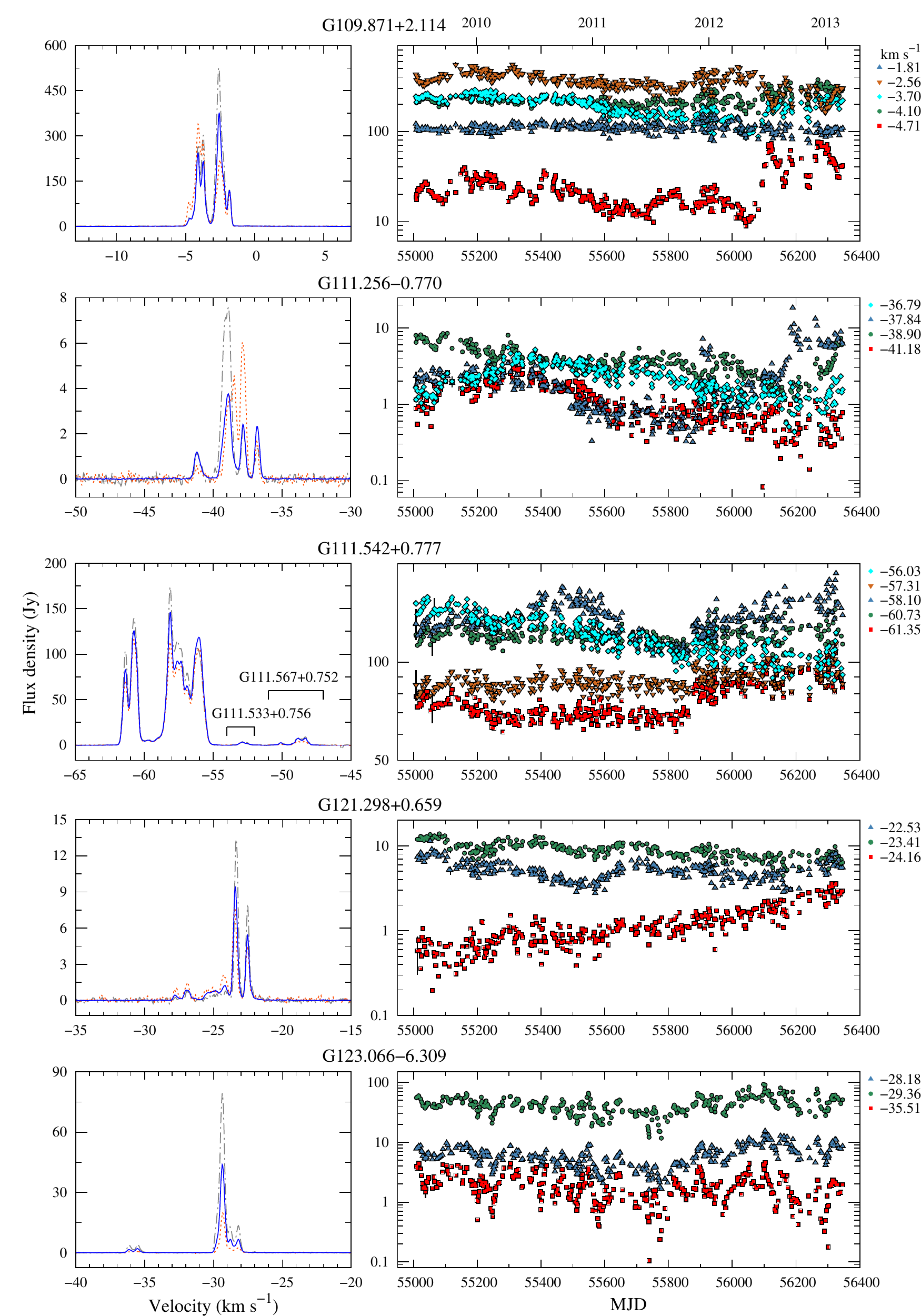}}
\contcaption{
\label{fig:continued}}
\end{figure*}

\begin{figure*}   
\resizebox{\hsize}{!}{\includegraphics[width=\textwidth,height=9in]{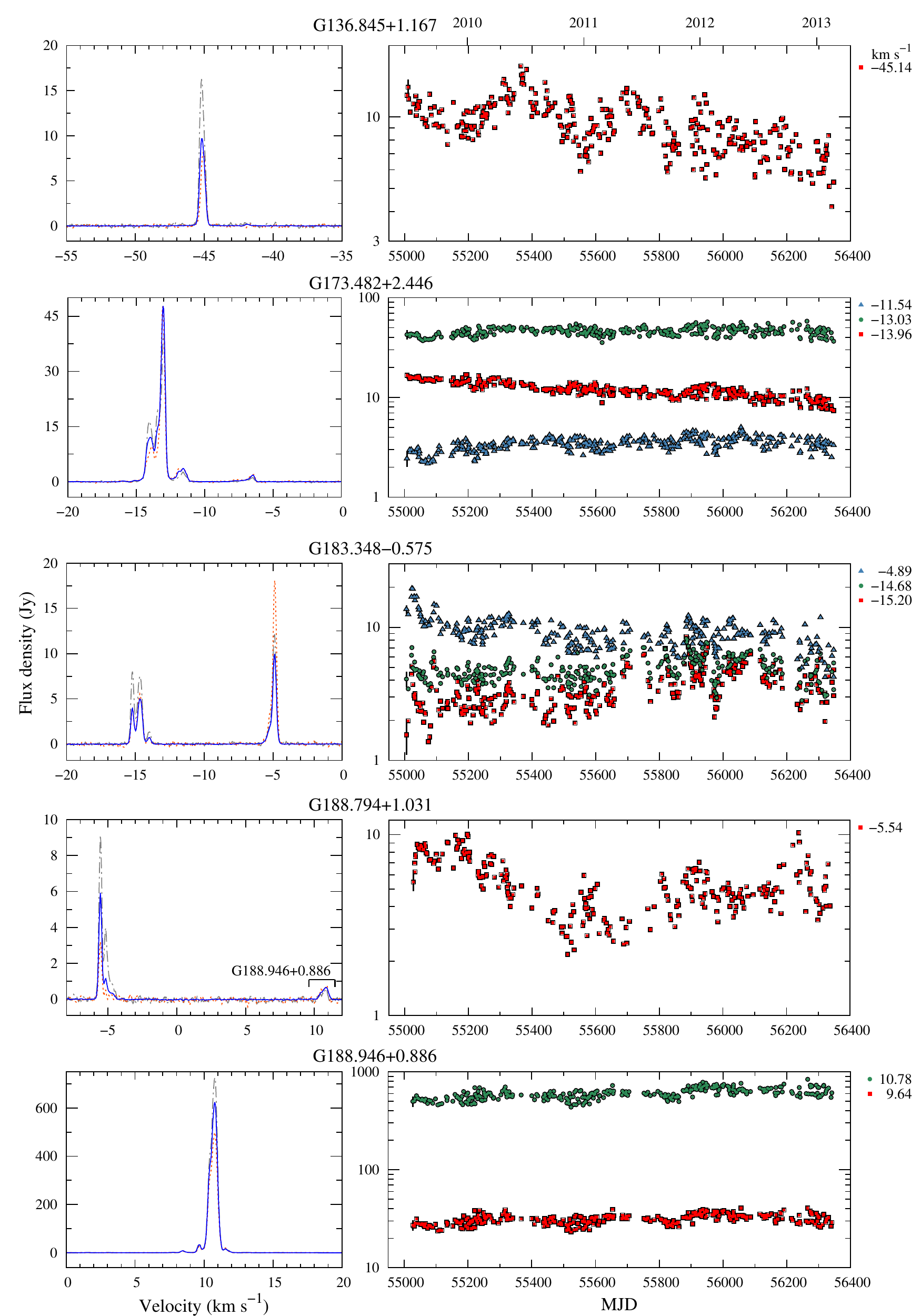}}
\contcaption{
\label{fig:continued}}
\end{figure*}

\begin{figure*}   
\resizebox{\hsize}{!}{\includegraphics[width=\textwidth,height=9in]{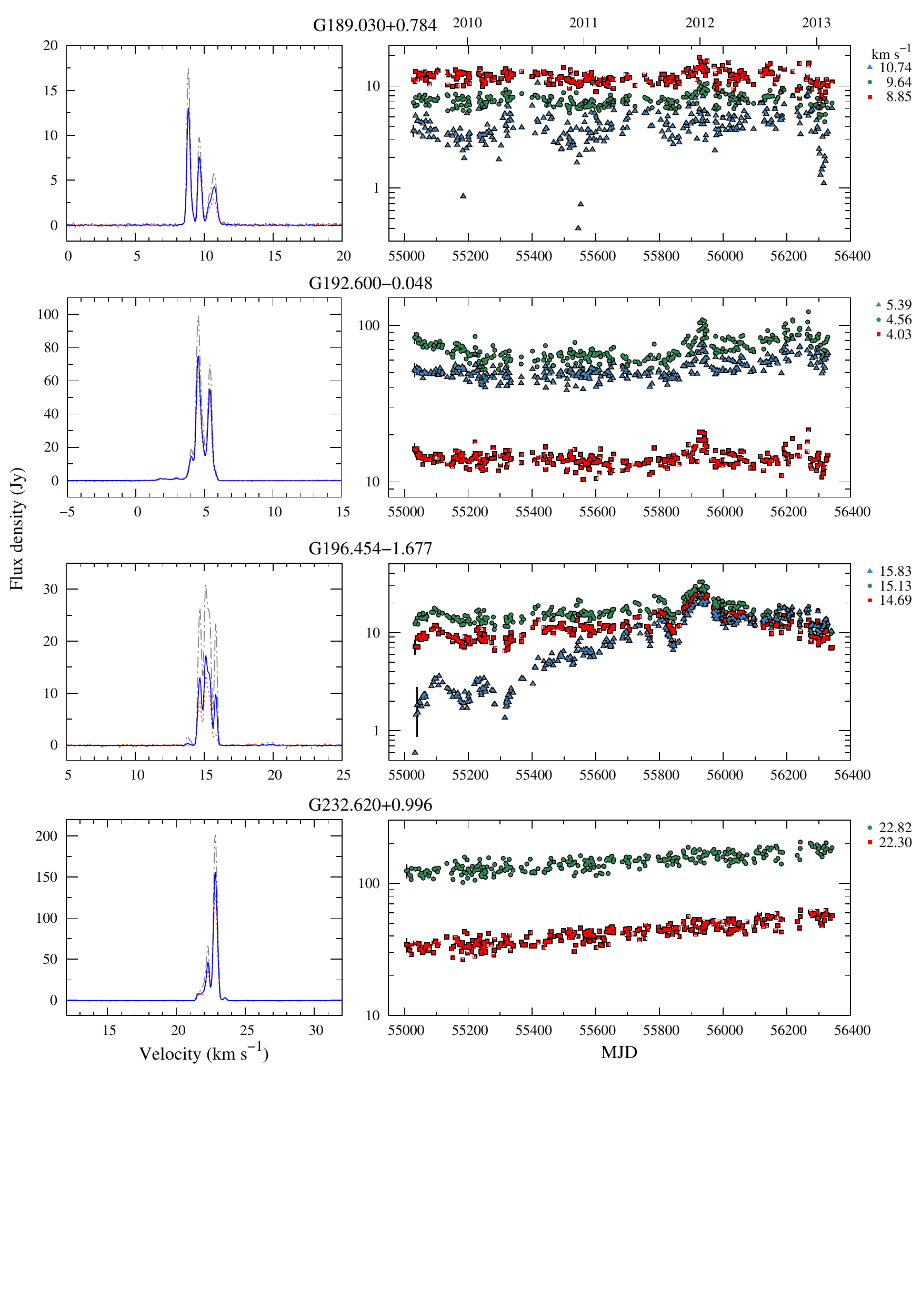}}
\contcaption{
\label{fig:continued}}
\end{figure*}
\clearpage

\begin{appendix}
 \section{}


 \begin{figure}  
\includegraphics[angle=0, scale=0.5]{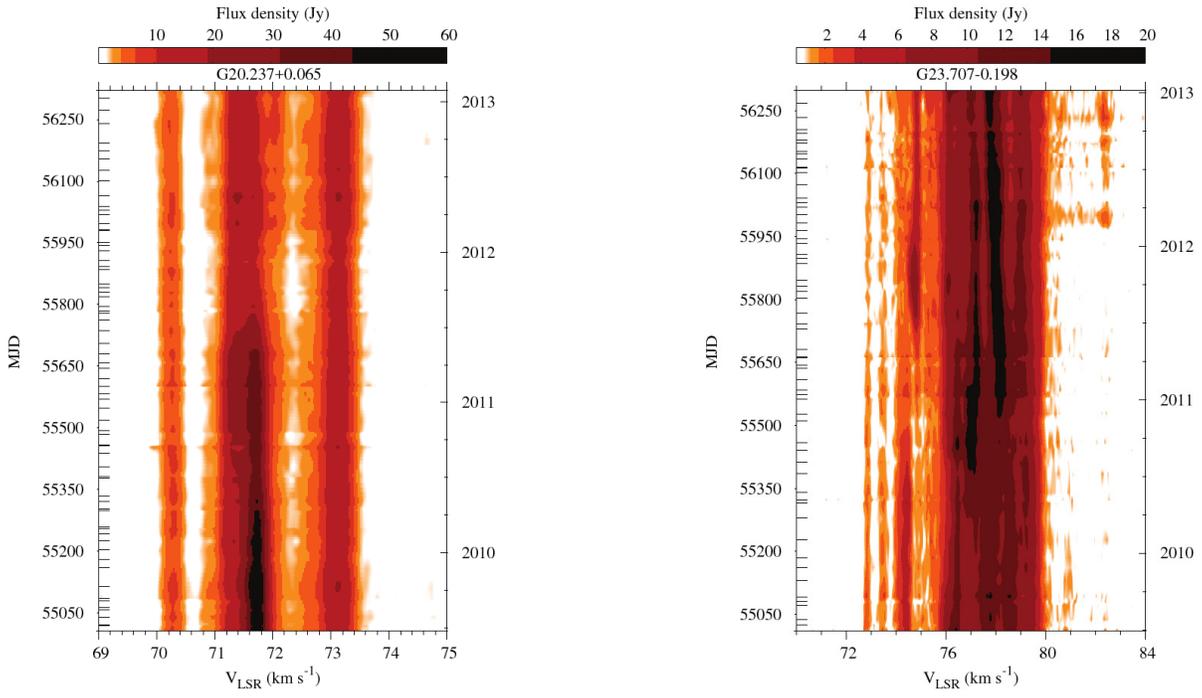}
\caption{Dynamic spectra for sources with complex variability. The velocity scale is relative to the local
standard of rest. The horizontal bars in the left coordinate correspond to the dates of the observed spectra.
\label{dyn-spectra}}
\end{figure}

\begin{figure}  

\includegraphics[angle=0, scale=0.5]{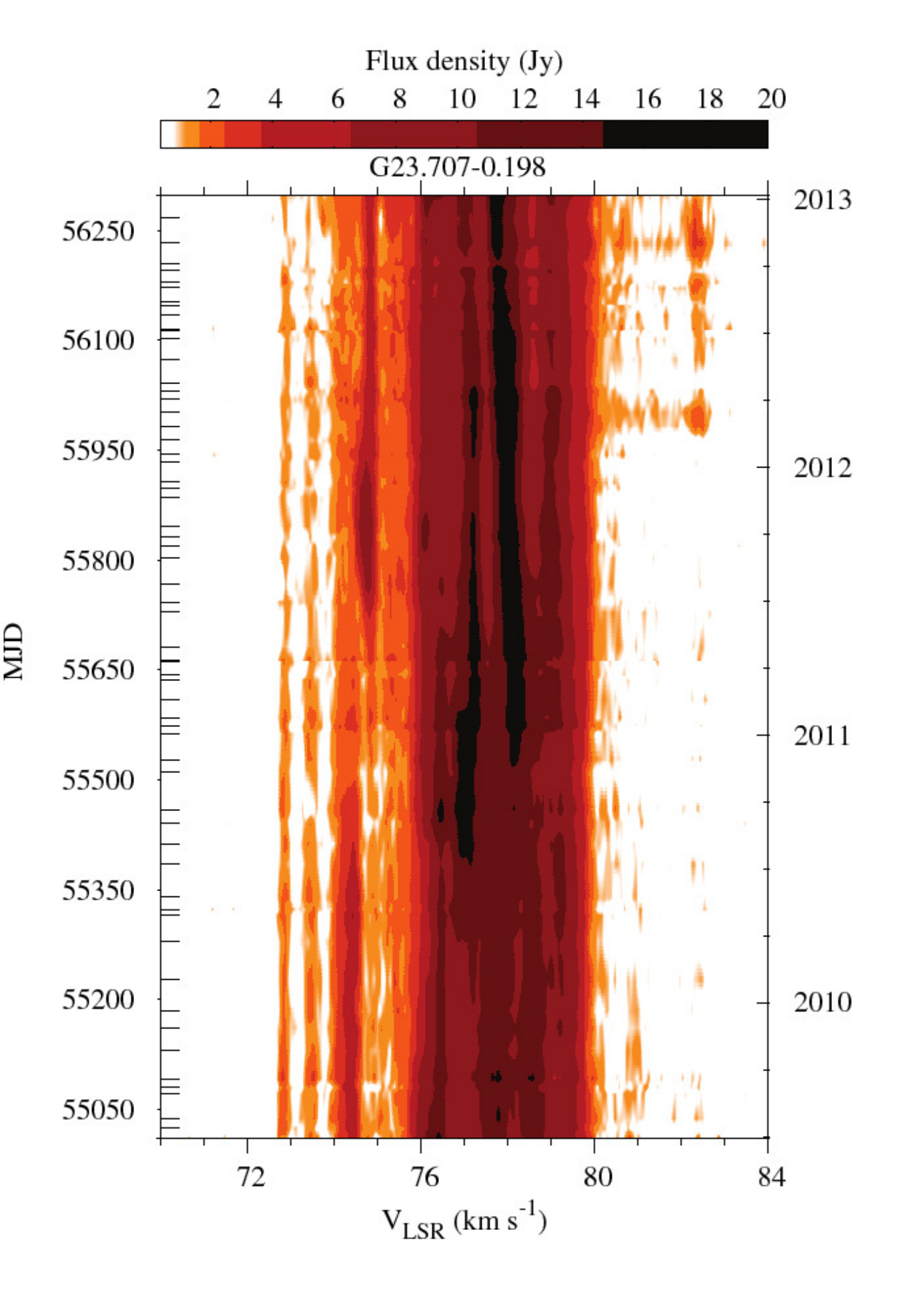}
\includegraphics[angle=0, scale=0.5]{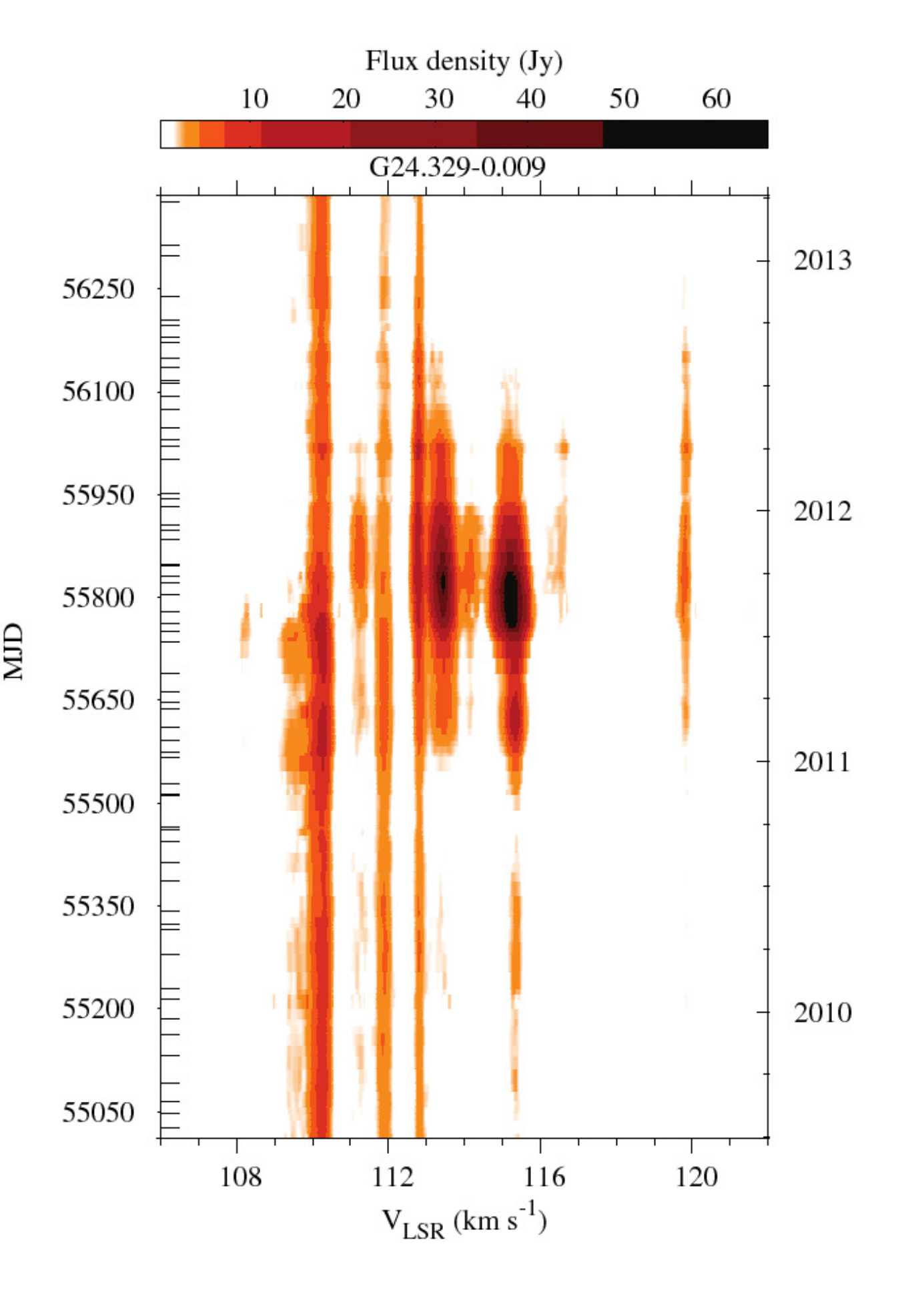}
\contcaption{
\label{fig:continued}}
\end{figure}

\begin{figure} 
\includegraphics[angle=0, scale=0.5]{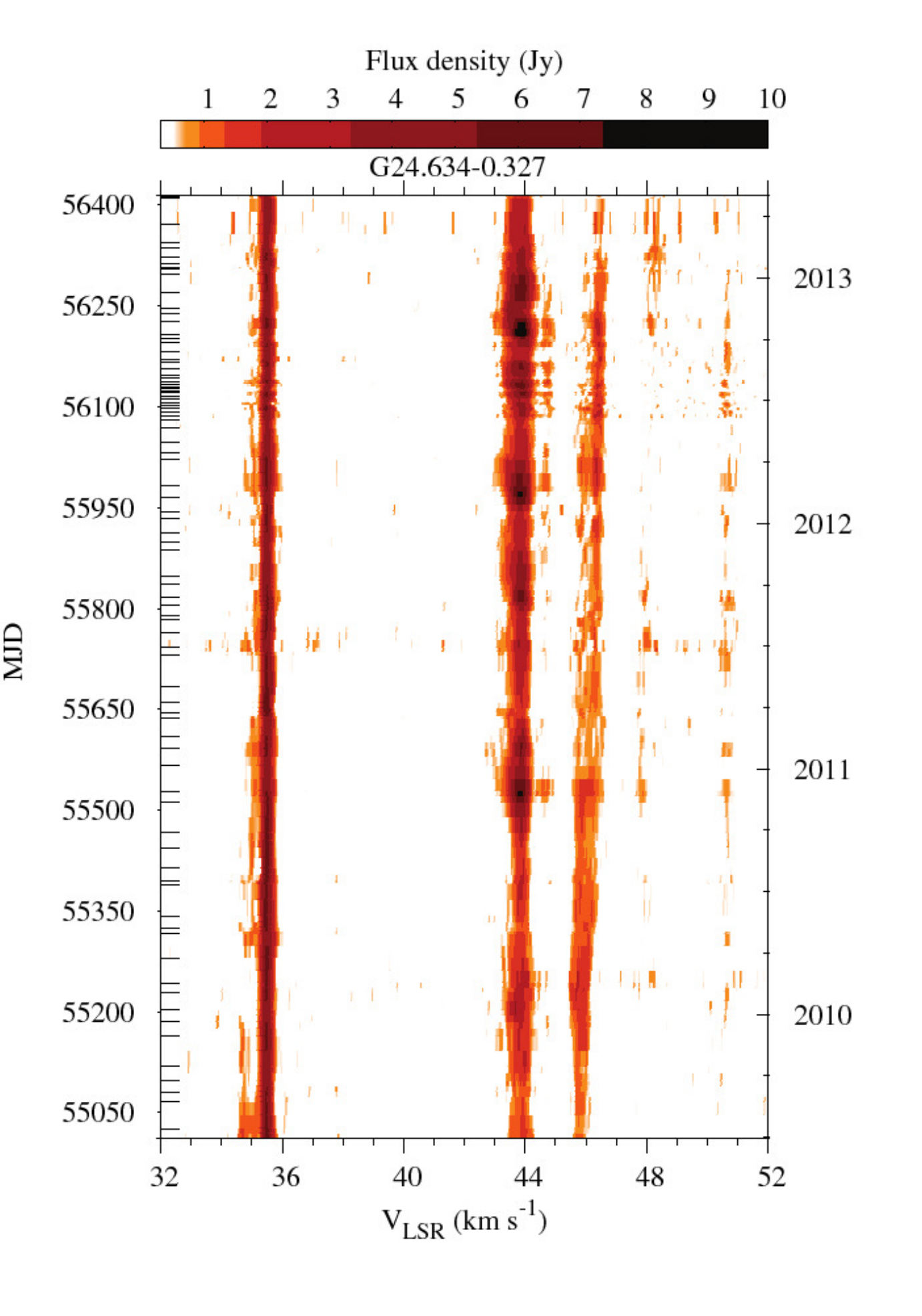}
\includegraphics[angle=0, scale=0.5]{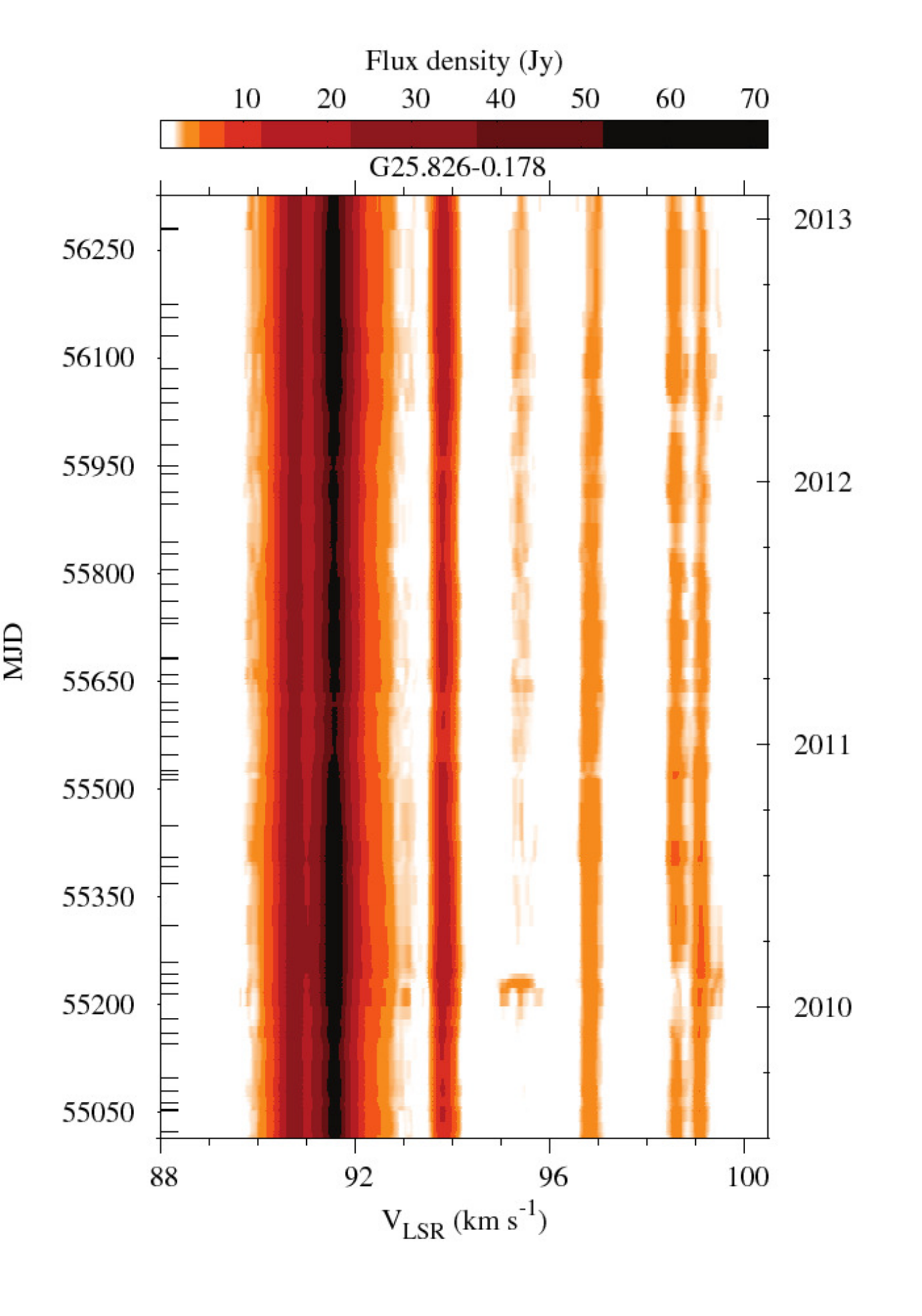}
\contcaption{
\label{fig:continued}}
\end{figure}

\begin{figure}   
\includegraphics[angle=0, scale=0.5]{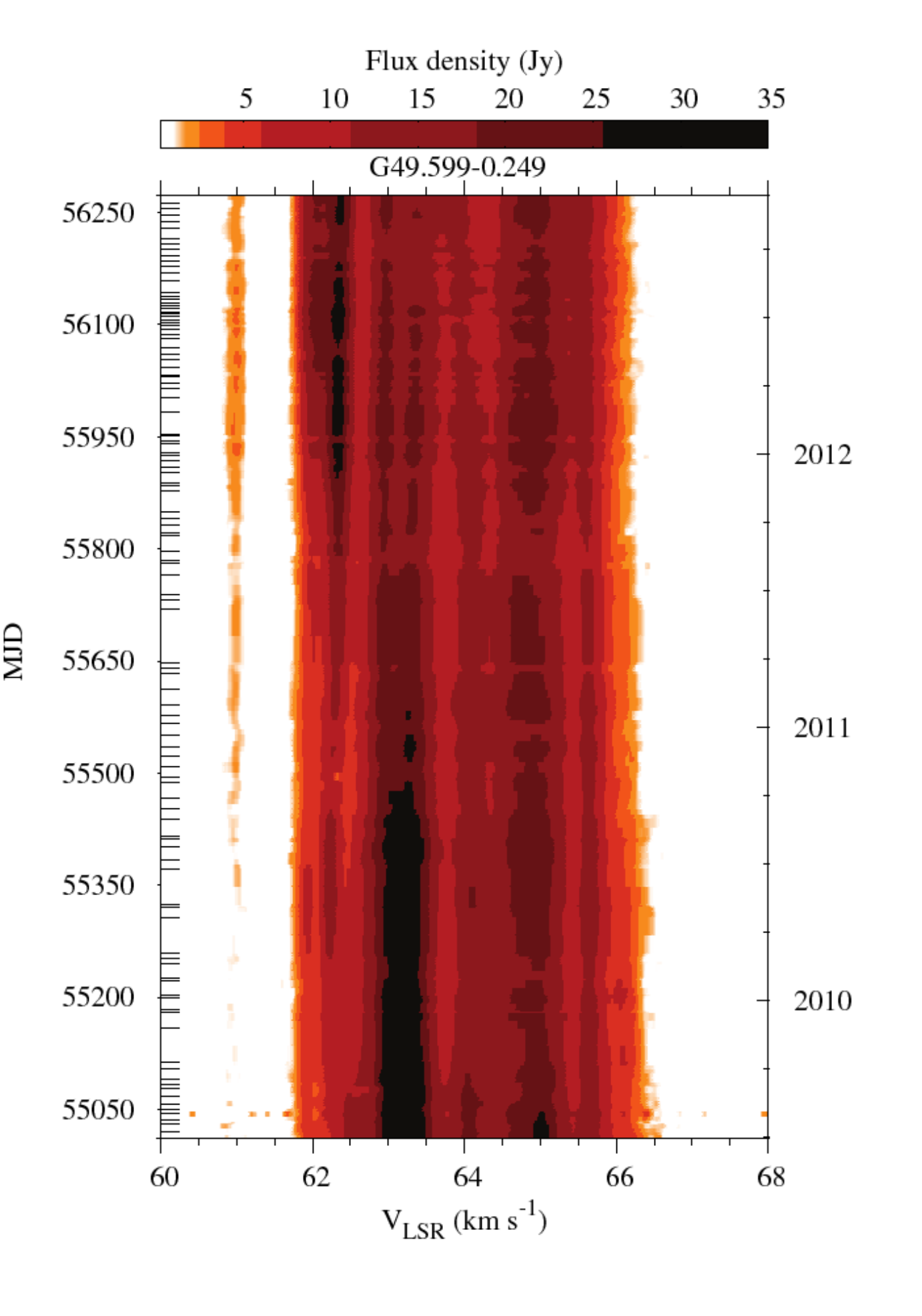}
\includegraphics[angle=0,scale=0.5]{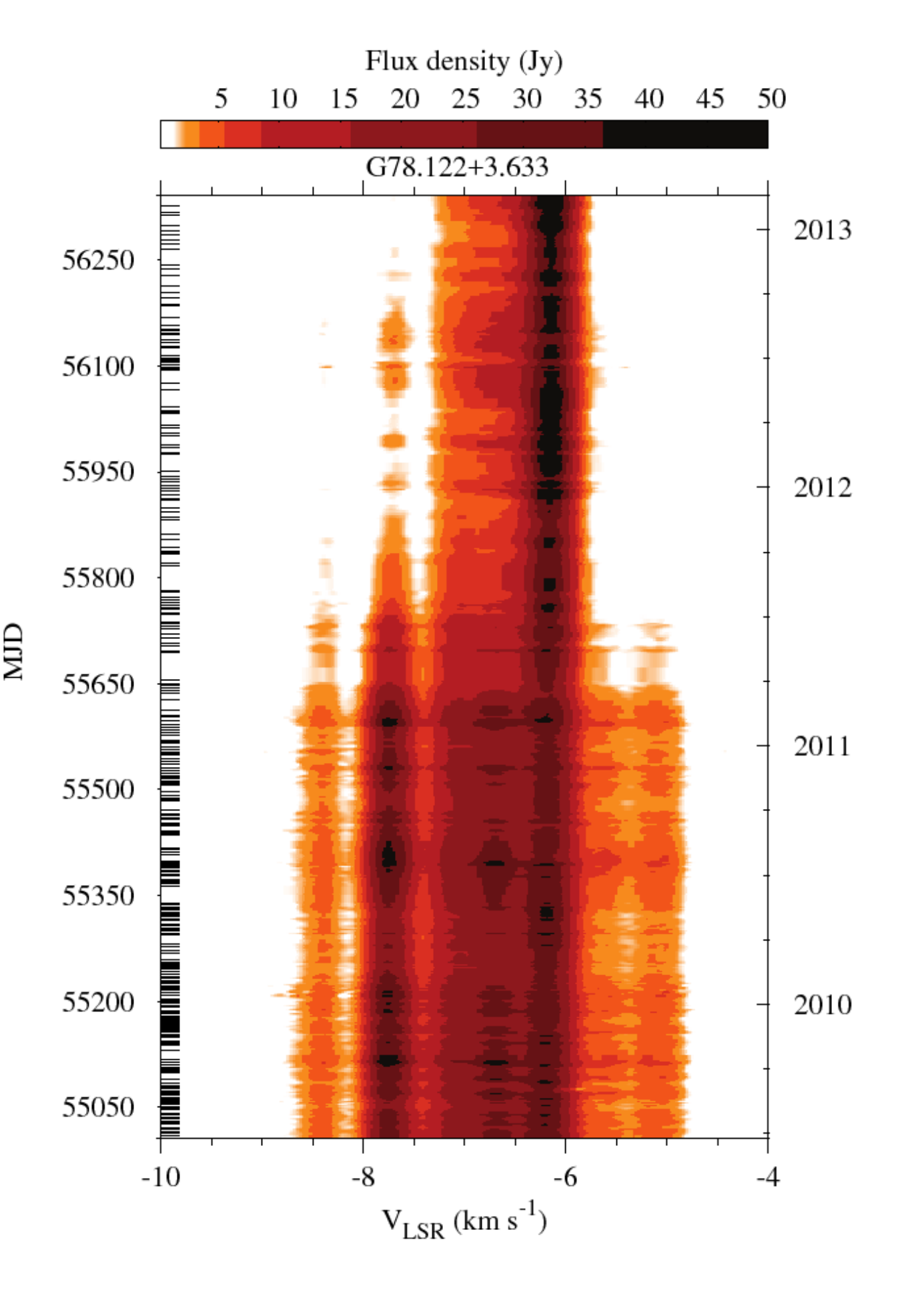}
\contcaption{
\label{fig:continued}}
\end{figure}



\clearpage 
\newpage 

\section{}
\begin{figure}   
\includegraphics[angle=0, scale=0.5]{dyn_g23p208-eps-converted-to.pdf}
\includegraphics[angle=0, scale=0.5]{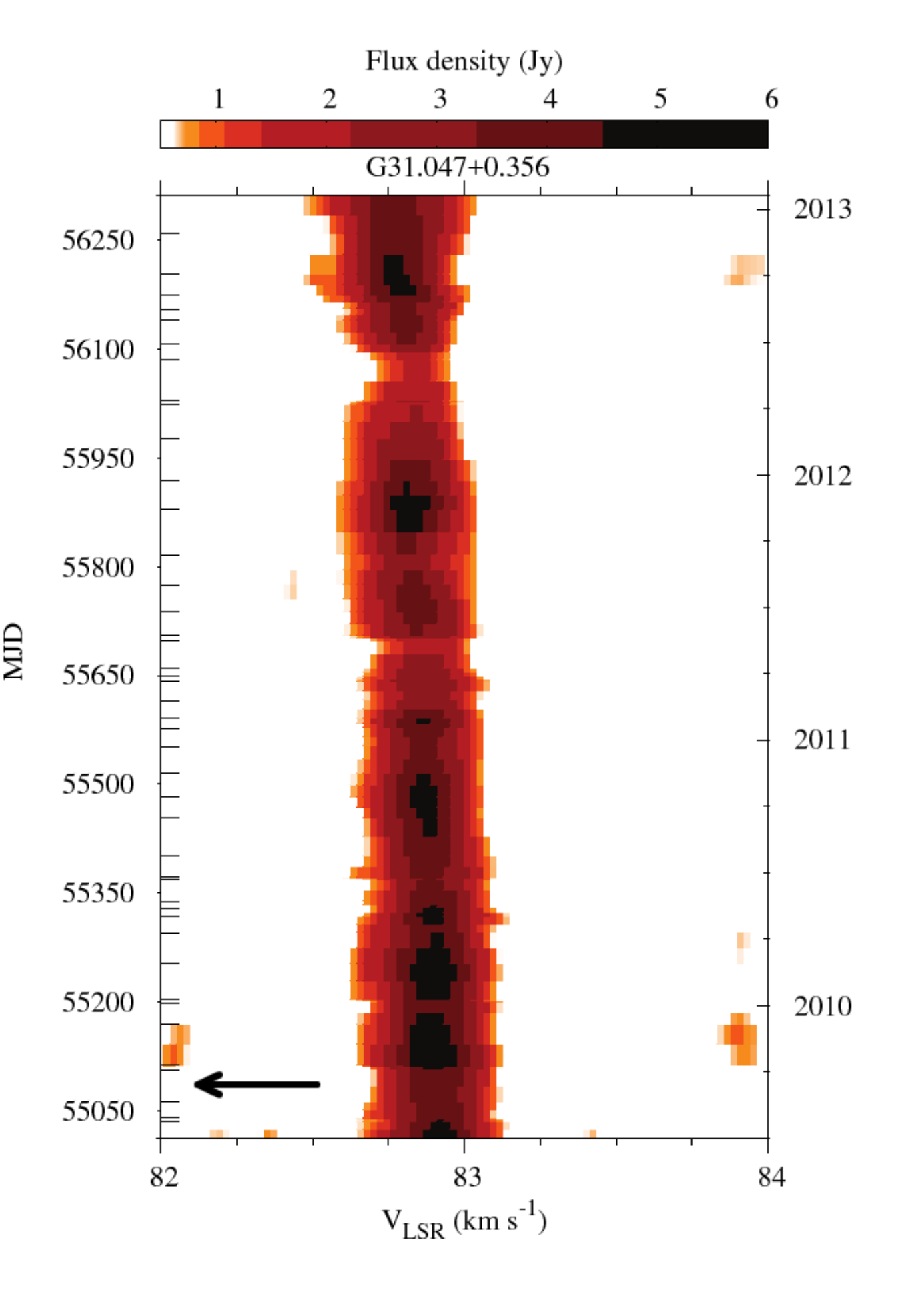}
\caption{Dynamic spectra for features with velocity drifts. The arrow indicates the systemic velocity for each source.
\label{dyn-drift}} 
\end{figure}

\begin{figure}   
\includegraphics[angle=0, scale=0.5]{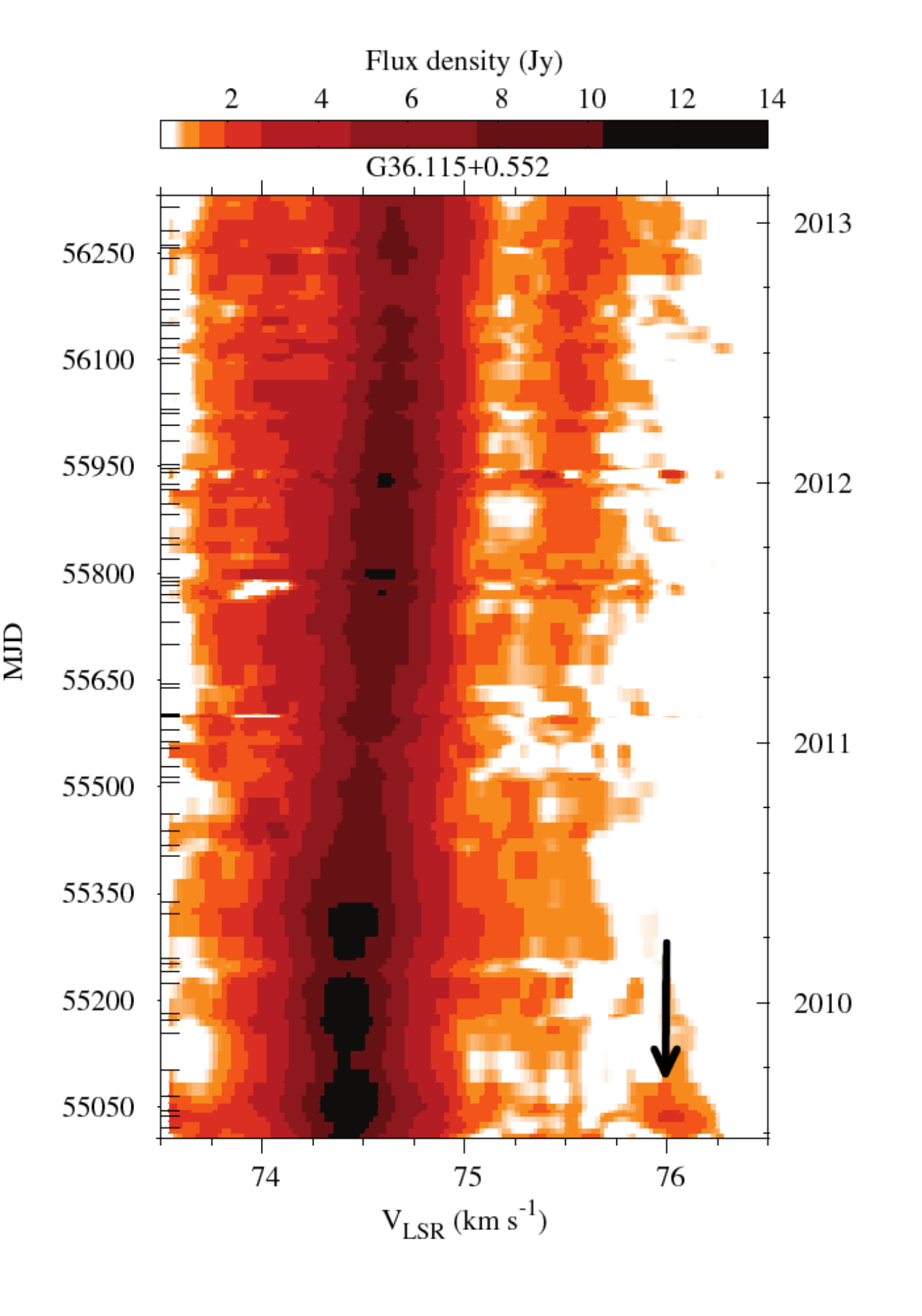}
\includegraphics[angle=0, scale=0.5]{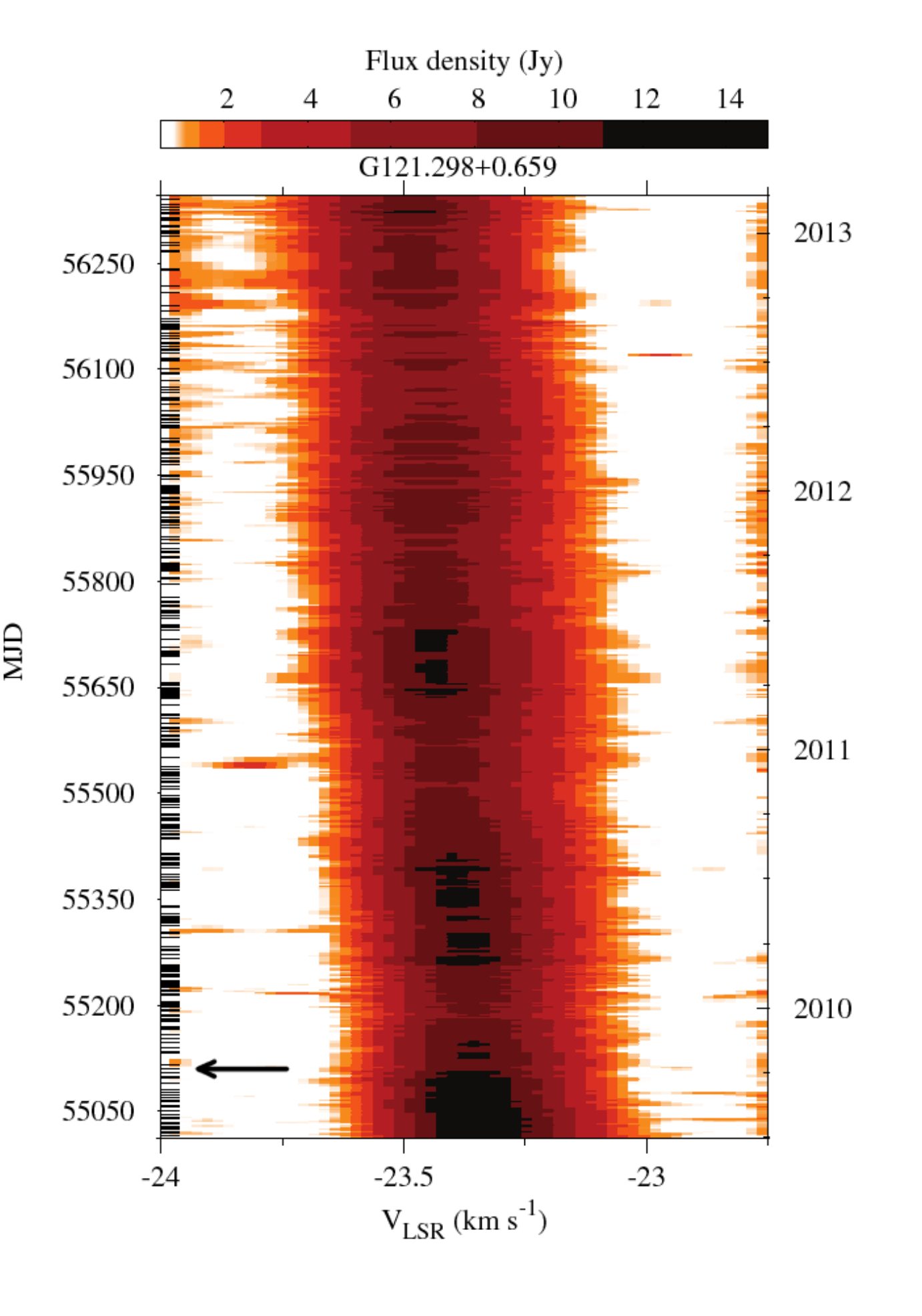}
\contcaption{
\label{fig:continued}}
\end{figure}
\end{appendix}


\end{document}